%% file: Cours_Cemracs_100726.tex
\documentclass[12pt]{article}
\usepackage[latin1]{inputenc}
\usepackage{color,graphicx,graphics,psfrag}
\usepackage{amsmath,amstext,amssymb,amsfonts,amscd}
\usepackage{subfigure}

\def\Box{\leavevmode\vbox{\hrule
     \hbox{\vrule\kern4pt\vbox{\kern4pt}%
           \vrule}\hrule}}
\def\blackbox{\leavevmode\vrule height 5pt width 4pt depth 0pt\relax}
\def\endproof{\null\hfill {$\blackbox$}\bigskip}

\newcounter{appendix}
\def\appendix{\advance\c@appendix by 1
   \def\thesection{\Alph{section}}
   \ifnum\c@appendix=1 \setcounter{section}{-1} \fi
   \@startsection {section}{1}{\z@}{-3.5ex plus -1ex minus
   -.2ex}{2.3ex plus .2ex}{\Large\bf}}

\def\paragraph#1{{\bf #1\ }}

\newtheorem{lemma}{Lemma}[section]

\newtheorem{definition}[lemma]{Definition}

\newtheorem{proposition}[lemma]{Proposition}

\newtheorem{remark}{Remark}[section]
\title{Asymptotic-Preserving Schemes for Fluid Models of Plasmas}
\author{P. Degond }
\date{}
\begin{document}

\maketitle

\vspace{0.5 cm}

\begin{center}
1-Université de Toulouse; UPS, INSA, UT1, UTM ;\\ 
Institut de Mathématiques de Toulouse ; \\
F-31062 Toulouse, France. \\
2-CNRS; Institut de Mathématiques de Toulouse UMR 5219 ;\\ 
F-31062 Toulouse, France.\\
email: pierre.degond@math.univ-toulouse.fr
\end{center}

\vspace{0.5 cm}
\begin{abstract}
These notes summarize a series of works related to the numerical approximation of plasma fluid problems. We construct so-called 'Asymptotic-Preserving' schemes which are valid for a large range of values (from very small to order unity) of the dimensionless parameters that appear in plasma fluid models. Specifically, we are interested in two parameters, the scaled Debye length which quantifies how close to quasi-neutrality the plasma is, and the scaled cyclotron period, which is inversely proportional to the magnetic field strength. We will largely focus on the ideas, in order to enable the reader to apply these concepts to other situations. 
\end{abstract}


\vskip 0.4cm

%%%%%%%%%%%%%%%%%%%%%%%%%%%%%%%%%%%%%%%%%%%%%%%%%%%%%%%%%%%%%%%%%%%%%%%%%%%%%%%%%%%%%%%%%%%%%%%%
%%%%%%%%%%%%%%%%%%%%%%%%%%%%%%%%%%%%%%%%%%%%%%%%%%%%%%%%%%%%%%%%%%%%%%%%%%%%%%%%%%%%%%%%%%%%%%%%
%%%%%%%%%%%%%%%%%%%%%%%%%%%%%%%%%%%%%%%%%%%%%%%%%%%%%%%%%%%%%%%%%%%%%%%%%%%%%%%%%%%%%%%%%%%%%%%%
%%%%%%%%%%%%%%%%%%%%%%%%%%%%%%%%%%%%%%%%%%%%%%%%%%%%%%%%%%%%%%%%%%%%%%%%%%%%%%%%%%%%%%%%%%%%%%%%

\pagebreak

$\mbox{ }$
\vspace{3cm}

\begin{center}
{\huge \bf Foreword}
\end{center}

\input{foreword}

%%%%%%%%%%%%%%%%%%%%%%%%%%%%%%%%%%%%%%%%%%%%%%%%%%%%%%%%%%%%%%%%%%%%%%%%%%%%%%%%%%%%%%%%%%%%%%%%
%%%%%%%%%%%%%%%%%%%%%%%%%%%%%%%%%%%%%%%%%%%%%%%%%%%%%%%%%%%%%%%%%%%%%%%%%%%%%%%%%%%%%%%%%%%%%%%%
%%%%%%%%%%%%%%%%%%%%%%%%%%%%%%%%%%%%%%%%%%%%%%%%%%%%%%%%%%%%%%%%%%%%%%%%%%%%%%%%%%%%%%%%%%%%%%%%
%%%%%%%%%%%%%%%%%%%%%%%%%%%%%%%%%%%%%%%%%%%%%%%%%%%%%%%%%%%%%%%%%%%%%%%%%%%%%%%%%%%%%%%%%%%%%%%%

\pagebreak

$\mbox{ }$
\vspace{2cm}

\begin{center}
{\huge \bf Introduction}
\end{center}

\vspace{1cm}

\input{intro}

%%%%%%%%%%%%%%%%%%%%%%%%%%%%%%%%%%%%%%%%%%%%%%%%%%%%%%%%%%%%%%%%%%%%%%%%%%%%%%%%%%%%%%%%%%%%%%%%
%%%%%%%%%%%%%%%%%%%%%%%%%%%%%%%%%%%%%%%%%%%%%%%%%%%%%%%%%%%%%%%%%%%%%%%%%%%%%%%%%%%%%%%%%%%%%%%%
%%%%%%%%%%%%%%%%%%%%%%%%%%%%%%%%%%%%%%%%%%%%%%%%%%%%%%%%%%%%%%%%%%%%%%%%%%%%%%%%%%%%%%%%%%%%%%%%
%%%%%%%%%%%%%%%%%%%%%%%%%%%%%%%%%%%%%%%%%%%%%%%%%%%%%%%%%%%%%%%%%%%%%%%%%%%%%%%%%%%%%%%%%%%%%%%%

\pagebreak

$\mbox{ }$
\vspace{5cm}

\begin{center}
{\huge \bf Part 1 }

\vspace{1.5cm}
{\huge \bf Quasineutrality}
\end{center}
\pagebreak

%
\input{quasineutrality}

%%%%%%%%%%%%%%%%%%%%%%%%%%%%%%%%%%%%%%%%%%%%%%%%%%%%%%%%%%%%%%%%%%%%%%%%%%%%%%%%%%%%%%%%%%%%%%%%
%%%%%%%%%%%%%%%%%%%%%%%%%%%%%%%%%%%%%%%%%%%%%%%%%%%%%%%%%%%%%%%%%%%%%%%%%%%%%%%%%%%%%%%%%%%%%%%%
%%%%%%%%%%%%%%%%%%%%%%%%%%%%%%%%%%%%%%%%%%%%%%%%%%%%%%%%%%%%%%%%%%%%%%%%%%%%%%%%%%%%%%%%%%%%%%%%
%%%%%%%%%%%%%%%%%%%%%%%%%%%%%%%%%%%%%%%%%%%%%%%%%%%%%%%%%%%%%%%%%%%%%%%%%%%%%%%%%%%%%%%%%%%%%%%%

\pagebreak

$\mbox{ }$
\vspace{5cm}

\begin{center}
{\huge \bf Part 2 }

\vspace{1.5cm}
{\huge \bf Large magnetic fields}
\end{center}
\pagebreak

%

\input{drift}

%
\input{aniso}

%%%%%%%%%%%%%%%%%%%%%%%%%%%%%%%%%%%%%%%%%%%%%%%%%%%%%%%%%%%%%%%%%%%%%%%%%%%%%%%%%%%%%%%%%%%%%%%%
%%%%%%%%%%%%%%%%%%%%%%%%%%%%%%%%%%%%%%%%%%%%%%%%%%%%%%%%%%%%%%%%%%%%%%%%%%%%%%%%%%%%%%%%%%%%%%%%
%%%%%%%%%%%%%%%%%%%%%%%%%%%%%%%%%%%%%%%%%%%%%%%%%%%%%%%%%%%%%%%%%%%%%%%%%%%%%%%%%%%%%%%%%%%%%%%%
%%%%%%%%%%%%%%%%%%%%%%%%%%%%%%%%%%%%%%%%%%%%%%%%%%%%%%%%%%%%%%%%%%%%%%%%%%%%%%%%%%%%%%%%%%%%%%%%

\pagebreak

$\mbox{ }$
\vspace{2cm}

\begin{center}
{\huge \bf Conclusion}
\end{center}

\vspace{1cm}

%
\input{conclu}

%%%%%%%%%%%%%%%%%%%%%%%%%%%%%%%%%%%%%%%%%%%%%%%%%%%%%%%%%%%%%%%%%%%%%%%%%%%%%%%%%%%%%%%%%%%%%%%%
%%%%%%%%%%%%%%%%%%%%%%%%%%%%%%%%%%%%%%%%%%%%%%%%%%%%%%%%%%%%%%%%%%%%%%%%%%%%%%%%%%%%%%%%%%%%%%%%
%%%%%%%%%%%%%%%%%%%%%%%%%%%%%%%%%%%%%%%%%%%%%%%%%%%%%%%%%%%%%%%%%%%%%%%%%%%%%%%%%%%%%%%%%%%%%%%%
%%%%%%%%%%%%%%%%%%%%%%%%%%%%%%%%%%%%%%%%%%%%%%%%%%%%%%%%%%%%%%%%%%%%%%%%%%%%%%%%%%%%%%%%%%%%%%%%

\pagebreak

%

\input{biblio_Cemracs}
%%%%%%%%%%%%%%%%%%%%%%%%%%%%%%%%%%%%%%%%%%%%%%%%%%%%%%%%%%%%%%%%
%%%%%%%%%%%%%%%%%%%%%%%%%%%%%%%%%%%%%%%%%%%%%%%%%%%%%%%%%%%%%%%%
\end{document}

%% file: foreword.tex
\vspace{2cm}
The material of these notes is the product of a research programme which has extended over several years and has involved a large number of collaborations. I would like to address special thanks to my collaborators Fabrice Deluzet, Giacomo Dimarco, Alexei Lozinski,  Marie-Hélène Vignal (Institut de Mathématiques de Toulouse), Stéphane Brull (Institut de Mathématiques de Bordeaux), N. Crouseilles, Eric Sonnedr\"ucker (IRMA, Strasbourg), J-G. Liu (Duke University), C. Negulescu (LATP, Marseille) and my current or former students and post-docs, Céline Parzani, Pierre Crispel, Jacek Narski, Laurent Navoret, Alexandre Mouton, Dominique Savelief (Institut de Mathématiques de Toulouse), Sever Hirstoaga (IRMA, Strasbourg), Afeintou Sangam (Laboratoire J-A. Dieudonné, Nice), An-Bang Sun (X'ian University) and Min Tang (Laboratoire J-L. Lions, Paris). 

\vspace{1cm}
The problems and questions which are dealt with in these notes have been strongly inspired by many years of collaboration and contracts with the Commissariat à l'Energie Atomique or the Centre National d'Etudes Spatiales, and I would like to thank Gérard Gallice, Jean Ovadia, Christian Tessieras (CEA-Cesta), Annalisa Ambroso (now at the company Areva), Anela Kumbaro, Pascal Omnes, Jacques Segré (CEA-Saclay), Gloria Falchetto, Xavier Garbet, Maurizio Ottaviani (CEA-Cadarache), Kim-Claire Le Than (CEA-DIF), Jacques Payan (CNES), as well as Franck Assous. I would also like to acknowledge support from the Fondation RTRA 'Sciences et Technologies Aéronautiques et Spatiales', under the 'Plasmax project', led by Florent Christophe (ONERA-Toulouse).

%% file: intro.tex
\section*{Numerical resolution of perturbation problems}
%\label{subsec_intro_num_res_perturb}

These notes are about the application of the 'Asymptotic-Preserving' methodology to construct schemes for plasma fluids problem which sustain large variations of some of the characteristic dimensionless parameters of the plasma. We will be specifically concerned with two of these parameters, the scaled Debye length on the one hand and the scaled cyclotron period on the other hand. The former quantifies how close to quasi-neutrality the plasma is while the latter measures the confinement effects due to the magnetic field.  

Let us consider a singular perturbation problem $P^\varepsilon$ whose solutions converge to those of a limit problem $P^0$ when the perturbation parameter $\varepsilon$ tends to zero. Usually, when $\varepsilon \ll 1$, standard numerical methods (like e.g. explicit methods in the case of time-dependent problems) break down. The reason is that the stability condition limits the allowed time step to a maximal value which depends on $\varepsilon$ and tends to zero when $\varepsilon \to 0$. In the case of hyperbolic problems, this problem occurs if one of the wave-speeds tends to infinity with $\varepsilon$. For instance, in the low Mach-number limit of compressible fluids, the acoustic wave speeds tend to infinity when the Mach-number tends to zero. 

In order to overcome such problems, the usual strategy consists in solving the limit problem $P^0$ instead of 
$P^\varepsilon$. For instance, in the case of the Low Mach-number limit, the incompressible Euler or Navier-Stokes equations will be solved. However, there are several difficulties with this strategy, which we now outline. The first one is that it supposes that $P^0$ has been previously determined and the second one is that it assumes that $P^0$ easy to solve. Both assumptions are by no means obvious. There are cases where the determination of the limit problem is difficult if not impossible. Even if $P^0$ is well-known, it usually involves equations of mixed type (for instance, the small Mach-number flow model involves a combination of equations of hyperbolic and elliptic character), where constraints (such as the divergence free constraint on the velocity) need to be enforced. The abundant literature on the Stokes and Navier-Stokes problem shows that enforcing this constraint is not an easy problem. 

The problem complexifies even more when the parameter $\varepsilon$ is not uniformly small. This sentence may sound a little awkward, since $\varepsilon$ is a number which should have a definite fixed value. However, $\varepsilon$ is usually a ratio of characteristic lengths which may vary in space and time. For instance, in a plasma sheath, where the density drops by orders of magnitude, the value of the Debye length changes dramatically. In other cases, such as in boundary layers, one must change the scaling length from say the size of the experiment, to the typical dimension of the boundary layer. Therefore, the definition of a uniform value for $\varepsilon$ is difficult, and it is more appropriate to view $\varepsilon$ as a local quantity. 

In such a situation, $\varepsilon$ may be small in some areas and order unity in other regions. Then, the use of the limit problem $P^0$ leads to wrong results in the regions where $\varepsilon$ is not small. To make a more accurate simulation, it is necessary to decompose the simulation domain into regions where $\varepsilon = O(1)$ and regions where $\varepsilon \ll 1$ and to solve $P^\varepsilon$ in the former and $P^0$ in the latter. However, the practical realization of this coupling strategy is very complex. 

Indeed, first, it requires to define the location where the transition from $P^\varepsilon$ to $P^0$ or vice-versa takes place. This is not an obvious question when $\varepsilon$ has a smooth rather than abrupt transition. The results then depend on the particular location of this transition. This drawback can be slightly circumvented by the use of a smooth transition (like fictitious mixture models in multiphase flows for instance). Nonetheless, the need to designate a specific region where the shift between the two models takes place is detrimental to the robustness and the reliability of the model.

Once the location of the interface or transition region between $P^\varepsilon$ and $P^0$ has been determined, the second problem to solve is the definition of the coupling strategy between the two models. Usually, $P^0$ involves some kind of reduction of information compared to $P^\varepsilon$. For instance, in the low-Mach number limit, the velocity field becomes divergence free, meaning that it depends on two independent scalar quantities instead of three like in compressible flows. At the interface, in the passage $P^\varepsilon \to P^0$, it is necessary to project the unknowns of $P^\varepsilon$ onto those of $P^0$, and vice versa, in the passage $P^0 \to P^\varepsilon$ the unknowns of $P^\varepsilon$ must be reconstructed from those of $P^0$. For instance, in the passage from compressible to low Mach-number regimes, the velocity must be projected onto a divergence free field. In the reverse transition, the irrotational part of the velocity must be reconstructed from a divergence free velocity.  

The answer to such questions is by no means obvious and the simulation results also depend on the choices of the projection-reconstruction operators. Connection conditions can be sought by solving an interior layer problem connecting the state variables of the $P^\varepsilon$ problem at $x=-\infty$ to those of the $P^0$ problem at $x = +\infty$ through a spatial rescaling of $P^\varepsilon$  in the direction normal to the interface. However, quite often, this analysis does not lead to a closed set of connection conditions. The interior layer problem itself carries some approximations because it involves a rescaling which leads to neglecting all derivatives in the tangential direction to the interface. 

Finally, supposing that the questions above have found a satisfactory answer, the whole strategy must be practically implemented. The mesh must be constrained to match the interface. Therefore, if the interface is not planar, which is very likely in a realistic case, an unstructured mesh must be used. Additionally, the interface location may have to evolve in time. This brings several additional questions. First, the motion strategy for the interface must be defined. What criterion will decide for this motion ? Second, if the interface is moved, the mesh must be moved accordingly, otherwise the matching with the interface will be lost. Moving meshes with time is complex, uneasy and costly, both in terms of CPU time and of development time.   

All these questions have led an increasing number of teams to look for schemes which are valid in both the $\varepsilon =O(1)$ and $\varepsilon \ll 1$ regimes. For instance, there is an increasing literature about so-called 'all-speed schemes' for compressible flows, which are valid for all values of the Mach number \cite{Bra_Nko_JCP09, Deg_Jin_Liu, Deg_Tan, Har_Ams_JCP71}. However, some precautions must be made because, a scheme may be stable in the $\varepsilon \ll 1$ limit and yet provide a wrong solution, i.e. a solution which is not consistent with the $P^0$ problem. The correct concept for doing so is the so-called Asymptotic-Preserving (AP) scheme, which is described in the next section.

\section*{Asymptotic-Preserving methodology}
%\label{subsec_intro_AP_method}

The concept of an Asymptotic-Preserving method has been introduced by S. Jin \cite{Jin} for transport in diffusive regimes. 
A scheme $P^{\varepsilon,h}$ for $P^\varepsilon$ with discretization parameters $h$ (standing for both the time and space steps) is called Asymptotic Preserving (or AP) if it is stable irrespective of how small the perturbation parameter $\varepsilon$ is, and if it leads to a scheme $P^{0,h}$ which is consistent with the limit problem $P^0$ when $\varepsilon$ tends to zero with fixed discretization parameters $h$. This property is illustrated in the commutative diagram below.

\medskip
$$ \begin{CD} 
P^{\varepsilon,h} @>{h \to 0}>>  P^{\varepsilon} \\
@VV{\varepsilon \to 0}V  @VV{\varepsilon \to 0}V\\
P^{0,h} @>{h \to 0}>> P^{0} 
\end{CD}
$$

\bigskip
AP schemes are extremely powerful tools as they allow the use of the same scheme to
discretize $P^\varepsilon$ and $P^0$, with fixed discretization parameters. This not possible if the AP property is not satisfied because, either the scheme develops an instability when the microscopic scale is under-resolved (i.e. when $\varepsilon$ is too small), or it is stable but not consistent with the limit problem $P^0$ and therefore, provides a wrong solution. By contrast, when the order of magnitude of $\varepsilon$ changes dramatically, the AP scheme realizes an automatic transition between $P^\varepsilon$ and $P^0$. There is no model change, no need to define a coupling strategy, an interface location, or divising complicated adaptive meshing strategies to follow the interface. Thanks to these properties, AP-schemes are extremely efficient, producing computational saving of several orders of magnitude. 

The literature about AP schemes is recent, yet increasingly abundant and applied to various contexts (see e.g.  \cite{Ben_Lem_Mie_JCP08, Bue_Cor_NumerMath07, Bue_Cor_M2AN02, Buet_Despres_JCP06, Car_Gou_Laf_JCP08, Fil_Jin_JCP10, Gos_Tos_NumerMath04, Klar_SINUM99, Lem_Mie_SISC08, McC_Low_JCP08,  Sea_Kla_JCP06}). 

In this document, we will illustrate a general methodology for devising AP schemes. This methodology is divided into two main steps. The first step is a 'Reformulation' step. Indeed, the passage $P^\varepsilon \to P^0$ leads to a change in the type, nature or simply expression of the equations which determine some of the unknowns. The reformulation step consists in finding an equivalent form $R^\varepsilon$ of problem $P^\varepsilon$ such that $R^\varepsilon$ explicitly appears as a perturbation of $P^0$. This concept may sound vague but we hope that the various illustrations below will convince the reader of its relevance. To some extent, the reformulation procedure consists in bringing the limit model $P^0$ into the model $P^\varepsilon$ 'by brute force'. 

Once the Reformulation step is passed, the Discretization step per se can start. The most obvious idea, which is to discretize the reformulated model $R^\varepsilon$ should actually be discarded. The reason is that $R^\varepsilon$ is usually quite complicated and it is not clear what a 'good' discretization of it is. Additionally, there are some structural properties which should be preserved by the discretization. For systems of conservation laws for instance, these properties could be a special conservative form or a special formula for the numerical viscosities. It is not clear at all how to enforce these properties on a direct discretization of $R^\varepsilon$. 

So, the discretization step proceeds as follows: Discretize $P^\varepsilon$ (and not $R^\varepsilon$) into a scheme $P^{\varepsilon,h}$ in such a way that the various manipulations which led to $R^\varepsilon$ in the continuous case can be performed at the discrete level. Then, the reformulation procedure, performed at the discrete level, permits the derivation of a reformulated scheme $R^{\varepsilon,h}$ which is consistent with the reformulated continuous model $R^\varepsilon$. Then, the scheme $R^{\varepsilon,h}$ appears as a perturbation of a scheme $R^{0,h}$
which is consistent with the limit problem $P^0$, in just the same way as the reformulated model  $R^\varepsilon$ appears as a perturbation of $P^0$. By this procedure, in the case of conservation laws, additional properties such as conservativity, or special choices of numerical viscosities can be imposed on $P^{\varepsilon,h}$ and can be carried over to $R^{\varepsilon,h}$ by the discrete reformulation step. The discrete reformulation step also leads to a form which is the most suited for numerical resolution. 

In other words, it is preferable to 'reformulate the discretization', rather than 'discretize the reformulation'. We will illustrate this motto on different examples throughout this document. 

We now make some comments. The reformulation procedure requires that the limit problem $P^0$ is well identified and well-posed. In the current state-of-the-art, there is no possibility of deriving AP schemes for models $P^\varepsilon$ whose limit $P^0$ is not well-identified and well-posed. 

The second remark is that, for the scheme $P^{\varepsilon,h}$ to be 'reformulable' into $R^{\varepsilon,h}$ and to have a limit $R^{0,h}$ when $\varepsilon \to 0$, some level of implicitness is required. The design of an AP scheme is just about determining which terms have to be evaluated implicitly in order to meet the requirements of the AP property. Sometimes, there is not a unique way to perform this goal (we will see an example below). In all cases, this key point is the most important and difficult part of the procedure. For this reason, in most of the cases, constructing an AP scheme is above-all a question of time discretization. This is why, in these notes, we will systematically devise the AP methodology first in the time semi-discrete setting, before applying it to the fully-discrete case.  

Then, the question is: why not choosing all terms fully implicit. Then, we are guaranteed to make an AP scheme. There are several issues. First, this is doing too much, because requesting a scheme to be AP is not asking for inconditional stability, but only for uniform stability with respect to a given parameter, which is weaker. The fully implicit scheme must be inverted. This requires an iterative procedure (because the problem is usually nonlinear) inside which a large linear system solver must be called. For systems of conservation laws, the linear operator to invert is issued from a first order differential system. Therefore, it is generally not positive definite and its inversion can be quite unstable. 

By contrast, in designing AP schemes it is usually possible to reduce the inversion of the implicit part to the inversion of an elliptic operator. This leads to a much more stable numerical resolution. In the case of systems of conservation laws in which the limit $\varepsilon \to 0$ corresponds to some characteristic speeds going to infinity, this amounts to solving a wave equation (for the fields associated to these characteristic speeds) by a fully implicit scheme. Therefore, the computation of the next time iterate involves the inversion of an elliptic equation corresponding to the stationary wave equation. 

When dealing with a fully implicit scheme, the iterative procedure needed to invert the nonlinear system is usually a Newton method, which is based on a linearization procedure. When the limit $\varepsilon \to 0$ corresponds to the enforcement of some nonlinear constraints, the linearization procedure may destroy the AP character of the scheme. This is the case for instance in the low Mach-number limit which is characterized by the condition that the gradient of the pressure should vanish. For real gases, the pressure is a nonlinear function of the state variables. Linearizing this condition by the Newton method may lead to large errors. Of course, if full convergence of the Newton iteration is reached, the scheme is AP. But this is never the case. The stopping criteria consist in checking if the residual of the iterations is below some threshold. But in the limit $\varepsilon \to 0$, the conditionning of the problem is very bad and checking on the residual does not guarantee a reasonable convergence level for the solution. Then, the solution is not only bad,  but also inconsistent with the nonlinear constraint characterizing the limit regime, because of the linearization in the Newton iterations. 

We hope that this discussion has been convincing that fully implicit schemes are not 'the' ultimate method for perturbation problems. Of course, fully implicit scheme may be valuable, but they have to be implemented with iterative methods which guarantee the AP character of the scheme even if the iterations are stopped before convergence. In our experience, going from an AP scheme to a fully implicit scheme only brings minor improvements in general.

\section*{Plan of these notes}
%\label{subsec_intro_AP_method}

In these notes, we plan to develop the AP methodologies on two different classes of asymptotic limits.

Part 1 will be devoted to the quasi-neutral limit in plasma fluid models. We will successively consider the one-fluid isentropic Euler-Poisson model and the one-fluid isentropic Euler-Maxwell models. The extension to two or more species, to full Euler models instead of isentropic models, and finally, to other variants of the quasi-neutral limit such as the case of the Euler-Poisson-Boltzmann model are briefly discussed at the end. The material of these two sections has been (or will be) published in \cite{Cri_Deg_Vig_CRAS05, Cri_Deg_Vig_07, Deg_Del_Sav_prep, Deg_Del_Liu_Sav_Vig_prep, Deg_Liu_Vig_08, Vignal_SIAP10}. The application of this strategy to kinetic models (Vlasov-Poisson equations) has been published in \cite{Bel_Cro_Deg_JSC09, Deg_Del_Nav_CRAS06, Deg_Del_Nav_JCP10}. The study of the transition from quasineutrality to non-quasineutrality in plasmas has been investigated in \cite{Cri_Deg_Vig_JCP05, Cri_Deg_M3AS_07, Deg_Par_Vig_MCM03, Deg_Par_Vig_MMS03, Let_Par_Vig_JCP07}, as well as in \cite{Fra_Ock_JPP70, Ha_Sle_CMP03, Rie_Dau_JAP99, Slemrod_JNLS01}. The quasi-neutral limit has been theoretically investigated in \cite{Cor_Gre_CPDE00, Gas_Hsi_Mar_Wan_JMAA02, Jun_Pen_AA01, Peng_M2AN01, Sle_Ste_JNLS01}.

Part 2 will be devoted to the drift-fluid limit in the Euler-Lorentz system, i.e. the isentropic Euler system of gas dynamics equations subjected to a Lorentz force. The scaling is such that the whole Lorentz force and the pressure force are large compared to the inertial force. In this case, we will show that two AP strategies can be divised. Both lead to the resolution of a strongly anisotropic diffusion operator corresponding to diffusion along the magnetic field lines. The material of this section has been published in \cite{Deg_Del_San_JCP09} and is the subject of works in preparation \cite{Bru_Deg_Del_prep, Deg_Del_Loz_Nar_Neg_prep2}. Related material for more general anisotropic diffusion equations or for the full Euler-Lorentz case is or will be available in \cite{Deg_Del_Neg_MMS10, Deg_Del_Loz_Nar_Neg_prep1, Bru_Del_Mou_prep, Deg_Del_Mou_prep}. The drift-fluid limit is the cornerstone of many physical models, see e.g. \cite{Dorland_Hammett_PhysfluidsB93, Falchetto_Ottaviani_PRL04, Garbet_PhPl01, Hammett_Dorland_PPCF93}. We refer the reader to the monograph \cite{Hazeltine_Meiss} for the physics of magnetic plasma confinement. The mathematical investigation of this limit is to the best of our knowledge, open.

In these notes, we focus on the concepts and ideas. We refer the reader to the bibliography given above for applications to practical cases and performance tests. All performance tests indicate that these schemes are more efficient than standard schemes by several orders of magnitude (for instance, a factor $10^4$ has been reached in the case of the drift-fluid limit, see \cite{Deg_Del_San_JCP09}).

%% file: quasineutrality.tex
%%%%%%%%%%%%%%%%%%%%%%%%%%%%%%%%%%%%%%%%%%%%%%%%%%%%%%%%%%%%%%%%%%%%%%%%%%%%%%%%%%%%%%%%%%%%%%%%
%%%%%%%%%%%%%%%%%%%%%%%%%%%%%%%%%%%%%%%%%%%%%%%%%%%%%%%%%%%%%%%%%%%%%%%%%%%%%%%%%%%%%%%%%%%%%%%%
%%%%%%%%%%%%%%%%%%%%%%%%%%%%%%%%%%%%%%%%%%%%%%%%%%%%%%%%%%%%%%%%%%%%%%%%%%%%%%%%%%%%%%%%%%%%%%%%
%%%%%%%%%%%%%%%%%%%%%%%%%%%%%%%%%%%%%%%%%%%%%%%%%%%%%%%%%%%%%%%%%%%%%%%%%%%%%%%%%%%%%%%%%%%%%%%%

%%%%%%%%%%%%%%%%%%%%%%%%%%%%%%%%%%%%%%%%%%%%%%%%%%%%%%%%%%%%%%%%%%%%%%%%%%%%%%%%%%%%%%%%%%%%%%%%
%%%%%%%%%%%%%%%%%%%%%%%%%%%%%%%%%%%%%%%%%%%%%%%%%%%%%%%%%%%%%%%%%%%%%%%%%%%%%%%%%%%%%%%%%%%%%%%%
\setcounter{equation}{0}
\section{The Euler-Poisson system}
\label{sec_QN_EP_1F}

%%%%%%%%%%%%%%%%%%%%%%%%%%%%%%%%%%%%%%%%%%%%%%%%%%%%%%%%%%%%%%%%%%%%%%%%%%%%%%%%%%%%%%%%%%%%%%%%
\subsection{Setting of the problem}
\label{subsec_QN_EP_1F_setting}

The One-Fluid Euler-Poisson (EP) model describes the plasma electrons through a system of isothermal or isentropic gas dynamics equations subjected to an electrostatic force. The electrostatic potential is related to the electron density through the Poisson equation. The ions are supposed to form a neutralizing background, i.e. they are steady and with uniform density. The EP model is written 
\begin{eqnarray}
& & \hspace{-1cm} \partial_t n + \nabla \cdot (nu) = 0, \label{1FEP_n} \\
& & \hspace{-1cm} m ( \partial_t (nu) + \nabla \cdot (nu \otimes u)) + \nabla p= e n \nabla \phi, \label{1FEP_u} \\
& & \hspace{-1cm} - \Delta \phi = \frac{e}{{\epsilon}_0} (n_i - n). \label{1FEP_phi}
\end{eqnarray}
Here, $n(x,t) \geq 0$, $u(x,t) \in {\mathbb R}^d $ and $\phi(x,t)\in {\mathbb R} $ stand for the electron density, electron velocity and electric potential respectively, and depend on the space-variable $x \in {\mathbb R}^d$ and on the time $t\geq0$. Strictly speaking, the ion density $n_i(x,t)$ itself satisfies a system of Euler equations. However, in this section, we suppose it uniform and constant in time for simplicity and ignore the question of the coupling of the electrons to the ions (see section \ref{sec_QN_extension}). The positive elementary charge is denoted by $e$ and the electron mass, by $m$. The electron pressure $p$ is supposed to be a given function of the density (e.g. $p(n) = k_B T n$ in the isothermal case, where $T$ is the electron temperature and $k_B$, the Boltzmann constant, or $p(n) = C n^\gamma$, where $\gamma >1$ and $C>0$ are given constants, in the isentropic case). $\epsilon_0$ refers to the vacuum permittivity. The operators $\nabla$, $\nabla \cdot $ and $\Delta$ are respectively the gradient, divergence and Laplace operators and $u \otimes u$ denotes the tensor product of the vector $u$ with itself. Finally, $d$ is the dimension ($d = 1, 2$, or $3$).

If the quasi-neutral assumption is made, the Poisson equation (\ref{1FEP_phi}) is replaced by the constraint of zero local charge: 
\begin{eqnarray}
& & \hspace{-1cm} n = n_i , \label{EP_qn}
\end{eqnarray}
In this context, the quasi-neutral Euler-Poisson model coincides with the incompressible Euler (IE) model: 
\begin{eqnarray}
& & \hspace{-1cm} \nabla \cdot u = 0, \label{IE_n} \\
& & \hspace{-1cm} \partial_t u + (u \cdot \nabla) u + \nabla \pi = 0 , \label{IE_u} 
\end{eqnarray}
together with (\ref{EP_qn}), with the hydrostatic pressure $\pi = - e m^{-1} \phi$. In deriving (\ref{IE_u}), we have used that for $u$ satisfying (\ref{IE_n}), $\nabla (u \otimes u) = (u \cdot \nabla) u$. 

The passage from EP to IE can be understood by a suitable scaling of the model, which highlights the role of the scaled Debye length:
\begin{eqnarray}
& & \hspace{-1cm} \lambda = \frac{\lambda_D}{x_0}, \quad \lambda_D = \left( \frac{\epsilon_0 k_B T}{e^2 n^*} \right)^{1/2}, \label{Debye} 
\end{eqnarray}
where $x_0$ is the typical size of the system under consideration. $\lambda_D$ measures the spatial scale associated with the electrostatic interaction between the particles. The dimensionless parameter $\lambda$ is usually small, which formalizes the fact that the electrostatic interaction occurs at spatial scales which are much smaller than the usual scales of interest. However, there are situations, for instance in boundary layers, or at the plasma-vacuum interface, where the electrostatic interaction scale must be taken into account. This means that the choice of the relevant macroscopic length $x_0$ may depend on the location inside the system and that in general, the parameter $\lambda$ may vary by orders of magnitude from one part of the domain to another one.

%%%%%%%%%%%%%%%%%%%%%%%%%%%%%%%%%%%%%%%%%%%%%%%%%%%%%%%%%%
\subsection{Scaling of the EP model and quasineutral limit}
\label{subsec_QN_EP_1F_scaling}

Let $x_0$, $t_0$, $u_0$, $p_0$, $\phi_0$ and $n_0$ be space, time, velocity, pressure, potential and density scales. Scaled position, time, velocity, pressure, potential and density are defined by $ \bar{x} = x/x_0$, $ \bar{t} = t/t_0$, $ \bar{u} = u/u_0$, $ \bar{p} (\bar n)= p(n)/p_0$, $ \bar{\phi} = - \phi/\phi_0$ and  $ \bar{n} = n/n_0$. We choose $x_0$ to be the typical size of the system (for instance an inter-electrode distance or the size of the vacuum chamber). We also choose $n_0=n_i$. We define a temperature scale by the relation $p_0 = n_0 k_B T_0$ and define the velocity scale as $u_0 = (k_B T_0 m^{-1})^{1/2}$. Finally, the potential scale is set to $\phi_0 = k_B T_0 /e$, the so-called thermal potential. 

Inserting this scaling and omitting the bars gives rise to the EP model in scaled form: 

\begin{definition}
The scaled Euler-Poisson (EP) model is:
\begin{eqnarray}
& & \hspace{-1cm}  \partial_t n^\lambda + \nabla \cdot ( n^\lambda u^\lambda )  = 0,  \label{S_1FEP_n}
\\
& & \hspace{-1cm}   \partial_t ( n^\lambda u^\lambda ) +  \nabla \cdot ( n^\lambda u^\lambda \otimes u^\lambda )  + \nabla p(n^\lambda) =  n^\lambda \nabla \phi^\lambda,  \label{S_1FEP_u}
\\
& & \hspace{-1cm}  - \lambda^2 \Delta \phi^\lambda  = 1 - n^\lambda. \label{S_1FEP_phi}
\end{eqnarray}
where $\lambda$ is the scaled Debye length (\ref{Debye}). 
\label{def_EP}
\end{definition}

\noindent
Formally passing to the limit $ \lambda \to 0 $ in this system and supposing that $n^\lambda \to n^0$, $u^\lambda \to u^0$, $\phi^\lambda \to \phi^0$ as smoothly as needed, we are led to the Icompressible Euler (IE) model: 

\begin{definition} 
The scaled Incompressible Euler (IE) model is: 
\begin{eqnarray}
& & \hspace{-1cm} \nabla \cdot u^0   = 0, \label{S_IE_n} \\
& & \hspace{-1cm}  \partial_t u^0 + ( u^0 \cdot \nabla ) u^0  =  \nabla \phi^0,  \label{S_IE_u} \\
& & \hspace{-1cm}  n^0 = 1.  \label{S_IE_phi} 
\end{eqnarray}
\label{def_IE}
\end{definition}

\noindent
Indeed, the first two equations are the scaled form of the Icompressible Euler (IE) model where the potential $\phi^0$ is the negative hydrostatic pressure and acts as the Lagrange multiplier of the incompressibility constraint (\ref{S_IE_n}). In order to resolve the constraint and find an equation for $\phi^0$, we take the divergence of (\ref{S_IE_u}) and insert the constraint (\ref{S_IE_n}). We find: 
\begin{eqnarray}
& & \hspace{-1cm}  \Delta \phi^0 = \nabla^2 : ( u^0 \otimes u^0)  .  \label{S_IE_phi_2} 
\end{eqnarray}
This equation is equivalent to (\ref{S_IE_n}) provided that (\ref{S_IE_n}) is satisfied initially. Therefore, we state:

\begin{definition} 
The Reformulated Incompressible Euler (RIE) model is: 
\begin{eqnarray}
& & \hspace{-1cm}  n^0 = 1.  \label{S_IE_phi_3} \\
& & \hspace{-1cm}  \partial_t u^0 + ( u^0 \cdot \nabla ) u^0  =  \nabla \phi^0,  \label{S_IE_u_2} \\
& & \hspace{-1cm}  \Delta \phi^0 = \nabla^2 : ( u^0 \otimes u^0)  .  \label{S_IE_n_2} 
\end{eqnarray}
\label{def_RIE}
\end{definition}

\begin{proposition}
The RIE model supplemented with eq. (\ref{S_IE_n}) at time $t=0$ is equivalent to the IE model. 
\label{prop_equiv_IE_RIE}
\end{proposition}

The question arises if there exists a Reformulated Euler-Poisson model in the same way as there exists a Reformulated Incompressible Euler model. Using the mass and momentum conservation equations, such a reformulation can be derived. Indeed, taking the time derivative of (\ref{S_1FEP_n}), the divergence of (\ref{S_1FEP_u}) and eliminating the second order time derivative of $n^\lambda$ by means of the Poisson equation (\ref{S_1FEP_phi}), we are led to the so-called 'Reformulated Poisson equation': 
\begin{eqnarray}
& & \hspace{-1cm}  - \nabla \cdot \left[ (n^\lambda + \lambda^2 \partial^2_t) \nabla \phi^\lambda \right] =   
- \nabla^2 : \left( n^\lambda u^\lambda \otimes u^\lambda + p(n^\lambda) \mbox{Id} \right)
,  \label{S_1FEP_phi_ref_0} 
\end{eqnarray}
We refer to the Reformulated Euler-Poisson (REP) model as the model consisting of the mass and momentum conservation equations complemented with the reformulated Poisson equation (\ref{S_1FEP_phi_ref_0}):

\begin{definition}
The Reformulated Euler-Poisson (REP) model is:
 \begin{eqnarray}
& & \hspace{-1cm}  \partial_t n^\lambda + \nabla \cdot ( n^\lambda u^\lambda )  = 0,  \label{S_1FEP_n_ref}
\\
& & \hspace{-1cm}   \partial_t ( n^\lambda u^\lambda ) +  \nabla \cdot ( n^\lambda u^\lambda \otimes u^\lambda )  + \partial_x p(n^\lambda) =  n^\lambda \nabla \phi^\lambda,  \label{S_1FEP_u_ref}
\\
& & \hspace{-1cm}  - \nabla \cdot \left[ (n^\lambda + \lambda^2 \partial^2_t) \nabla \phi^\lambda \right] =   
- \nabla^2 : \left( n^\lambda u^\lambda \otimes u^\lambda + p(n^\lambda) \mbox{Id} \right)
.  \label{S_1FEP_phi_ref} 
\end{eqnarray}
\end{definition}

\noindent
A solution of the REP model is a solution of the original EP model if and only if the initial data satisfy the Poisson equation in its original form (\ref{S_1FEP_phi}) together with its time derivative. Indeed, (\ref{S_1FEP_phi_ref_0}) is a second order differential equation in time and requires Cauchy data for $\phi^0$ and its time-derivative at $t=0$. The equivalence between the EP and REP models is summarized in: 

\begin{proposition} 
The REP model is equivalent to the original EP model (\ref{S_1FEP_n}), (\ref{S_1FEP_u}), (\ref{S_1FEP_phi}), provided that the Poisson equation in its original form (\ref{S_1FEP_phi}) and its time derivative are satisfied at initial time. 
\label{prop_REP}
\end{proposition}

\noindent
Note however that, at the discrete level, schemes based on the REP model may differ from those based on the EP model. It clearly appears, that (\ref{S_1FEP_phi_ref}) is a singular perturbation of (\ref{S_IE_phi_2}) when $\lambda \to 0$ and consequently that the REP model formally tends to the RIE model in this limit. Therefore, using the REP formulation appears as a good strategy to devise schemes for the EP model which are consistent with the IE model in the limit $\lambda \to 0$. In the next section, we propose a time semi-discrete scheme based on this strategy.

%%%%%%%%%%%%%%%%%%%%%%%%%%%%%%%%%%%%%%%%%%%%%%%%%%%%%%%%%%
\subsection{Time-semi-discretization and AP property}
\label{subsec_QN_EP_1F_time}

We denote by $ \delta  $ the time step. For any function $g(x,t)$, we denote by $ g^m (x) $ an approximation of $g(x,t^m)$  at time $ t^m = m \delta $. The classical time-semi-discretization is based on the EP model. We propose a new time-semi-discretization based on the REP model.

In the classical method, the force term in the momentum equation is taken implicitly. S. Fabre \cite{Fabre_JCP_101_445} has shown that this implicitness is needed for the stability of the scheme (an explicit treatment of the force term leads to an inconditionally unstable scheme). Additionally, this implicitness still gives rise to an explicit resolution, since the mass conservation can be used to update the density, then the Poisson equation is used to update the potential, and finally the resulting potential is inserted in the momentum equation to update the velocity. However, stability is subjected to a condition of the type $\delta = 0(\lambda)$ and the method breaks down in the quasineutral limit $\lambda \to 0$. 

\begin{definition}
The classical time-semi-discretization of the EP model is: 
\begin{eqnarray}
& & \hspace{-1cm}  \delta^{-1} (n^{\lambda, m+1} - n^{\lambda, m}) + 
  \nabla \cdot ( n^{\lambda, m} u^{\lambda, m} ) = 0, \label{CD_1FEP_n} \\
& & \hspace{-1cm}   \delta^{-1} ( n^{\lambda, m+1} u^{\lambda, m+1} - n^{\lambda, m} u^{\lambda, m})
  + \nabla \cdot ( n^{\lambda, m} u^{\lambda, m} \otimes u^{\lambda, m} ) + \nabla p(n^{\lambda, m})  = \nonumber \\
& & \hspace{9cm}   =  n^{\lambda, m+1} \nabla \phi^{\lambda, m+1}, \label{CD_1FEP_u} \\
& & \hspace{-1cm}  - \lambda^2 \Delta \phi^{\lambda, m+1}  =
  1 - n^{\lambda, m+1}. \label{CD_1FEP_phi}
\end{eqnarray}
\label{def_EP_SD_class}
\end{definition}

\noindent
This scheme cannot be AP. Indeed, the limit $\lambda \to 0$ with fixed $\delta$ leads to 
\begin{eqnarray}
& & \hspace{-1cm}  \delta^{-1} (n^{0, m+1} - n^{0, m}) + 
  \nabla \cdot ( n^{0, m} u^{0, m} ) = 0, \label{CD_1FEP_n_l=0} \\
& & \hspace{-1cm}   \delta^{-1} ( n^{0, m+1} u^{0, m+1} - n^{0, m} u^{0, m})
  + \nabla \cdot ( n^{0, m} u^{0, m} \otimes u^{0, m} ) + \nabla p(n^{0, m})=
\nonumber \\
& & \hspace{9cm}   =    n^{0, m+1} \nabla \phi^{0, m+1}, \label{CD_1FEP_u_l=0} \\
& & \hspace{-1cm}  n^{0, m+1} = 1, \label{CD_1FEP_phi_l=0}
\end{eqnarray}
which does not provide a valid recursion for computing the unknowns at time $t^{m+1}$ knowing them at time $t^m$. Indeed, (\ref{CD_1FEP_phi_l=0}), once inserted into (\ref{CD_1FEP_n_l=0}), provides a divergence constraint on $u^{0, m}$, which is impossible to fulfill since $u^{0, m}$ is a datum from the previous time step.

A cure for this deficiency is to evaluate the mass flux implicitly. Simultaneously, we can see that the density in the force term can be taken explicit. This leads to the following 

\begin{definition}
The AP time Semi-Discretization of the EP model (or SD-EP scheme) is: 
\begin{eqnarray}
& & \hspace{-1cm}  \delta^{-1} (n^{\lambda, m+1} - n^{\lambda, m}) + 
  \nabla \cdot ( n^{\lambda, m+1} u^{\lambda, m+1} ) = 0, \label{AP_1FEP_n} \\
& & \hspace{-1cm}   \delta^{-1} ( n^{\lambda, m+1} u^{\lambda, m+1} - n^{\lambda, m} u^{\lambda, m})
  + \nabla \cdot ( n^{\lambda, m} u^{\lambda, m} \otimes u^{\lambda, m} ) + \nabla p(n^{\lambda, m})  =
\nonumber \\
& & \hspace{9cm}   =    n^{\lambda, m} \nabla \phi^{\lambda, m+1}, \label{AP_1FEP_u} \\
& & \hspace{-1cm}  - \lambda^2 \Delta \phi^{\lambda, m+1}  =
  1 - n^{\lambda, m+1}. \label{AP_1FEP_phi}
\end{eqnarray}
\label{def_EP_SP_AP}
\end{definition}

\noindent
Using the discrete momentum eq. (\ref{AP_1FEP_u}) and the Poisson eq. (\ref{AP_1FEP_phi}) respectively, $n^{\lambda, m+1}$ $u^{\lambda, m+1}$ and $n^{\lambda, m+1}$ can be eliminated from the mass equation (\ref{AP_1FEP_n}). This leads to: 
\begin{eqnarray}
& & \hspace{-1cm}  - \nabla \cdot \left( ( \delta^2 n^{\lambda, m} + \lambda^2) \nabla \phi^{\lambda, m+1} \right) = 
1 - n^{\lambda, m} +  \delta \nabla \cdot ( n^{\lambda, m} u^{\lambda, m} ) \nonumber\\
& & \hspace{3cm} - \delta^2 \nabla \cdot \left( \nabla \cdot ( n^{\lambda, m} u^{\lambda, m} \otimes u^{\lambda, m} ) + \nabla p(n^{\lambda, m}) \right)  . \label{AP_1FEP_phi_ref}
\end{eqnarray}
This elliptic equation for $\phi^{\lambda, m+1}$ is a discrete analog of the reformulated Poisson eq. (\ref{S_1FEP_phi_ref}). Indeed, using (\ref{AP_1FEP_n}) between times $m-1$ and $m$ and (\ref{AP_1FEP_phi}), we can write it as 
\begin{eqnarray}
& & \hspace{-1cm}  - \nabla \cdot \left( ( \delta^2 n^{\lambda, m} + \lambda^2) \nabla \phi^{\lambda, m+1} \right) =  \lambda^2 ( - 2 \Delta \phi^{\lambda, m} + \Delta \phi^{\lambda, m-1} ) \nonumber\\
& & \hspace{3cm} - \delta^2 \nabla^2 : \left( n^{\lambda, m} u^{\lambda, m} \otimes u^{\lambda, m}  +  p(n^{\lambda, m}) \mbox{Id}  \right)  ,  \label{AP_1FEP_phi_ref_2}
\end{eqnarray}
where the discretization of the second order time-derivative of $\Delta \phi$ appears. We summarize this in the 

\begin{definition}
The Reformulated time Semi-Discretization of the EP model (RSD-EP scheme) is: 
\begin{eqnarray}
& & \hspace{-1cm}  \delta^{-1} (n^{\lambda, m+1} - n^{\lambda, m}) + 
  \nabla \cdot ( n^{\lambda, m+1} u^{\lambda, m+1} ) = 0, \label{AP_1FEP_n_1} \\
& & \hspace{-1cm}   \delta^{-1} ( n^{\lambda, m+1} u^{\lambda, m+1} - n^{\lambda, m} u^{\lambda, m})
  + \nabla \cdot ( n^{\lambda, m} u^{\lambda, m} \otimes u^{\lambda, m} ) + \nabla p(n^{\lambda, m})  =
\nonumber \\
& & \hspace{9cm}   =    n^{\lambda, m} \nabla \phi^{\lambda, m+1}, \label{AP_1FEP_u_1} \\
& & \hspace{-1cm}  - \nabla \cdot \left( ( \delta^2 n^{\lambda, m} + \lambda^2) \nabla \phi^{\lambda, m+1} \right) = 
1 - n^{\lambda, m} +  \delta \nabla \cdot ( n^{\lambda, m} u^{\lambda, m} ) \nonumber\\
& & \hspace{3cm} - \delta^2 \nabla \cdot \left( \nabla \cdot ( n^{\lambda, m} u^{\lambda, m} \otimes u^{\lambda, m} ) + \nabla p(n^{\lambda, m}) \right)  . \label{AP_1FEP_phi_ref_1}
\end{eqnarray}
\label{def_RSD_EP}
\end{definition}

\begin{proposition}
The RSD-EP scheme is equivalent to the SD-EP scheme.
\label{prop_equiv_SDEP-RSDEP}
\end{proposition}

\noindent
Formally passing to the limit $ \lambda \to 0 $ with fixed $\delta$ in this scheme leads to the following scheme:

\begin{proposition}
We consider a sequence of solutions of the RSD-EP scheme whose initial conditions are well-prepared, i.e. satisfy \, $n^{\lambda, 0} - 1 = O(\lambda^2)$, \, $\nabla \cdot (n^{\lambda, 0} u^{\lambda, 0}) = O(\lambda^2)$. Then, 
the formal $\lambda \to 0$ of this sequence of solutions satisfies the following scheme:  
\begin{eqnarray}
& & \hspace{-1cm}  n^{0, m+1}  = 1, \label{AP_1FIE_n_2} \\
& & \hspace{-1cm}   \delta^{-1} ( u^{0, m+1} - u^{0, m})
  + \nabla \cdot ( u^{0, m} \otimes u^{0, m} )  =
\nabla \phi^{0, m+1}, \label{AP_1FIE_u_2} \\
& & \hspace{-1cm}  - \Delta \phi^{0, m+1} =   -  \nabla^2 : ( u^{0, m} \otimes u^{0, m} ). \label{AP_1FEP_phi_ref_l=0}
\end{eqnarray}
This scheme (the Reformulated Semi-discrete Incompressible Euler scheme or RSD-IE scheme) is consistent with the IE model. 
\label{prop_RSD_IE}
\end{proposition}

\noindent
{\bf Proof:} 
Leting $\lambda \to 0$ in the RSD-EP scheme leads to: 
\begin{eqnarray}
& & \hspace{-1cm}  \delta^{-1} (n^{0, m+1} - n^{0, m}) + 
  \nabla \cdot ( n^{0, m+1} u^{0, m+1} ) = 0, \label{AP_1FIE_n} \\
& & \hspace{-1cm}   \delta^{-1} ( n^{0, m+1} u^{0, m+1} - n^{0, m} u^{0, m})
  + \nabla \cdot ( n^{0, m} u^{0, m} \otimes u^{0, m} ) + \nabla p(n^{0, m})  =
\nonumber \\
& & \hspace{9cm}   =    n^{0, m} \nabla \phi^{0, m+1}, \label{AP_1FIE_u} \\
& & \hspace{-1cm}  - \delta^2 \nabla \cdot \left( n^{0, m}  \nabla \phi^{0, m+1} \right) = 
1 - n^{0, m} +  \delta \nabla \cdot ( n^{0, m} u^{0, m} ) \nonumber\\
& & \hspace{3cm} - \delta^2 \nabla \cdot \left( \nabla \cdot ( n^{0, m} u^{0, m} \otimes u^{0, m} ) + \nabla p(n^{0, m}) \right)  . \label{AP_1FIE_phi_ref}
\end{eqnarray}
Taking the combination \, $\delta \times (\ref{AP_1FIE_n}) - \delta^2 \times \nabla \cdot (\ref{AP_1FIE_u}) + (\ref{AP_1FIE_phi_ref})$ \, leads to (\ref{AP_1FIE_n_2}) for $m \geq 0$. With the well-prepared initial condition assumption, we have $n^{0, m}  = 1$ for all $m \geq 0$ and consequently, using (\ref{AP_1FIE_n}) again, to $\nabla \cdot ( n^{0, m+1} u^{0, m+1} ) = 0$ for all $m \geq 0$. With the well-prepared initial condition again, we deduce that $\nabla \cdot ( n^{0, m} u^{0, m} ) = 0$ for all $m \geq 0$. Then, (\ref{AP_1FIE_phi_ref}) reduces to (\ref{AP_1FEP_phi_ref_l=0})
for $m \geq 0$, while (\ref{AP_1FIE_u}) reduces to (\ref{AP_1FIE_u_2}) for $m \geq 0$. 
This shows that the scheme is consistent with the RIE model. \endproof

\begin{remark}
The well-prepared initial condition hypothesis on $n^{\lambda, 0}$ is satisfied as soon as the first iterate $m=0$ satisfies the Poisson eq. $- \lambda^2 \Delta \phi^{\lambda, 0}  = 1 - n^{\lambda, 0}$ which is a natural condition for a solution of the Euler-Poisson problem. If the well-prepared initial condition hypothesis on $u^{\lambda, 0}$ is not satisfied, the RSD-IE scheme is still satisfied but starting at iterate $m=1$. If none of the well-prepared initial condition hypotheses is satisfied, the RSD-IE scheme is satisfied starting at iterate $m=2$. 
\label{rem_RSD-IE_well_prepared}
\end{remark}

\noindent
This proposition and the previous ones show that, in the $\lambda \to 0$ limit, the SD-EP scheme is consistent with the Incompressible Euler model. Therefore, the SD-EP is AP, provided that its stability condition is independent of $\lambda$ when $\lambda \to 0$ (Asymptotic Stability property). This will be shown, at least in a linearized setting, by the stability analysis below. We note that the computational cost of the AP-scheme is the same as for the classical scheme, the Poisson equation being replaced by the elliptic equation (\ref{AP_1FEP_phi_ref}), which has roughly the same computational cost.
This AP scheme has been proposed initially in \cite{Cri_Deg_Vig_CRAS05} and \cite{Cri_Deg_Vig_07}.

%%%%%%%%%%%%%%%%%%%%%%%%%%%%%%%%%%%%%%%%%%%%%%%%%%%%%%%%%%
\subsection{Linearized stability analysis of the time-semi-discretization}
\label{subsec_QN_EP_1F_stability}

The linearized EP model will be considered about the state defined by $n^\lambda = 1$, $u^\lambda = 0$, $\phi^\lambda = 0$ (which is obviously a stationary solution). We will also restrict ourselves to a one-dimensional situation. Expanding $n^\lambda = 1 + \varepsilon \tilde n^\lambda$, $u^\lambda = \varepsilon \tilde u^\lambda$ and $\phi^\lambda = \varepsilon \tilde \phi^\lambda$, with $\varepsilon \ll 1$ being the intensity of the perturbation to the stationary state, and retaining only the linear terms in $\varepsilon$, we find the linearized EP model: 
\begin{eqnarray}
& & \hspace{-1cm} \partial_t \tilde n^\lambda + \partial_x \tilde u^\lambda = 0, \label{LEP_n} \\
& & \hspace{-1cm} \partial_t \tilde u^\lambda + T \partial_x \tilde n^\lambda = \partial_x \tilde \phi^\lambda , \label{LEP_u} \\
& & \hspace{-1cm} \lambda ^2 \partial^2_x \tilde \phi^\lambda = \tilde n^\lambda , \label{LEP_phi} 
\end{eqnarray}
with $T = p'(1)$. Introducing $\hat n^\lambda$, $\hat u^\lambda$, $\hat \phi^\lambda$, the partial Fourier transforms of $\tilde n^\lambda$, $\tilde u^\lambda$, $\tilde \phi^\lambda$ with respect to $x$, we are led to the system of ODE's: 
\begin{eqnarray}
& & \hspace{-1cm} \partial_t \hat n^\lambda + i \xi \hat u^\lambda = 0, \label{FLEP_n} \\
& & \hspace{-1cm} \partial_t \hat u^\lambda + i T \xi \hat n^\lambda = i \xi \hat \phi^\lambda , \label{FLEP_u} \\
& & \hspace{-1cm} - \lambda ^2 \xi^2 \hat \phi^\lambda = \hat n^\lambda  , \label{FLEP_phi} 
\end{eqnarray}
where $\xi$ is the Fourier dual variable to $x$. The general solution of this model takes the form
\begin{eqnarray}
& & \hspace{-1cm} \left( \begin{array}{c} \hat n^\lambda \\ \hat u^\lambda \end{array} \right) = \sum_{\pm} e^{s^\lambda_\pm t} \left( \begin{array}{c} \hat n^\lambda_\pm \\ \hat u^\lambda_\pm \end{array} \right), 
\label{EPB_gen_sol}
\end{eqnarray}
with
\begin{eqnarray}
& & \hspace{-1cm} s^\lambda_\pm  = \pm i ( T \xi^2 + \lambda^{-2} ), 
\label{EP_s}
\end{eqnarray}
and $n^\lambda_\pm$, $u^\lambda_\pm$ are given functions of $\xi$, fixed by the initial conditions of the problem. In particular, since $s^\lambda_\pm $ are pure imaginary numbers, the $L^2$ norm of the solution is preserved with time. When $\lambda \to 0$, then $|s^\lambda_\pm| \to \infty$, which means that the limit $\lambda \to 0$ is an oscillatory limit. We will see that the AP-scheme averages out these fast oscillations.

The goal of this section is to analyze the linearized stability properties of both the classical and SD-EP schemes. We want to show that the classical scheme requires a CFL condition of the type $\delta = O(\lambda)$ while the stability condition for the SD-EP scheme is independent of $\lambda$ when $\lambda \to 0$. This last property is known as 'Asymptotic-Stability' and is a component of the Asymptotic-Preserving property (see introduction). Indeed, the faculty of letting $\lambda \to 0$ in the scheme with fixed $\delta$ is possible only if the stability condition of the scheme is independent of $\lambda$ in this limit. We will prove $L^2$-stability uniformly with respect to $\lambda$ for the linearized problem (\ref{FLEP_n})-(\ref{FLEP_phi}).  

In general, time-semi-discretizations of hyperbolic problems are inconditionally unstable. This is easily verified on the discretization (\ref{FLEP_n})-(\ref{FLEP_u}) if $\hat \phi$ is set to $0$. This is because the skew adjoint operator $\partial_x$ has the same effect as a centered space-differencing. For fully discrete schemes, stability is obtained at the price of decentering, i.e. adding numerical viscosity. To mimic the effect of this viscosity, in the present section, we will consider the linearized Viscous Euler-Poisson (VEP) model, which consists of the linearized EP model (\ref{LEP_n})-(\ref{LEP_phi}) with additional viscosity terms (in this section, we drop the tildes for notational convenience): 
\begin{eqnarray}
& & \hspace{-1cm} \partial_t  n^\lambda + \partial_x  u^\lambda - \beta \partial_x^2 n^\lambda = 0, \label{LVEP_n} \\
& & \hspace{-1cm} \partial_t  u^\lambda + T \partial_x \tilde n^\lambda - \beta \partial_x^2 u^\lambda = \partial_x  \phi^\lambda , \label{LVEP_u} \\
& & \hspace{-1cm} \lambda ^2 \partial^2_x  \phi^\lambda =  n^\lambda . \label{LVEP_phi} 
\end{eqnarray}
where $\beta$ is a numerical viscosity coefficient. We keep in mind that, in the spatially discretized case, $\beta$ proportional to the mesh size $h$: 
\begin{equation}
\beta = c h , \label{beta}
\end{equation}
with the constant $c$ to be specified later on. 

The classical and SD-EP time discretizations are given by 
\begin{eqnarray}
& & \hspace{-1cm} \delta^{-1}   (n^{\lambda,m+1} - n^{\lambda,m})  + \partial_x  u^{\lambda,m+a} - \beta \partial_x^2 n^{\lambda,m} = 0, \label{DLVEP_n} \\
& & \hspace{-1cm} \delta^{-1}   (u^{\lambda,m+1} - u^{\lambda,m}) + T \partial_x  n^{\lambda,m} - \beta \partial_x^2 u^{\lambda,m} = \partial_x  \phi^{\lambda,m+1} , \label{DLVEP_u} \\
& & \hspace{-1cm} \lambda ^2 \partial^2_x  \phi^{\lambda,m+1} =  n^{\lambda,m+1} , \label{DLVEP_phi} 
\end{eqnarray}
with $a=0$ for the classical scheme and $a=1$ for the SD-EP scheme. Passing to Fourier space with $\xi$ being the dual variable to $x$, and eliminating $\hat \phi^{\lambda,m+1}$, we find the following recursion relations:  
\begin{eqnarray}
& & \hspace{-1cm} \delta^{-1}   (\hat n^{\lambda,m+1} - \hat n^{\lambda,m})  + i \xi \hat  u^{\lambda,m+a} + \beta \xi^2 \hat n^{\lambda,m} = 0, \label{FDLVEP_n} \\
& & \hspace{-1cm} \delta^{-1}   (\hat u^{\lambda,m+1} -  \hat u^{\lambda,m}) + i T \xi  \hat n^{\lambda,m} +  \frac{i}{\lambda^2 \xi}  \hat n^{\lambda,m+1} + \beta \xi^2 \hat u^{\lambda,m} = 0 , \label{FDLVEP_u} 
\end{eqnarray}

The characteristic equations for thus recursion formula is: 
\begin{eqnarray}
& & 
q^2 - 2 q \left( 1 -\beta \xi^2 \delta - \frac{\delta^2}{2 \lambda^2} \right) + (1 -\beta \xi^2 \delta)^2 + T \xi^2 \delta^2 = 0  .
\label{CD_EP_char_eq}
\end{eqnarray}
for the classical scheme ($a=0$) and
\begin{eqnarray}
& & 
( 1 + \frac{\delta^2}{\lambda^2} ) q^2 - 2 q \left( 1 - \beta \xi^2 \delta - \frac{1}{2} T \xi^2 \delta^2 \right) + (1 - \beta \xi^2 \delta)^2 = 0   .
\label{AP_EP_char_eq}
\end{eqnarray}
for the SD-EP scheme ($a=1$), where $q$ is the characteristic root. Each of these quadratic equations has two roots $q^\lambda_\pm(\xi)$ which provide the two independent solutions of the corresponding recursion formulas. Their most general solution is of the form
\begin{eqnarray}
& & \hspace{-1cm} \left( \begin{array}{c} \hat n^{\lambda,m}(\xi) \\ \hat u^{\lambda,m}(\xi) \end{array} \right) = \sum_{\pm} (q^\lambda_\pm(\xi))^m \left( \begin{array}{c} \hat n^\lambda_\pm(\xi) \\ \hat u^\lambda_\pm(\xi) \end{array} \right), \quad \forall m \in {\mathbb N}, 
\label{FDLEP_general}
\end{eqnarray}
where $n^\lambda_\pm(\xi)$ and $u^\lambda_\pm(\xi)$ depend on the initial condition only. A necessary and sufficient condition for $L^2$ stability is that $|q^\lambda_\pm(\xi)| <1$. However, requesting this condition for all $\xi \in {\mathbb R}$ is too restrictive. To account for the effect of a spatial discretization in this analysis, we must restrict the range of admissible Fourier wave-vectors $\xi$ to the interval $[-\frac{\pi}{h}, \frac{\pi}{h}]$. Indeed, a space discretization of step $h$ cannot represent wave-vectors of magnitude larger than $\frac{\pi}{h}$. This motivates the following definition of stability: 

\begin{definition}
The scheme is stable if and only if
\begin{equation}
|q^\lambda_\pm(\xi)|\leq 1, \quad \forall \xi \quad \mbox{ such that } \quad |\xi| < \frac{\pi}{h}.
\label{stab_cnd}
\end{equation}
\label{def_stab}
\end{definition}

\noindent
Now, our goal is to find sufficient conditions on $\delta$ such that either schemes are stable. More precisely, In \cite{Deg_Liu_Vig_08}, the following result is proven: 

\begin{proposition}
We assume that the numerical viscosity $\beta$ satisfies (\ref{beta}). 

\noindent
(i) We assume that $\delta$ satisfies the CFL condition $\delta \leq C_0 h$ with $C_0 = \frac{2c}{T + c^2 \pi^2}$. Then, there exist two constants $0 < C_1 < C'_1$ and a constant $\lambda_1 >0$ (depending on $c$, $T$), such that the classical scheme is stable provided that $\delta \leq C_1 \lambda $ and unstable if $\delta \geq C'_1 \lambda$, for all $\lambda$, $0 < \lambda \leq \lambda_1$. 

\noindent
(ii) There exists a constant $C_2>0 $  (depending on $c$, and $T$ but not on $\lambda$), such that the SD-EP scheme is stable for all $\lambda >0$ provided that $\delta \leq C_2 h $. 
\label{prop_EP_stab}
\end{proposition}

\noindent
The first statement of this theorem says that, under the CFL condition for the hydrodynamic part of the system, a necessary and sufficient condition for the stability of the classical scheme is that $\delta \leq 0(\lambda)$. This confirms that the classical scheme is not Asymptotically-Stable and cannot be AP. 

The second statement shows that the stability condition of the SD-EP scheme is independent of $\lambda$ in the limit $\lambda \to 0$ (and actually, whatever value $\lambda$ has). Thus the Asymptotic Stability property is fulfilled. The stability is not unconditional: a CFL condition for the hydrodynamic part is still required. However, the goal of an AP scheme is not to provide unconditional stability, but simply Asymptotic Stability with respect to a given parameter. This more restricted stablity requirement allows us to select the terms that need to be evaluated implicitly, and leads to schemes which are more easily resolved. 

Adding more implicitness in the scheme does not necessarily improve the AP-ness of the scheme, and sometimes a degradation is observed. Indeed, fully implicit schemes require nonlinear solvers which furnish only an approximate solution of the scheme. They may perform poorly in highly ill-conditionned situations; The stopping test in case of iterative solvers may also be not as accurate as expected. The result is that the AP property can actually be missed by fully implicit solvers. It can be retrieved at the price of some specific AP pre-conditionning. 

The linearized stability analysis does not provide a complete proof of the AP property. For this purpose, energy estimates on the fully non-linear system should be proved. Some partial answers can be found in \cite{Deg_Liu_Vig_08}. The stability analysis can also be developed about a stationary state with non-zero velocity, with a similar conclusion. We refer the reader to \cite{Deg_Liu_Vig_08} for more detail.

%%%%%%%%%%%%%%%%%%%%%%%%%%%%%%%%%%%%%%%%%%%%%%%%%%%%%%%%%%
\subsection{Full (space-time) discretization: enforcing the Gauss law}
\label{subsec_QN_EP_1F_space}

%%%%%%%%%%%%%%%%%%%%%%%%%%%%%%%%%%%%%%%%%%%%%%%%%%%%%%%%%%
\subsubsection{General framework and classical fully-discrete scheme}
\label{subsubsec_QN_EP_1F_space_class}

We now discuss some issues related to the spatial discretisation. For this purpose, we restrict ourselves to a one-dimensional framework but all the considerations here extend straightforwardly to any dimension. We introduce a spatial discretization with a uniform mesh of step $ h $ and we denote by $C_k$ the cell $[(k-1/2) h, (k+1/2)h]$ and $x_k = kh$, with $k \in {\mathbb Z}$. Like in usual first-order shock capturing schemes, the fluid unknowns $n$ and $u$ are approximated by piecewise constant functions within the cell $C_k$ and represented by cell-centered values $ n|_{k}^{m}$, $ u|_{k}^{m}$ at time $t^m = m \delta$. The electric potential $\phi$ is also approximated by cell-centered values  $ \phi|_{k}^{m}$. The discretization of the hydrodynamic part is performed by means of a first order shock capturing scheme. 
We denote by $f_{n}|_{k+1/2}^{m} $ and $f_{u}|_{k+1/2}^{m}$ the numerical fluxes for the mass and  momentum conservation equations respectively, at time $ t^m = m \delta $ and at the cell interface $x_{k+1/2}$. Similarly, $E|_{k+1/2}^{m}$ denotes the approximation of the electric field $E = - \partial_x \phi$ at $x_{k+1/2}$ and $ t^m $. 

The classical scheme is defined as follows: 

\begin{definition}
The fully-discretized classical scheme for the Euler-Poisson model is: 
\begin{eqnarray}
& & \hspace{-1cm}  \delta^{-1} (n|_k^{m+1} - n|_k^m) + h^{-1} (f_n|_{k+1/2}^{m} - f_n|_{k-1/2}^{m})  = 0, \label{1FEP_n_fd} \\
& & \hspace{-1cm} \delta^{-1}  ((nu)|_k^{m+1} - (nu)|_k^m) + h^{-1} (f_{u}|_{k+1/2}^{m} - f_{u}|_{k-1/2}^{m})  =   - n|_k^{m+1} \tilde E|_k^{m+1} , \label{1FEP_u_fd} \\
& & \hspace{-1cm} \lambda^2 h^{-1} (E|_{k+1/2}^{m+1} -E|_{k-1/2}^{m+1}) = 1 - n|_k^{m+1},  \label{1FEP_E_fd} \\
& & \hspace{-1cm} E|_{k+1/2}^{m+1} = - h^{-1} (\phi|_{k+1}^{m+1} - \phi|_{k}^{m+1}), \quad   \tilde E|_k^{m+1} = \frac{1}{2} (E|_{k+1/2}^{m+1} + E|_{k-1/2}^{m+1}) .  \label{1FEP_phi_fd}
\end{eqnarray}
The numerical fluxes $f_n|_{k+1/2}^{m}$ and $f_{u}|_{k+1/2}^{m}$ are given by:
\begin{eqnarray}
& & \hspace{-1cm}  \left( \begin{array}{c} f_n|_{k+1/2}^{m} \\ f_{u}|_{k+1/2}^{m} \end{array} \right) = \frac{1}{2} \left\{ \left( \begin{array}{c}   (nu)|_k^m \\ \left(nu^2 + p(n) \right)|_k^m \end{array} \right) 
+ \left( \begin{array}{c}   (nu)|_{k+1}^m \\ \left(nu^2 + p(n) \right)|_{k+1}^m \end{array} \right) \right.  + 
\nonumber \\
& & \hspace{6cm} + \left. \mu_{k+1/2}^m \, \left( \begin{array}{c}  n|_k^m - n|_{k+1}^m  \\ (nu)|_k^m - (nu)|_{k+1}^m  \end{array} \right) \right\}, 
\label{1FEP_hydro_flux}
\end{eqnarray}
where $\mu_{k+1/2}^m$ is the viscosity matrix. 
\label{def_FDEP_class}
\end{definition}

\noindent
Here, we will not discuss the choice of the viscosity matrix (see \cite{God_Rav_96, Leveque_02, Toro_99} for a discussion of this point). For instance, in the simple case of a Local Lax-Friedrichs (LLF) scheme \cite{Leveque_02,Rusanov} $\mu_{k+1/2}^m$ corresponds to a local evaluation of the wave speeds (see e.g. \cite{Cri_Deg_Vig_07} for more details). 

We note that the electric field is implicit in the momentum conservation eq. As shown by Fabre \cite{Fabre_JCP_101_445}, this is a necessary condition for conditional stability.  In spite of this implicitness, the scheme can be solved explicitly. The sequence of updates goes as follows: first the mass conservation eq. (\ref{1FEP_n_fd}) is used to compute $n^{m+1}$. Then (\ref{1FEP_E_fd}) and (\ref{1FEP_phi_fd}) are combined into the discrete Poisson equation 
\begin{eqnarray}
& & \hspace{-1cm} - \lambda^2 h^{-2} (\phi|_{k+1}^{m+1} - 2 \phi|_{k}^{m+1} + \phi|_{k-1}^{m+1}) = 1 - n|_k^{m+1},  \label{1FEP_phi_fd2} 
\end{eqnarray}
the inversion of which allows us to compute $\phi^{m+1}$ (provided suitable boundary conditions are specified. We will not discuss this point here). Finally, with the momentum balance eq. (\ref{1FEP_u_fd}) we find the value of $(nu)|_k^{m+1}$. The stability condition for this scheme combines a CFL condition for the hydrodynamic part and a condition of the type $\delta \leq O(\lambda)$ \cite{Fabre_JCP_101_445}. Therefore, this scheme cannot be AP.

%%%%%%%%%%%%%%%%%%%%%%%%%%%%%%%%%%%%%%%%%%%%%%%%%%%%%%%%%%
\subsubsection{AP fully discrete scheme}
\label{subsubsec_QN_EP_1F_space_AP}

The AP-scheme is defined as follows: 

\begin{definition}
The Fully Discretized AP scheme for the Euler-Poisson model (FD-EP scheme) is: 
\begin{eqnarray}
& & \hspace{-1cm}  \delta^{-1} (n|_k^{m+1} - n|_k^m) + h^{-1} (\tilde f_n|_{k+1/2}^{m+1} - \tilde f_n|_{k-1/2}^{m+1})  = 0, \label{1FEP_n_fd_ap} \\
& & \hspace{-1cm} \delta^{-1}  ((nu)|_k^{m+1} - (nu)|_k^m) + h^{-1} (f_{u}|_{k+1/2}^{m} - f_{u}|_{k-1/2}^{m})  =   - n|_k^{m} \tilde E|_k^{m+1} , \label{1FEP_u_fd_ap} \\
& & \hspace{-1cm} \lambda^2 h^{-1} (E|_{k+1/2}^{m+1} -E|_{k-1/2}^{m+1}) = 1 - n|_k^{m+1},  \label{1FEP_E_fd_ap} \\
& & \hspace{-1cm} E|_{k+1/2}^{m+1} = - h^{-1} (\phi|_{k+1}^{m+1} - \phi|_{k}^{m+1}), \quad   \tilde E|_k^{m+1} = \frac{1}{2} (E|_{k+1/2}^{m+1} + E|_{k-1/2}^{m+1}) . \label{1FEP_phi_fd_ap}
\end{eqnarray}
The momentum flux $f_{u}|_{k+1/2}^{m}$ is given by (\ref{1FEP_hydro_flux}) while the mass flux is given by: 
\begin{eqnarray}
& & \hspace{-1cm}  \tilde f_n|_{k+1/2}^{m+1} = f_n|_{k+1/2}^{m} - \frac{\delta}{2} (n|_{k+1}^m + n|_{k}^m)  \, E|_{k+1/2}^{m+1}   - \frac{\delta h^{-1}}{2} (f_{u}|_{k+3/2}^{m} - f_{u}|_{k-1/2}^{m}) 
. \label{1FEP_fn_m+1_fd2_mod} 
\end{eqnarray}
\label{def_FDEP}
\end{definition}

\noindent
Compared to the classical scheme, the FD-EP scheme exhibits a modified mass flux (\ref{1FEP_fn_m+1_fd2_mod}),  and uses an explicit evaluation of the density appearing in front of the electric field in the momentum equation (\ref{1FEP_u_fd_ap}). We now provide the rationale for the use of the mass flux (\ref{1FEP_fn_m+1_fd2_mod}). In the following discussion, we will restrict ourselves to an LLF (or Rusanov) scalar numerical viscosity for simplicity, but the discussion could be extended straightforwardly to any kind of shock-capturing method. We leave the details to the reader.  We show that the flux (\ref{1FEP_fn_m+1_fd2_mod}) can be interpreted as some minor modification of a flux $\tilde {\tilde f}_n|_{k+1/2}^{m+1}$  in which the central part is implicit: 
\begin{eqnarray}
& & \hspace{-1cm}  \tilde {\tilde f}_n|_{k+1/2}^{m+1} = \frac{1}{2} \left[ (nu_x)|_k^{m+1} + (nu_x)|_{k+1}^{m+1} + \mu_{k+1/2}^m \, \left( n|_k^m - n|_{k+1}^m \right) \right] .
\label{1FEP_fn_m+1_fd} 
\end{eqnarray}
Indeed, starting from the definition (\ref{1FEP_fn_m+1_fd}) and inserting (\ref{1FEP_u_fd_ap}) into (\ref{1FEP_fn_m+1_fd}), we can write: 
\begin{eqnarray}
& & \hspace{-1cm}  \tilde {\tilde f}_n|_{k+1/2}^{m+1} = f_n|_{k+1/2}^{m} - \frac{\delta}{4} \left[ n|_{k+1}^m \, E|_{k+3/2}^{m+1} + (n|_{k+1}^m + n|_{k}^m)  \, E|_{k+1/2}^{m+1} + n|_{k}^m  \, E|_{k-1/2}^{m+1}  \right] \nonumber \\
& & \hspace{7cm} - \frac{\delta h^{-1}}{2} (f_{u}|_{k+3/2}^{m} - f_{u}|_{k-1/2}^{m}) 
. \label{1FEP_fn_m+1_fd2} 
\end{eqnarray}
This flux involves an average of $E^{m+1}$ over three neighbouring mesh points. In order to reduce the numerical diffusion, we replace (\ref{1FEP_fn_m+1_fd2}) by (\ref{1FEP_fn_m+1_fd2_mod}).  Therefore, (\ref{1FEP_fn_m+1_fd2_mod}) is, up to a lumping of the electric field average, what results from an implicit evaluation of the central part of the mass flux. It is an order $O(\delta)$ modification of the explicit flux, but this simple modification is crucial in making the scheme AP. Formula (\ref{1FEP_fn_m+1_fd2_mod}) can be used irrespective of the expression of the explicit fluxes $f_n|_{k+1/2}^{m}$, $f_u|_{k+1/2}^{m}$ and is therefore valid whatever choice of numerical viscosity is made.

We now show that the FD-EP scheme can be reformulated in a consistent way to the REP model.

%%%%%%%%%%%%%%%%%%%%%%%%%%%%%%%%%%%%%%%%%%%%%%%%%%%%%%%%%%
\subsubsection{Reformulation}
\label{subsubsec_QN_EP_1F_space_AP_reform}

\begin{definition}
The Reformulated Fully Discretized AP scheme for the Euler-Poisson model (RFD-EP scheme) is: 
\begin{eqnarray}
& & \hspace{-1cm}  \delta^{-1} (n|_k^{m+1} - n|_k^m) + h^{-1} (\tilde f_n|_{k+1/2}^{m+1} - \tilde f_n|_{k-1/2}^{m+1})  = 0, \label{1FEP_n_fd_ap_2} \\
& & \hspace{-1cm} \delta^{-1}  ((nu)|_k^{m+1} - (nu)|_k^m) + h^{-1} (f_{u}|_{k+1/2}^{m} - f_{u}|_{k-1/2}^{m})  =   - n|_k^{m} \tilde E|_k^{m+1} , \label{1FEP_u_fd_ap_2} \\
& & \hspace{-1cm} - h^{-2} \left[ (\lambda^2 + \frac{\delta^2}{2} (n|_{k+1}^{m} + n|_{k}^{m}) ) (\phi|_{k+1}^{m+1} - \phi|_{k}^{m+1})  \right. \nonumber \\
& & \hspace{5cm} \left.  - (\lambda^2 + \frac{\delta^2}{2} (n|_{k}^{m} + n|_{k-1}^{m}) ) (\phi|_{k}^{m+1} - \phi|_{k-1}^{m+1})  \right] \nonumber \\
& & \hspace{3cm} 
=  1 - n|_k^{m} + h^{-1} \delta  (f_n|_{k+1/2}^{m} - f_n|_{k-1/2}^{m}) \nonumber \\
& & \hspace{3.5cm} 
 - \frac{ h^{-2} \delta^2}{2} (f_{u}|_{k+3/2}^{m} - f_{u}|_{k+1/2}^{m} - f_{u}|_{k-1/2}^{m} + f_{u}|_{k-3/2}^{m}) 
.  \label{1FEP_E_fd4} \\
& & \hspace{-1cm} E|_{k+1/2}^{m+1} = - h^{-1} (\phi|_{k+1}^{m+1} - \phi|_{k}^{m+1}),  \quad   \tilde E|_k^{m+1} = \frac{1}{2} (E|_{k+1/2}^{m+1} + E|_{k-1/2}^{m+1}), \label{1FEP_phi_fd_ap_2}
\end{eqnarray}
with $\tilde f_n|_{k+1/2}^{m+1}$ and $f_{u}|_{k+1/2}^{m}$ given by (\ref{1FEP_fn_m+1_fd2_mod}) and by (\ref{1FEP_hydro_flux}) respectively.  
\label{def_RFDEP}
\end{definition}

\medskip
\begin{proposition}
The FD-EP and RFD-EP schemes are equivalent.
\label{prop_equiv_FDEP-RFDEP}
\end{proposition}

\noindent
{\bf Proof: } We first insert
(\ref{1FEP_n_fd_ap}) into (\ref{1FEP_E_fd_ap}) and get
\begin{eqnarray}
& & \hspace{-1cm} h^{-1} \left[ ( \lambda^2 E|_{k+1/2}^{m+1} -  \delta \tilde f_n|_{k+1/2}^{m+1} ) - 
( \lambda^2 E|_{k-1/2}^{m+1} -  \delta \tilde f_n|_{k-1/2}^{m+1} ) \right] =  1 - n|_k^{m}.  \label{1FEP_E_fd2} 
\end{eqnarray}
Using (\ref{1FEP_fn_m+1_fd2_mod}), we have: 
\begin{eqnarray}
& & \hspace{-1cm} \lambda^2 E|_{k+1/2}^{m+1} -  \delta \tilde f_n|_{k+1/2}^{m+1} = (\lambda^2 + \frac{\delta^2}{2} (n|_{k+1}^{m} + n|_{k}^{m}) ) E|_{k+1/2}^{m+1} \nonumber \\
& & \hspace{5cm} - \delta  \left[
f_n|_{k+1/2}^{m} - \frac{\delta h^{-1}}{2} (f_{u}|_{k+3/2}^{m} - f_{u}|_{k-1/2}^{m}) \right]. 
 \label{1FEP_E_fd3} 
\end{eqnarray}
Then, inserting (\ref{1FEP_E_fd3}) and (\ref{1FEP_phi_fd_ap}) into (\ref{1FEP_E_fd2}) leads to (\ref{1FEP_E_fd4}). \endproof

\noindent
Eq. (\ref{1FEP_E_fd4}) is a discrete elliptic equation which allows to find $\phi^{m+1}$, provided boundary conditions (the same as for the Poisson equation) are specified. It clearly appears as a consistent spatial discretization of (\ref{AP_1FEP_phi_ref}). The time update for the AP-scheme follows a different sequence from that of the classical scheme. We first solve for $\phi^{m+1}$ by inverting the discrete elliptic equation (\ref{1FEP_E_fd4}). Then, $u^{m+1}$ and $n^{m+1}$ can be explicitly computed by successively using (\ref{1FEP_u_fd_ap_2}) and (\ref{1FEP_n_fd_ap_2}). 

Although not explicitly solved, the Gauss equation is satisfied exactly. More precisely, we have: 

\begin{proposition}
The solution of the RFD-EP scheme satisfies the discrete Gauss eq. (\ref{1FEP_E_fd_ap}) exactly. 
\label{prop_EP_Gauss}
\end{proposition}

\noindent
{\bf Proof:} It follows the steps of the proof of proposition \ref{prop_equiv_FDEP-RFDEP} backwards. Inserting the first equation (\ref{1FEP_phi_fd_ap_2}) into (\ref{1FEP_E_fd4}), and having (\ref{1FEP_fn_m+1_fd2_mod}) in mind, we see that the latter is equivalent to 
\begin{eqnarray}
& & \hspace{-1cm} h^{-1} \lambda^2 \left( E|_{k+1/2}^{m+1} - E|_{k-1/2}^{m+1}  \right) =  1 - n|_k^{m} + h^{-1} \delta  (\tilde f_n|_{k+1/2}^{m+1} - \tilde f_n|_{k-1/2}^{m+1}) 
.  \label{1FEP_E_fd5} 
\end{eqnarray}
Inserting the discrete mass balance eq. (\ref{1FEP_n_fd_ap_2}) into (\ref{1FEP_E_fd5}) leads to the discrete Gauss equation (\ref{1FEP_E_fd_ap}).  \endproof

\noindent
We now investigate the limit $\lambda \to 0$. We define:

%%%%%%%%%%%%%%%%%%%%%%%%%%%%%%%%%%%%%%%%%%%%%%%%%%%%%%%%%%
\subsubsection{$\lambda \to 0$ limit and AP property}
\label{subsubsec_QN_EP_1F_space_AP_lim_l_to_0}

\begin{proposition}
(i) We consider a sequence of solutions of the RFD-EP scheme whose initial conditions are well-prepared, i.e. satisfy \, $n^{\lambda, 0} - 1 = O(\lambda^2)$. Then, the formal $\lambda \to 0$ of this sequence of solutions satisfies the following scheme:  \begin{eqnarray}
& & \hspace{-1cm}  n|_k^{m+1} = 1, \label{1FIE_n_fd_ap_3} \\
& & \hspace{-1cm} \delta^{-1}  (u|_k^{m+1} - u|_k^m) + h^{-1} (f_{u}|_{k+1/2}^{m} - f_{u}|_{k-1/2}^{m})  =   -  \tilde E|_k^{m+1} , \label{1FIE_u_fd_ap_3} \\
& & \hspace{-1cm} - h^{-2} \left( \phi|_{k+1}^{m+1} - 2 \phi|_{k}^{m+1} +  \phi|_{k-1}^{m+1} \right) =  
- \frac{ h^{-2}}{2} (f_{u}|_{k+3/2}^{m} - f_{u}|_{k+1/2}^{m} - f_{u}|_{k-1/2}^{m} + f_{u}|_{k-3/2}^{m}) 
\nonumber \\
& & \hspace{7cm} 
+ h^{-1} \delta^{-1}  (f_n|_{k+1/2}^{m} - f_n|_{k-1/2}^{m}) ,  
\label{1FIE_E_fd1}  \\
& & \hspace{-1cm} E|_{k+1/2}^{m+1} = - h^{-1} (\phi|_{k+1}^{m+1} - \phi|_{k}^{m+1}),  \quad   \tilde E|_k^{m+1} = \frac{1}{2} (E|_{k+1/2}^{m+1} + E|_{k-1/2}^{m+1}), \label{1FIE_phi_fd_ap_3}
\end{eqnarray}
with 
\begin{eqnarray}
& & \hspace{-1cm}  \left( \begin{array}{c} f_n|_{k+1/2}^{m} \\ f_{u}|_{k+1/2}^{m} \end{array} \right) = \frac{1}{2} \left\{ \left( \begin{array}{c}   u|_k^m + u|_{k+1}^m \\ u^2 |_k^m + u^2 |_{k+1}^m  \end{array} \right)  + \mu_{k+1/2}^m \, \left( \begin{array}{c}  0  \\ u|_k^m - u|_{k+1}^m  \end{array} \right) \right\} , 
\label{1FEP_hydro_flux_n=1}
\end{eqnarray}
$\mu_{k+1/2}^m$ being the viscosity matrix. 

\noindent
(ii) We denote by:
\begin{eqnarray*}
& & \hspace{-1cm}
 \left( \begin{array}{c} D_n|_{k+1/2}^m \\ D_u|_{k+1/2}^m  \end{array} \right) =  \mu_{k+1/2}^m \, \left( \begin{array}{c}  0  \\ u|_k^m - u|_{k+1}^m  \end{array} \right).
\end{eqnarray*}
Then (\ref{1FIE_E_fd1}) is equivalent to 
\begin{eqnarray}
& & \hspace{-1cm} - h^{-2} \left( \phi|_{k+1}^{m+1} - 2 \phi|_{k}^{m+1} +  \phi|_{k-1}^{m+1} \right) =    - \frac{ h^{-2} }{2} (f_{u}|_{k+3/2}^{m} - f_{u}|_{k+1/2}^{m} - f_{u}|_{k-1/2}^{m} + f_{u}|_{k-3/2}^{m}) \nonumber \\
& & \hspace{2cm} 
-   h^{-2} \left( \phi|_{k+1}^{m} - 2 \phi|_{k}^{m} +  \phi|_{k-1}^{m} \right) +   \frac{ h^{-2}}{4} \left( \phi|_{k+2}^{m} - 2 \phi|_{k}^{m} +  \phi|_{k-2}^{m} \right)\nonumber \\
& & \hspace{2cm} + h^{-1} \delta^{-1} ( D_n|_{k+1/2}^{m+1} - D_n|_{k-1/2}^{m+1} - D_n|_{k+1/2}^{m} +  D_n|_{k-1/2}^{m+1})
,  \label{1FIE_E_fd5}
\end{eqnarray}
for $m \geq 1$. If the additional 'well-prepared' initial condition hypothesis $h^{-1} ( u|_{k+1}^{0} - u|_{k-1}^{0}) = O(\lambda^2)$ is made, eq. (\ref{1FIE_E_fd5}) holds true for $m = 0$ without the last two terms. Then, this scheme (the Reformulated Fully-Discrete Incompressible Euler scheme or RFD-IE scheme) is consistent with the RIE model, as soon as there exist two constants $C_1$ and $C_2$ such that and $C_1 h \leq \delta \leq C_2 h$. If the viscosity matrix is such that $D_n|_{k+1/2}^m = 0$ for all $k$, $m$, then the RFD-IE scheme is consistent wit the RIE model without any condition on the time step. 
\label{prop_RFD-IE}
\end{proposition}

\noindent
{\bf Proof:} (i) Taking the limit $\lambda \to 0$ in the RFD-EP scheme leads to: 
\begin{eqnarray}
& & \hspace{-1cm}  \delta^{-1} (n|_k^{m+1} - n|_k^m) + h^{-1} (\tilde f_n|_{k+1/2}^{m+1} - \tilde f_n|_{k-1/2}^{m+1})  = 0, \label{1FIE_n_fd_ap_2} \\
& & \hspace{-1cm} \delta^{-1}  ((nu)|_k^{m+1} - (nu)|_k^m) + h^{-1} (f_{u}|_{k+1/2}^{m} - f_{u}|_{k-1/2}^{m})  =   - n|_k^{m} \tilde E|_k^{m+1} , \label{1FIE_u_fd_ap_2} \\
& & \hspace{-1cm} - \frac{ h^{-2} \delta^2}{2}  \left[  (n|_{k+1}^{m} + n|_{k}^{m})  (\phi|_{k+1}^{m+1} - \phi|_{k}^{m+1})  - (n|_{k}^{m} + n|_{k-1}^{m})  (\phi|_{k}^{m+1} - \phi|_{k-1}^{m+1})  \right] = \nonumber \\
& & \hspace{3cm} 
=  1 - n|_k^{m} + h^{-1} \delta  (f_n|_{k+1/2}^{m} - f_n|_{k-1/2}^{m}) \nonumber \\
& & \hspace{3.5cm} 
 - \frac{ h^{-2} \delta^2}{2} (f_{u}|_{k+3/2}^{m} - f_{u}|_{k+1/2}^{m} - f_{u}|_{k-1/2}^{m} + f_{u}|_{k-3/2}^{m}) 
.  \label{1FIE_E_fd4} \\
& & \hspace{-1cm} E|_{k+1/2}^{m+1} = - h^{-1} (\phi|_{k+1}^{m+1} - \phi|_{k}^{m+1}),  \quad   \tilde E|_k^{m+1} = \frac{1}{2} (E|_{k+1/2}^{m+1} + E|_{k-1/2}^{m+1}), \label{1FIE_phi_fd_ap_2}
\end{eqnarray}
We now perform the same calculations as for the proof of proposition \ref{prop_EP_Gauss}. Inserting the first equation (\ref{1FIE_phi_fd_ap_2}) into (\ref{1FIE_E_fd4}), and having (\ref{1FEP_fn_m+1_fd2_mod}) in mind, we see that the latter is equivalent to 
\begin{eqnarray}
& & \hspace{-1cm}  0  =  1 - n|_k^{m} + h^{-1} \delta  (\tilde f_n|_{k+1/2}^{m+1} - \tilde f_n|_{k-1/2}^{m+1}) 
.  \label{1FIE_E_fd6} 
\end{eqnarray}
Inserting the discrete mass balance eq. (\ref{1FIE_n_fd_ap_2}) into (\ref{1FIE_E_fd6}) leads to (\ref{1FIE_n_fd_ap_3}). With the well-prepared initial condition assumption, we deduce that $n|_k^m = 1$ for all $m \geq 0$. Inserting this information into (\ref{1FIE_u_fd_ap_2}) and (\ref{1FEP_hydro_flux}) leads to (\ref{1FIE_u_fd_ap_3}) and (\ref{1FEP_hydro_flux_n=1}). For the momentum flux, we have removed the constant $p(1)$ because only differences of fluxes at successive mesh points are used and the constant $p(1)$ is cancelled in this process. Then, for $m \geq 1$, (\ref{1FIE_E_fd4}) implies (\ref{1FIE_E_fd1}). This ends the proof of point (i).  

\medskip
\noindent
(ii) Taking half the difference of (\ref{1FIE_u_fd_ap_3}) for $k+1$ and $k-1$, we obtain: 
\begin{eqnarray}
& & \hspace{-1cm} \delta^{-1}  ((f_n|_{k+1/2}^{m+1} - f_n|_{k-1/2}^{m+1}) - (f_n|_{k+1/2}^{m} - f_n|_{k-1/2}^{m}))  \nonumber \\
& & \hspace{2cm} - \frac{\delta^{-1}}{2} \left[ ( D_n|_{k+1/2}^{m+1} - D_n|_{k-1/2}^{m+1}) - (D_n|_{k+1/2}^{m} -  D_n|_{k-1/2}^{m+1}) \right]  + \nonumber \\
& & \hspace{2cm} + \frac{h^{-1}}{2} ((f_{u}|_{k+3/2}^{m} - f_{u}|_{k+1/2}^{m}) - (f_{u}|_{k-1/2}^{m} - f_{u}|_{k-3/2}^{m}) ) = \nonumber \\
& & \hspace{5cm} =  \frac{h^{-1}}{4} \left( \phi|_{k+2}^{m+1} - 2 \phi|_{k}^{m+1} +  \phi|_{k-2}^{m+1} \right) , \label{1FIE_u_fd_ap_5} 
\end{eqnarray}
Multiplying this equation by $ h^{-1}$ and subtracting (\ref{1FIE_E_fd1}) to it leads to 
\begin{eqnarray}
& & \hspace{-1cm}  \delta^{-1} h^{-1}  (f_n|_{k+1/2}^{m+1} - f_n|_{k-1/2}^{m+1}) = -   h^{-2} \left( \phi|_{k+1}^{m+1} - 2 \phi|_{k}^{m+1} +  \phi|_{k-1}^{m+1} \right) + \nonumber \\
& & \hspace{2cm} +   \frac{ h^{-2}}{4} \left( \phi|_{k+2}^{m+1} - 2 \phi|_{k}^{m+1} +  \phi|_{k-2}^{m+1} \right) \nonumber \\
& & \hspace{2cm} + \frac{h^{-1} \delta^{-1}}{2} \left[ ( D_n|_{k+1/2}^{m+1} - D_n|_{k-1/2}^{m+1}) - (D_n|_{k+1/2}^{m} -  D_n|_{k-1/2}^{m+1}) \right]
,  \label{1FIE_u_fd_ap_4} 
\end{eqnarray}
for all $m \geq 0$. Inserting (\ref{1FIE_u_fd_ap_4}) into (\ref{1FIE_E_fd1}) leads to (\ref{1FIE_E_fd5}) for $m \geq 1$. For $m=0$, the well prepared initial condition hypothesis on the velocity leads to the same eq. (\ref{1FIE_E_fd5}) without the last three terms. Now, the second and third terms of (\ref{1FIE_E_fd5}) are $O(h^2)$ approximations of $\partial^2_x \phi(x_k,t^m)$. Therefore, their difference is $O(h^2)$. 

We now turn to the last term at the right-hand side of (\ref{1FIE_E_fd5}). If the viscosity matrix is such that $D_n|_{k+1/2}^{m} = 0$ for all $k$, $m$, then, this term is identically zero and what precedes shows that (\ref{1FIE_E_fd5}) is a consistent approximation of (\ref{S_IE_n_2}). If this is not the case, we must estimate the last term at the right-hand side of (\ref{1FIE_E_fd5}). For this purpose, let us denote by $D_n(x,t)$ a function interpolating $D_n|_{k+1/2}^{m}$ in time and space. Then, Taylor's expansion up to second order about $(x_k, t^{m+1/2})$ shows that 
\begin{eqnarray}
& & \hspace{-1cm}  ( D_n|_{k+1/2}^{m+1} - D_n|_{k-1/2}^{m+1}) - (D_n|_{k+1/2}^{m} -  D_n|_{k-1/2}^{m+1}) = O(h^2 + \delta^2) || \partial^2 D_n ||_{\infty} = \nonumber \\
& & \hspace{9cm}  = O(h(h^2 + \delta^2)) ,  \label{1FIE_interpol_D} 
\end{eqnarray}
because the numerical viscosity is $O(h)$. Then, the third term at the right-hand side of (\ref{1FIE_E_fd5}) is $O(\delta + h^2 \delta^{-1})$, which is $O(h)$ if there exist two constants $C_1$ and $C_2$ such that and $C_1 h \leq \delta \leq C_2 h$. Under this condition, (\ref{1FIE_E_fd5}) is still a consistent approximation of (\ref{S_IE_n_2}).

That eqs. (\ref{1FIE_n_fd_ap_3}), (\ref{1FIE_u_fd_ap_3}) are consistent approximations of (\ref{S_IE_phi_3}) and (\ref{S_IE_u_2}) respectively is obvious. We conclude that the RFD-IE scheme is consistent with the IE model. This ends the proof of the proposition. \endproof

\noindent 
The previous proposition shows that the Fully-Discrete FD-EP scheme is AP. The assumption that $D_n|_{k+1/2}^{m} = 0$ for all $k$, $m$, is satisfied if the viscosity matrix is a scalar, like in the Rusanov scheme, or more generally, if the artificial viscosity applying to the density equation only involves the density. For instance, taking any arbitrary viscosity matrix, and replacing the line corresponding to the density by only a diagonal term would also satisfy this requirement. If this condition is not satisfied, then 
the AP property requires some "inverse CFL condition" $C_1 h \leq \delta$ to hold.

%%%%%%%%%%%%%%%%%%%%%%%%%%%%%%%%%%%%%%%%%%%%%%%%%%%%%%%%%%
\subsection{Euler-Poisson problem: conclusion}
\label{subsec_QN_EP_1F_conclusion}

In this section, we have provided an Asymptotic-Preserving (AP) scheme for the one-fluid Euler-Poisson problem in the quasineutral limit, i.e. when the scaled Debye length tends to zero. In this limit, the Euler-Poisson problem reduces to the Incompressible Euler problem. To construct such a scheme, the basic idea is to derive a reformulation of the Poisson equation which is not singular when $\lambda \to 0$. The AP scheme is therefore designed to mimic this reformulated Poisson equation. This is done by conveniently choosing the terms that must be evaluated implicitly. The AP scheme can be based on any classical shock capturing scheme. It consists in perturbing the standard hydrodynamic fluxes by corrective terms which are of the order of the time step or the mesh step, and which therefore do not emperish the consistency of the scheme. But in the limit $\lambda \to 0$, they do provide consistent approximations of the Incompressible Euler equations. A linearized stability analysis has been reviewed. It confirms that the stability condition of the AP-scheme is independent of $\lambda$ when $\lambda \to 0$. However, a nonlinear analysis of the stability of the scheme is still lacking.

%%%%%%%%%%%%%%%%%%%%%%%%%%%%%%%%%%%%%%%%%%%%%%%%%%%%%%%%%%%%%%%%%%%%%%%%%%%%%%%%%%%%%%%%%%%%%%%%
%%%%%%%%%%%%%%%%%%%%%%%%%%%%%%%%%%%%%%%%%%%%%%%%%%%%%%%%%%%%%%%%%%%%%%%%%%%%%%%%%%%%%%%%%%%%%%%%
\setcounter{equation}{0}
\section{The Euler-Maxwell system}
\label{sec_QN_EM_1F}

%%%%%%%%%%%%%%%%%%%%%%%%%%%%%%%%%%%%%%%%%%%%%%%%%%%%%%%%%%
\subsection{The Euler-Maxwell system and its quasi-neutral limit}
\label{subsec_1F_EMS}

%%%%%%%%%%%%%%%%%%%%%%%%%%%%%%%%%%%%%%%%%%%%%%%%%%%%%%%%%%
\subsubsection{The scaled Euler-Maxwell system}
\label{subsubsec_1F_EMS}

The one-fluid Euler-Maxwell (EM) system consists of the mass and momentum balance equations for the electron fluid coupled to the Maxwell equations. For simplicity, we state the problem in an already scaled form. Let $n^\lambda (x,t) \geq 0$ and $u^\lambda(x,t) \in {\mathbb R}^d $ stand for the electron density and electron velocity respectively, where $\lambda >0$ is the scaled Debye length (\ref{Debye}). They depend on the space-variable $x \in {\mathbb R}^d$ and on the time $t\geq0$. The electron pressure $p=p(n)$ is supposed to be a given function of $n$ (isentropic assumption) for simplicity. We assume that the dimension $d=3$ for this presentation. The electric field  $E^\lambda(x,t) \in {\mathbb R}^d $ and the magnetic field $B^\lambda(x,t) \in {\mathbb R}^d $ are the two components of the electro-magnetic field. The electric sources are the electrical charge $\rho^\lambda(x,t) \in {\mathbb R}$ and the electric current $j^\lambda(x,t) \in {\mathbb R}^d $. We assume a neutralizing background of immobile ions of constant density equal to $1$ in scaled units, and of average velocity $0$. With these definitions, the scaled Euler-Maxwell model is defined as follows:

\begin{definition}
The scaled Euler-Maxwell (EM) model is:
\begin{eqnarray}
& & \hspace{-1cm} \partial_t n^\lambda + \nabla \cdot (n^\lambda u^\lambda) = 0, \label{S1F_n} \\
& & \hspace{-1cm}  \partial_t (n^\lambda u^\lambda) + \nabla \cdot (n^\lambda u^\lambda \otimes u^\lambda) + \nabla p(n^\lambda) = - n^\lambda (E^\lambda+ u^\lambda \times B^\lambda), \label{S1F_u}\\
& & \hspace{-1cm} \partial_t B^\lambda + \nabla \times E^\lambda = 0, \label{S1F_B} \\
& & \hspace{-1cm} \lambda^2 \partial_t E^\lambda - \nabla \times B^\lambda =  j^\lambda := n^\lambda u^\lambda  , \label{S1F_E} \\
& & \hspace{-1cm} \nabla \cdot B^\lambda = 0, \label{S1F_divB} \\
& & \hspace{-1cm} \lambda^2 \nabla \cdot E^\lambda = \rho^\lambda := 1 - n^\lambda, \label{S1F_divE} 
\end{eqnarray}
\label{def_EM}
\end{definition}

\noindent
Eqs (\ref{S1F_n}), (\ref{S1F_u}) are the mass and momentum balance equations for the electrons. The right-hand side of (\ref{S1F_u}) is the Lorentz force, which depends on the electro-magnetic field $(E^\lambda,B^\lambda)$. The latter is a solution of the Maxwell equations (\ref{S1F_B})-(\ref{S1F_divE}). Classically, (\ref{S1F_B}), (\ref{S1F_E}) are respecctively the Faraday and Ampere equations, and provide the time evolution of the electro-magnetic field.  The Gauss and Ampere equations (\ref{S1F_divE}) and (\ref{S1F_E}) respectively involve the electric sources $(\rho^\lambda,j^\lambda)$ which depend on the hydrodynamical quantities $n^\lambda$ and $u^\lambda$. A lot more physics could be considered. For instance, instead of an isentropic gas equation-of-state, we could consider a full hydrodynamic model including an evolution equation for the gas total energy. We have also neglected electron-ion collisions which otherwise would introduce a friction term in (\ref{S1F_u}). Interactions with a neutral gas component could also be introduced. The present setting retains the features which will be important for the forthcoming discussion, but the concepts can easily be extended to more complex physics. 

Eqs. (\ref{S1F_divB}), (\ref{S1F_divE}) are constraints which are satisfied at all times provided they are satisfied at initial time. When dealing with approximations of the Euler-Maxwell system, we will have to check that the proposed schemes are actually consistent with discrete versions of these constraints. So, from now on, we will separate these constraints from the main system. That they are satisfied will be a property of the proposed approximations. 

The scaling of this problem as well as the material contained in this section is developed in detailed in \cite{Deg_Del_Sav_prep}. Here, we briefly summarize the scaling hypotheses. The scaling of the hydrodynamic unknowns is the same as in the electrostatic case (see section \ref{subsec_QN_EP_1F_scaling}). In addition, the charge and current density scales are fixed to $\rho_0 = e n_0$ and $j_0 = e n_0 u_0$. The electric field scale is fixed in a consistant way with the electrostatic case, namely $E_0 = p_0 / (e x_0 n_0)$. The magnetic field scale is fixed by $B_0 = E_0 / u_0$. With this choice of scales, two dimensionless parameters remain: the scaled Debye length $\lambda$ and the ratio of the velocity scale to the speed of light: 
\begin{eqnarray}
& & \hspace{-1cm} \alpha = \frac{u_0}{c}, \quad \lambda = \left( \frac{\epsilon_0 k_B T}{e^2 n_0 x_0^2} \right)^{1/2} .  \label{alpha_beta_lambda} 
\end{eqnarray}
We link $\alpha$ and $\lambda$ in such a way that the resulting limit model keeps the largest number of terms (the so-called least-degeneracy principle). This principle gives $\alpha = \lambda$. This collection of scaling hypotheses and principles gives rise to the scaled Euler-Maxwell model as written above.

%%%%%%%%%%%%%%%%%%%%%%%%%%%%%%%%%%%%%%%%%%%%%%%%%%%%%%%%%%
\subsubsection{The $\lambda \to 0$ limit and the Quasi-Neutral Euler-Maxwell system}
\label{subsubsec_1F_QNEM}

We now investigate the formal limit  $\lambda \to 0$ of the scaled EM system. 

\begin{proposition}
The formal limit $\lambda \to 0$ of the scaled EM system is the Quasi-Neutral Euler-Maxwell (QN-EM) system:  
\begin{eqnarray}
& & \hspace{-1cm} \nabla \cdot u^0 = 0, \label{QN1F_n} \\
& & \hspace{-1cm} \partial_t u^0 + \nabla \cdot ( u^0 \otimes u^0)  = - (E^0+ u^0 \times B^0), \label{QN1F_u}\\
& & \hspace{-1cm} \partial_t B^0 + \nabla \times E^0 = 0, \label{QN1F_B} \\
& & \hspace{-1cm}  - \nabla \times B^0 =  u^0  , \label{QN1F_E} \\
& & \hspace{-1cm} \nabla \cdot B^0 = 0, \label{QN1F_divB} \\
& & \hspace{-1cm} n^0=1, \label{QN1F_divE} 
\end{eqnarray}
\label{prop_QNEM}
\end{proposition}

\noindent 
{\bf Proof:} The formal limit of the EM system is: 
\begin{eqnarray}
& & \hspace{-1cm} \partial_t n^0 + \nabla \cdot (n^0 u^0) = 0, \label{QN1F_n_1} \\
& & \hspace{-1cm}  \partial_t (n^0 u^0) + \nabla \cdot (n^0 u^0 \otimes u^0) + \nabla p(n^0) = - n^0 (E^0+ u^0 \times B^0), \label{QN1F_u_1}\\
& & \hspace{-1cm} \partial_t B^0 + \nabla \times E^0 = 0, \label{QB1F_B_1} \\
& & \hspace{-1cm} - \nabla \times B^0 = n^0 u^0  . \label{QN1F_E_1} 
\end{eqnarray}
By assumption, the sequence of initial data satisfies  the constraints (\ref{S1F_divB}), (\ref{S1F_divE}) and consequently $n^0$ and $B^0$ satisfy (\ref{QN1F_divB}), (\ref{QN1F_divE}) at time $t=0$. From the divergence of (\ref{QN1F_E_1}), we deduce (\ref{QN1F_n}) and that $\partial_t n^0 = 0$. Therefore, (\ref{QN1F_divE}) is satisfied at all times. The other equations (\ref{QN1F_u}), (\ref{QN1F_B}), (\ref{QN1F_E}) follow immediately. \endproof

\medskip
\noindent
In this model, the divergence free constraint on $u^0$ is a consequence of (\ref{QN1F_E}), while the divergence free constraint on $B^0$ is a consequence of (\ref{QN1F_B}) (and of the divergence free initial data). The time evolutions of $u^0$ and $B^0$ are constrained by (\ref{QN1F_E}). $E^0$ is the Lagrange multiplier of this constraint. If we resolve this constraint, we find the following model:

\begin{definition}
The reformulated QN-EM model (or RQN-EM model) is:
\begin{eqnarray}
& & \hspace{-1cm} \nabla \cdot u^0 = 0, \label{QN1F_n_ref} \\
& & \hspace{-1cm}  \partial_t u^0 + \nabla \cdot ( u^0 \otimes u^0)  = - (E^0+ u^0 \times B^0), \label{QN1F_u_ref}\\
& & \hspace{-1cm} \partial_t B^0 + \nabla \times E^0 = 0, \label{QN1F_B_ref} \\
& & \hspace{-1cm}  \nabla \times ( \nabla \times E^0) + E^0 =  - \nabla \cdot ( u^0 \otimes u^0)   -  u^0 \times B^0, \label{QN1F_E_ref} \\
& & \hspace{-1cm} \nabla \cdot B^0 = 0, \label{QN1F_divB_ref} \\
& & \hspace{-1cm} n^0=1, \label{QN1F_divE_ref} 
\end{eqnarray}
\label{def_RQNEM}
\end{definition}

\begin{proposition}
The RQN-EM model is equivalent to the QN-EM model provided that the Ampere eq. (\ref{QN1F_E}) is satisfied at time $t=0$. 
\label{prop_equiv_QNEM-RQNEM}
\end{proposition}

\noindent
{\bf Proof:} Let $(n^0,u^0,E^0,B^0)$ be a solution of the QN-EM model. Taking the curl of (\ref{QN1F_B}), adding it to (\ref{QN1F_u}) and using (\ref{QN1F_E}) to cancel the time-derivatives leads to (\ref{QN1F_E_ref}). Conversely, if $(n^0,u^0,E^0,B^0)$ is a solution of the RQN-EM model, by proceeding in the same way backwards, we easily find that $\partial_t (\nabla \times B^0 + u^0) = 0$. Therefore,  (\ref{QN1F_E}) is satisfied for all times as soon as it is satisfied initially. \endproof

\noindent
Eq. (\ref{QN1F_E_ref}) is a well-posed elliptic equation for $E^0$ (provided suitable boundary conditions are given, such as perfectly conducting or absorbing boundary conditions). In  the QN-EM model, the hyperbolic character of the Maxwell equations is lost: $E^0$ adjusts to the variations of $B^0$ instantaneously. If the initial conditions of the EM model do not satisfy (\ref{QN1F_E}), an initial layer occurs, during which high frequency oscillations are produced. The QN-EM model produces some kind of time averaging of these high frequency oscillations. 

\begin{remark}
If we neglect the inertia of the electrons, which amounts to removing the drift term in the momentum equation (\ref{QN1F_u}), the QN-EM model reduces to:  
\begin{eqnarray*}
& & \hspace{-1cm} \partial_t B^0 + \nabla \times E^0 = 0,  \\
& & \hspace{-1cm} u^0 = - \nabla \times B^0   , \\
& & \hspace{-1cm} E^0+ u \times B^0=0, \\
& & \hspace{-1cm} \nabla \cdot B^0 = 0, 
\end{eqnarray*}
which is the so-called Electron-MagnetoHydrodynamics (EMH) system \cite{Gor_Kin_Rud_94}. Here, we do not make any assumption about the electron time scales, which leads to a slightly more complex dynamics. 
\label{rem_emh}
\end{remark}

\noindent 
In the limit $\lambda \to 0$, a change of type of the Maxwell models occurs: it shifts from hyperbolic to elliptic. This is the signataure that the EM model is a singularly perturbed problem in the limit $\lambda \to 0$. In the process of building an AP scheme, the first step is to reformulate the EM model in such a way that this singular perturbation character appears more explicitly. This leads us to introduce the following:

\begin{definition}
The Reformulated Euler-Maxwell (REM) model is:
\begin{eqnarray}
& & \hspace{-1cm} \partial_t n^\lambda + \nabla \cdot (n^\lambda u^\lambda) = 0, \label{RS1F_n} \\
& & \hspace{-1cm}  \partial_t (n^\lambda u^\lambda) + \nabla \cdot (n^\lambda u^\lambda \otimes u^\lambda) + \nabla p(n^\lambda) = - n^\lambda (E^\lambda+ u^\lambda \times B^\lambda), \label{RS1F_u}\\
& & \hspace{-1cm} \partial_t B^\lambda + \nabla \times E^\lambda = 0, \label{RS1F_B} \\
& & \hspace{-1cm}  \lambda^2 \partial^2_t E^\lambda + \nabla \times ( \nabla \times E^\lambda) + n^\lambda E^\lambda = - \nabla \cdot (n^\lambda u^\lambda \otimes u^\lambda) - \nabla p(n^\lambda)  - n^\lambda  u^\lambda \times B^\lambda,  \label{RS1F_E} \\
& & \hspace{-1cm} \nabla \cdot B^\lambda = 0, \label{RS1F_divB} \\
& & \hspace{-1cm} \lambda^2 \nabla \cdot E^\lambda = \rho^\lambda := 1 - n^\lambda, \label{RS1F_divE} 
\end{eqnarray}
\label{def_REM}
\end{definition}

\begin{proposition}
The REM model is equivalent to the EM model provided that the Ampere eq. (\ref{S1F_E}) is satisfied at time $t=0$. The formal limit $\lambda \to 0$ of the REM model is the RQN-EM model. 
\label{prop_equiv_EM-REM}
\end{proposition}

\noindent
{\bf Proof:} We proceed like for the proof of proposition \ref{prop_equiv_QNEM-RQNEM}. We take the curl of (\ref{S1F_B}), add it to (\ref{S1F_u}), and use (\ref{S1F_E}) to eliminate the time derivatives of $nu$ and $B$. This leads to (\ref{RS1F_E}). Conversely, proceeding backwards leads to $\partial_t (\lambda^2 \partial_t E^\lambda - \nabla \times B^\lambda - n^\lambda u^\lambda) = 0$. Therefore, (\ref{S1F_E}) is satisfied for all times as soon as it is satisfied initially. The second sentence of the proposition is obvious. \endproof

\noindent
Eq. (\ref{RS1F_E}) is a wave equation for $E^\lambda$ with wave-speed $\lambda^{-1}$. The condition that (\ref{S1F_E}) must be satisfied initially provides the Cauchy datum on $\partial_t E$ requested by this time second order problem. The use of the REM model preferably to the EM model, in conjuction with an implicit time discretization of (\ref{RS1F_E}), is the key for the build-up of an AP scheme for the EM model in the quasi-neutral limit $\lambda \to 0$, as we will see in the next sections.

%%%%%%%%%%%%%%%%%%%%%%%%%%%%%%%%%%%%%%%%%%%%%%%%%%%%%%%%%%
\subsection{Time Semi-Discretization, AP property and linearized stability}
\label{subsec_1F_time}

%%%%%%%%%%%%%%%%%%%%%%%%%%%%%%%%%%%%%%%%%%%%%%%%%%%%%%%%%%
\subsubsection{Time-Semi-discretization}
\label{subsubsec_1F_time}

The classical scheme for Euler-Maxwell equations uses a semi-implicit discretization of the Maxwell equations otherwise the scheme for the Maxwell part is inconditionally unstable. The stability requirement of S. Fabre \cite{Fabre_JCP_101_445} extended to the electromagnetic case also requests the Lorentz force in the momentum equation to be implicit. As shown below, this classical scheme requires a CFL condition of the type $\delta \leq O(\lambda)$. The AP scheme will require two additional levels of implicitness: the first one is an implicit evaluation of the mass flux, like in the electrostatic case (see section \ref{subsec_QN_EP_1F_time}). The second one is a totally implicit scheme for the Maxwell part. 
The classical and AP schemes can be put in a unified framework in the definition below

\begin{definition}
Introducing the following discretizations: 
\begin{eqnarray}
& & \hspace{-1cm} \delta^{-1} (n^{\lambda, m+1} - n^{\lambda, m}) + \nabla \cdot (n^{\lambda, m+a} u^{\lambda, m+a}) = 0, \label{DS1F_n} \\
& & \hspace{-1cm}  \delta^{-1} ( n^{\lambda, m+1} u^{\lambda, m+1} - n^{\lambda, m} u^{\lambda, m}) + \nabla \cdot (n^{\lambda, m} u^{\lambda, m} \otimes u^{\lambda, m}) + \nabla p(n^{\lambda, m}) = \nonumber \\
& & \hspace{5cm} - (n^{\lambda, m+1-a} E^{\lambda, m+1}+ n^{\lambda, m} u^{\lambda, m} \times B^{\lambda, m}), \label{DS1F_u}\\
& & \hspace{-1cm} \delta^{-1} (B^{\lambda, m+1} - B^{\lambda, m}) + \nabla \times E^{\lambda, m+b} = 0, \label{DS1F_B} \\
& & \hspace{-1cm} \lambda^2 \delta^{-1} (E^{\lambda, m+1} - E^{\lambda, m}) - \nabla \times B^{\lambda, m+c} =  n^{\lambda, m+a} u^{\lambda, m+a}  , \label{DS1F_E} 
\end{eqnarray}
with $a$, $b$ and $c$ taking the values $0$ or $1$, we define: 
\begin{itemize}
\item[(i)] The classical time semi-discrete scheme: $(a,b,c) = (0,1,0)$ or $(0,0,1)$. We will consider the case $(a,b,c) = (0,0,1)$ (i.e. the implicitness is in the Ampere eq.) to be specific. 
\item[(ii)] The AP Semi-Discrete Euler-Maxwell scheme (SD-EM): $(a,b,c) = (1,1,1)$. 
\end{itemize}
\label{def_SDEM}
\end{definition}

\noindent
For both schemes, the discrete Gauss equation and the divergence free constraint on $B$ are satisfied: 

\begin{proposition}
For both the classical and SD-EM schemes, we have: 
\begin{eqnarray}
& & \hspace{-1cm} \nabla \cdot B^{\lambda, m} = 0, \label{DS1F_divB} \\
& & \hspace{-1cm} \lambda^2 \nabla \cdot E^{\lambda, m} = 1 - n^{\lambda, m}, \label{DS1F_divE} 
\end{eqnarray}
for all $m \in {\mathbb N}$, assuming that these equations are satisfied initially (i.e. assuming that the initial data satisfy (\ref{DS1F_divB}), (\ref{DS1F_divE}) with $m=0$).  
\label{prop_SDEM_Gauss}
\end{proposition}

\noindent
{\bf Proof:} Taking the curl of (\ref{DS1F_B}), we find that $\nabla \cdot B^{\lambda, m}=\nabla \cdot B^{\lambda, 0}$ for all $m$. Eq. (\ref{DS1F_divB}) follows from the assumption on the initial condition. To prove (\ref{DS1F_divE}), we take the divergence of (\ref{DS1F_E}) and eliminate $\nabla \cdot (n^{\lambda, m+a} u^{\lambda, m+a})$ thanks to (\ref{DS1F_n}). We find that  $\lambda^2 \nabla \cdot E^{\lambda, m} - 1 + n^{\lambda, m} = \lambda^2 \nabla \cdot E^{\lambda, 0} - 1 + n^{\lambda, 0}$. Again, (\ref{DS1F_divE}) follows from the assumption on the initial condition. \endproof

\noindent
We note that the mass flux in the mass conservation equation and the current in the Ampere equation must have the same degree of implicitness in order to guarantee the consistency with the Gauss equation. For the SD-EM scheme, it is convenient to use an explicit evaluation of the density in the Lorentz force (\ref{DS1F_u}), because this reduces the complexity of the inversion of the implicit scheme. This choice does not restrict the AP-character of the scheme nor does it change its linearized stability properties. 
The classical scheme cannot be AP because, when taking the limit $\lambda \to 0$, it does not provide a valid recursion for the computation of the variables at time $m+1$, knowing their values at time $m$. By contrast, the SD-EM scheme does provide a valid recursion.

%%%%%%%%%%%%%%%%%%%%%%%%%%%%%%%%%%%%%%%%%%%%%%%%%%%%%%%%%%
\subsubsection{Reformulation and AP property}
\label{subsubsec_1F_time_refor}

The SD-EM scheme can be reformulated into the following scheme: 

\begin{definition}
The Reformulated Semi-Discrete Euler-Maxwell scheme (RSD-EM \\
scheme) is: 
\begin{eqnarray}
& & \hspace{-1cm} \delta^{-1} (n^{\lambda, m+1} - n^{\lambda, m}) + \nabla \cdot (n^{\lambda, m+1} u^{\lambda, m+1}) = 0, \label{RDS1F_n} \\
& & \hspace{-1cm}  \delta^{-1} ( n^{\lambda, m+1} u^{\lambda, m+1} - n^{\lambda, m} u^{\lambda, m}) + \nabla \cdot (n^{\lambda, m} u^{\lambda, m} \otimes u^{\lambda, m}) + \nabla p(n^{\lambda, m}) = \nonumber \\
& & \hspace{5cm} - (n^{\lambda, m} E^{\lambda, m+1}+ n^{\lambda, m} u^{\lambda, m} \times B^{\lambda, m}), \label{RDS1F_u}\\
& & \hspace{-1cm} \delta^{-1} (B^{\lambda, m+1} - B^{\lambda, m}) + \nabla \times E^{\lambda, m+1} = 0, \label{RDS1F_B} \\
& & \hspace{-1cm} (\lambda^2 + \delta^2 n^{\lambda, m}) E^{\lambda, m+1} + \delta^2 \nabla \times (\nabla \times E^{\lambda, m+1}) = \nonumber \\
& & \hspace{1cm} = \lambda^2 E^{\lambda, m} +  \delta \nabla \times B^{\lambda, m} + \delta n^{\lambda, m} u^{\lambda, m} \nonumber \\
& & \hspace{2cm} - \delta^2 ( \nabla \cdot (n^{\lambda, m} u^{\lambda, m} \otimes u^{\lambda, m}) + \nabla p(n^{\lambda, m}) +  n^{\lambda, m} u^{\lambda, m} \times B^{\lambda, m})  , \label{RDS1F_E} 
\end{eqnarray}
\label{def_RSDEM}
\end{definition}

\begin{proposition}
The SD-EM and RSD-EM schemes are equivalent.
\label{equiv_SDEM-RSDEM}
\end{proposition}

\noindent
{\bf Proof:} From the SD-EM scheme (see definition \ref{def_SDEM}, with $(a,b,c) = (1,1,1)$), we insert (\ref{DS1F_B}) and (\ref{DS1F_u}) into (\ref{DS1F_E}) and get (\ref{RDS1F_E}). Proceeding backwards, we get the equivalence of the two schemes. \endproof

\noindent
Now, we investigate the limit $\lambda \to 0$. We first state:

\begin{proposition}
The formal limit $\lambda \to 0$ of the RSD-EM scheme (with 'well-prepared' initial data satisfying  the constraints (\ref{DS1F_divB}), (\ref{DS1F_divE}) at $m=0$ for all $\lambda$, and such that their limits satisfy $\nabla \times B^{0, 0} + u^{0, 0} = 0$) is the following Reformulated Semi-Discrete Quasi-neutral Euler-Maxwell scheme (RSD-QN scheme): 
\begin{eqnarray}
& & \hspace{-1cm} \nabla \cdot (u^{0, m+1}) = 0, \label{RDQN1F_n} \\
& & \hspace{-1cm}  \delta^{-1} ( u^{0, m+1} - u^{0, m}) + \nabla \cdot (u^{0, m} \otimes u^{0, m})  = \nonumber \\
& & \hspace{5cm} - (E^{0, m+1}+ u^{0, m} \times B^{0, m}), \label{RDQN1F_u}\\
& & \hspace{-1cm} \delta^{-1} (B^{0, m+1} - B^{0, m}) + \nabla \times E^{0, m+1} = 0, \label{RDQN1F_B} \\
& & \hspace{-1cm} E^{0, m+1} + \nabla \times (\nabla \times E^{0, m+1}) =  - ( \nabla \cdot ( u^{0, m} \otimes u^{0, m}) +  u^{0, m} \times B^{0, m})  , \label{RDQN1F_E} \\
& & \hspace{-1cm} \nabla \cdot B^{0, m+1} = 0, \label{RDQN1F_divB} \\
& & \hspace{-1cm} n^{0, m} = 1.  \label{RDQN1F_divE} 
\end{eqnarray}
\label{prop_RSDQN}
\end{proposition}

\noindent
{\bf Proof:} The formal $\lambda \to 0$ of the RSD-EM scheme is: 
\begin{eqnarray}
& & \hspace{-1cm} \delta^{-1} (n^{0, m+1} - n^{0, m}) + \nabla \cdot (n^{0, m+1} u^{0, m+1}) = 0, \label{RDS1F_n_l=0} \\
& & \hspace{-1cm}  \delta^{-1} ( n^{0, m+1} u^{0, m+1} - n^{0, m} u^{0, m}) + \nabla \cdot (n^{0, m} u^{0, m} \otimes u^{0, m}) + \nabla p(n^{0, m}) = \nonumber \\
& & \hspace{5cm} - (n^{0, m} E^{0, m+1}+ n^{0, m} u^{0, m} \times B^{0, m}), \label{RDS1F_u_l=0}\\
& & \hspace{-1cm} \delta^{-1} (B^{0, m+1} - B^{0, m}) + \nabla \times E^{0, m+1} = 0, \label{RDS1F_B_l=0} \\
& & \hspace{-1cm} \delta n^{0, m} E^{0, m+1} + \delta \nabla \times (\nabla \times E^{0, m+1}) = \nabla \times B^{0, m} +  n^{0, m} u^{0, m} \nonumber \\
& & \hspace{2cm} - \delta ( \nabla \cdot (n^{0, m} u^{0, m} \otimes u^{0, m}) + \nabla p(n^{0, m}) +  n^{0, m} u^{0, m} \times B^{0, m})  . \label{RDS1F_E_l=0} 
\end{eqnarray}
Now, taking the divergence of (\ref{RDS1F_u_l=0}) and substracting $\delta^{-1}$ times the divergence of (\ref{RDS1F_E_l=0}) leads to \, \, $\nabla \cdot (n^{0, m+1} u^{0, m+1}) = 0$ and to \, $n^{0, m+1} - n^{0, m} = 0$. Using that the sequence of initial data satisfies (\ref{DS1F_divE}) for all $\lambda$, we deduce that $ n^{0, m} $ satisfies the initial condition  $n^{0, 0} = 1$ from which $n^{0, m} = 1$ for all $m$ follows. Inserting this into (\ref{RDS1F_n_l=0}) through (\ref{RDS1F_E_l=0}) leads to (\ref{RDQN1F_n}) , (\ref{RDQN1F_u}), (\ref{RDQN1F_B}) and to: 
\begin{eqnarray}
& & \hspace{-1cm} \delta E^{0, m+1} + \delta \nabla \times (\nabla \times E^{0, m+1}) = \nabla \times B^{0, m} + u^{0, m} \nonumber \\
& & \hspace{4cm} - \delta ( \nabla \cdot ( u^{0, m} \otimes u^{0, m}) +  u^{0, m} \times B^{0, m})  .  \label{RDQN1F_E_1} 
\end{eqnarray}
But, adding (\ref{RDQN1F_u}) to the curl of (\ref{RDQN1F_B}) and substracting $\delta^{-1}$ times (\ref{RDQN1F_E_1}) leads to $\nabla \times B^{0, m} + u^{0, m} = 0$ for all $m \in {\mathbb N}^*$. By assumption, the initial condition also satisfies this condition, with $m=0$. Inserting it into (\ref{RDQN1F_E_1}) leads to (\ref{RDQN1F_E}). The divergence free constraint on $B^{0, m+1}$ follows from taking the divergence of (\ref{RDQN1F_B}) and the divergence free assumption on the initial data. \endproof

\begin{proposition}
The RSD-QN scheme is consistent with the RQN-EM model. 
\label{limit_RSDEM-RSDQN}
\end{proposition}

\noindent
{\bf Proof:} Obvious by comparing the RSD-QN scheme and the RQN-EM model. \endproof

\noindent
Propositions \ref{prop_RSDQN} and \ref{limit_RSDEM-RSDQN} show that, in the limit $\lambda \to 0$, the SD-EM scheme is consistent with the Quasi-Neutral Euler-Maxwell model. In other words, the SD-EM scheme is AP, provided that it is Asmptotically stable. In the section below, we show that it is so, at least in the linearized setting.

%%%%%%%%%%%%%%%%%%%%%%%%%%%%%%%%%%%%%%%%%%%%%%%%%%%%%%%%%%
\subsubsection{Stability analysis}
\label{subsubsec_1F_EM_stability}

We will not provide the details of the stability analysis, because it follows the same principles as exposed in section \ref{subsec_QN_EP_1F_stability}. The details can be found in \cite{Deg_Del_Sav_prep}. Again, to mimic the effect of a space decentering in the hydrodynamic part of the model, we consider a viscous Euler-Maxwell model, with viscosity $\beta$ satisfying (\ref{beta}). We also consider a linearized model about the state $n^\lambda = 1$, $u^\lambda = 0$, $E^\lambda = 0$, $B^\lambda = 0$ and proceed to an $L^2$ stability analysis in Fourier space. Again, to mimic the effect of the space discretization, we restrict ourselves to Fourier modes satisfying $|\xi| \leq \pi/h$, where $h$ is underlying spatial discretization. Again, the Fourier transform of the solution to the linearized viscous EM system can be written in the form (\ref{FDLEP_general}). We use definition \ref{def_stab} for stability and we say that the scheme is Asymptotically Stable if it is stable under a condition on the time-step $\delta$ independent of $\lambda$ when $\lambda \to 0$. In \cite{Deg_Del_Sav_prep}, we prove: 

\begin{proposition}
(i) The Classical Schemes is not Asymptotically Stable.

\noindent
(ii) The SD-EM scheme is Asymptotically Stable, i.e. it is stable under the CFL condition $\delta \leq \Gamma h$ where $\Gamma$ is a constant independent when $\lambda$ for $\lambda \to 0$. 
\label{prop_stab}
\end{proposition}

%%%%%%%%%%%%%%%%%%%%%%%%%%%%%%%%%%%%%%%%%%%%%%%%%%%%%%%%%%
\subsection{Spatial discretization: enforcing the Gauss law}
\label{subsec_1F_space}

%%%%%%%%%%%%%%%%%%%%%%%%%%%%%%%%%%%%%%%%%%%%%%%%%%%%%%%%%%
\subsubsection{Classical fully-discrete scheme}
\label{subsubsec_1F_EM_spatial_class}

We present the spatial discretization in the one-dimensional framework. In this framework, we keep the $x$ and $y$ components of the velocity and of the electric field, and the $z$ component of the magnetic field. 

The discretization follows the same general principles as for the Euler-Poisson case (see section \ref{subsec_QN_EP_1F_space}). The discrete unknowns are $ n|_{k}^{m}$, $ u_x|_{k}^{m}$, $ u_y|_{k}^{m}$, $ E_x|_{k+1/2}^{m} $,  $ B_z|_{k+1/2}^{m}$,  $ E_y|_{k}^{m}$. We denote by $f_{n}|_{k+1/2}^{m} $, $f_{u_{x}}|_{k+1/2}^{m}$, $f_{u_{y}}|_{k+1/2}^{m} $ the explicit hydrodynamic fluxes for the mass and $x$ and $y$-components of the momentum conservation equations respectively, while the implicit mass flux of the AP scheme will be denoted by $\tilde f_{n}|_{k+1/2}^{m+1} $

\begin{definition}
The classical fully discrete scheme is: 
\begin{eqnarray}
& & \hspace{-1cm}  \delta^{-1} (n|_k^{m+1} - n|_k^m) + h^{-1} (f_n|_{k+1/2}^{m} - f_n|_{k-1/2}^{m})  = 0, \label{1F_n_fd_class} \\
& & \hspace{-1cm} \delta^{-1}  ((nu_x)|_k^{m+1} - (nu_x)|_k^m) + h^{-1} (f_{u_x}|_{k+1/2}^{m} - f_{u_x}|_{k-1/2}^{m})  = \nonumber \\
& & \hspace{5cm}  =  - n|_k^{m+1} \tilde E_x|_k^{m+1}  - (nu_y)|_k^m \,  \tilde B_z|_k^m, \label{1F_qx_fd_class} \\
& & \hspace{-1cm} \delta^{-1}  ((nu_y)|_k^{m+1} - (nu_y)|_k^m) + h^{-1} (f_{u_y}|_{k+1/2}^{m} - f_{u_y}|_{k-1/2}^{m})  = \nonumber \\
& & \hspace{5cm}  =  - n|_k^{m+1} E_y|_k^{m+1}  + (nu_x)|_k^m \, \tilde  B_z|_k^m, \label{1F_qy_fd_class} \\
& & \hspace{-1cm}  \delta^{-1} (B_z|_{k+1/2}^{m+1} - B_z|_{k+1/2}^m) + h^{-1} (E_y|_{k+1}^{m} - E_y|_{k}^{m+b}) = 0, \label{1F_faraday_fd_class} \\
& & \hspace{-1cm}  \lambda^2 \delta^{-1} (E_x|_{k+1/2}^{m+1} - E_x|_{k+1/2}^m) = f_n|_{k+1/2}^{m}, \label{1F_amperex_fd_class} \\
& & \hspace{-1cm}  \lambda^2 \delta^{-1} (E_y|_k^{m+1} - E_y|_k^m) + h^{-1} (B_z|_{k+1/2}^{m+1} - B_z|_{k-1/2}^{m+1}) = (n u_y)|_k^{m}, \label{1F_amperey_fd_class}  
\end{eqnarray}
where 
\begin{eqnarray}
& & \hspace{-1cm}   \tilde B_z|_k^m  = \frac{1}{2} (B_z|_{k+1/2}^{m} + B_z|_{k-1/2}^m) , \quad \tilde E_x|_k^{m+1} = \frac{1}{2} (E_x|_{k+1/2}^{m+1} + E_x|_{k-1/2}^{m+1}) .  \label{interp_Bz_Ex_class} 
\end{eqnarray}
The numerical fluxes $f_n|_{k+1/2}^{m}$, $f_{u_x}|_{k+1/2}^{m}$ and $f_{u_y}|_{k+1/2}^{m}$ are given by:
\begin{eqnarray}
& & \hspace{-1cm}  \left( \begin{array}{c} f_n|_{k+1/2}^{m} \\ f_{u_x}|_{k+1/2}^{m} \\ f_{u_y}|_{k+1/2}^{m} \end{array} \right) = \frac{1}{2} \left\{ \left( \begin{array}{c}   (nu_x)|_k^m \\ \left(nu_x^2 + p(n) \right)|_k^m \\ \left(nu_x u_y \right)|_k^m 
\end{array} \right) 
+ \left( \begin{array}{c}   (nu_x)|_{k+1}^m \\ \left(nu_x^2 + p(n) \right)|_{k+1}^m \\ \left( nu_x u_y \right)|_{k+1}^m  \end{array} \right) \right. + 
\nonumber \\
& & \hspace{6cm} \left. + \mu_{k+1/2}^m \, \left( \begin{array}{c}  n|_k^m - n|_{k+1}^m  \\ (nu_x)|_k^m - (nu_x)|_{k+1}^m  \\ (nu_y)|_k^m - (nu_y)|_{k+1}^m  \end{array} \right) \right\}, 
\label{1FEM_hydro_flux}
\end{eqnarray}
where $\mu_{k+1/2}^m$ is the viscosity matrix. 
\label{def_FDEM_class}
\end{definition}

\noindent
Again, we do not make any specific choice of a viscosity matrix and refer to the cited literature for typical expressions of the viscosity matrix. Like in the semi-discrete scheme, the classical scheme cannot be AP, because, in the limit $\lambda \to 0$, it does not provide a valid recursion to compute the unknowns at the time-step $m+1$. We now describe the AP fully discrete scheme.

%%%%%%%%%%%%%%%%%%%%%%%%%%%%%%%%%%%%%%%%%%%%%%%%%%%%%%%%%%
\subsubsection{AP fully-discrete scheme}
\label{subsubsec_1F_EM_spatial_AP}

\begin{definition}
The AP Fully-Discrete Euler-Maxwell scheme (FD-EM scheme) is:
\begin{eqnarray}
& & \hspace{-1cm}  \delta^{-1} (n|_k^{m+1} - n|_k^m) + h^{-1} (\tilde f_n|_{k+1/2}^{m+1} - \tilde f_n|_{k-1/2}^{m+1})  = 0, \label{1F_n_fd} \\
& & \hspace{-1cm} \delta^{-1}  ((nu_x)|_k^{m+1} - (nu_x)|_k^m) + h^{-1} (f_{u_x}|_{k+1/2}^{m} - f_{u_x}|_{k-1/2}^{m})  = \nonumber \\
& & \hspace{5cm}  =  - n|_k^{m} \tilde E_x|_k^{m+1}  - (nu_y)|_k^m \,  \tilde B_z|_k^m, \label{1F_qx_fd} \\
& & \hspace{-1cm} \delta^{-1}  ((nu_y)|_k^{m+1} - (nu_y)|_k^m) + h^{-1} (f_{u_y}|_{k+1/2}^{m} - f_{u_y}|_{k-1/2}^{m})  = \nonumber \\
& & \hspace{5cm}  =  - n|_k^{m} E_y|_k^{m+1}  + (nu_x)|_k^m \, \tilde  B_z|_k^m, \label{1F_qy_fd} \\
& & \hspace{-1cm}  \delta^{-1} (B_z|_{k+1/2}^{m+1} - B_z|_{k+1/2}^m) + h^{-1} (E_y|_{k+1}^{m+1} - E_y|_{k}^{m+1}) = 0, \label{1F_faraday_fd} \\
& & \hspace{-1cm}  \lambda^2 \delta^{-1} (E_x|_{k+1/2}^{m+1} - E_x|_{k+1/2}^m) = \tilde f_n|_{k+1/2}^{m+1}, \label{1F_amperex_fd} \\
& & \hspace{-1cm}  \lambda^2 \delta^{-1} (E_y|_k^{m+1} - E_y|_k^m) + h^{-1} (B_z|_{k+1/2}^{m+1} - B_z|_{k-1/2}^{m+1}) = (n u_y)|_k^{m+1}, \label{1F_amperey_fd}  
\end{eqnarray}
where 
\begin{eqnarray}
& & \hspace{-1cm}   \tilde B_z|_k^m  = \frac{1}{2} (B_z|_{k+1/2}^{m} + B_z|_{k-1/2}^m) , \quad \tilde E_x|_k^{m+1} = \frac{1}{2} (E_x|_{k+1/2}^{m+1} + E_x|_{k-1/2}^{m+1}) .   \label{interp_Bz_Ex} 
\end{eqnarray}
The numerical fluxes $f_{u_x}|_{k+1/2}^{m}$ and $f_{u_y}|_{k+1/2}^{m}$ are given by (\ref{1FEM_hydro_flux}), while the implicit flux $\tilde f_n|_{k+1/2}^{m+1}$ is given by: 
\begin{eqnarray}
& & \hspace{-1cm}  \tilde f_n|_{k+1/2}^{m+1} = f_n|_{k+1/2}^{m} - \frac{\delta}{2} (n|_{k+1}^m + n|_{k}^m)  \, E_x|_{k+1/2}^{m+1}  \nonumber \\
& & \hspace{0cm} - \frac{\delta h^{-1}}{2} (f_{u_x}|_{k+3/2}^{m} - f_{u_x}|_{k-1/2}^{m}) - \frac{\delta}{2} ( (nu_y)|_k^m  + (nu_y)|_{k+1}^m ) \, B_z|_{k+1/2}^m  
. \label{1F_fn_m+1_fd2_mod} 
\end{eqnarray}
\label{def_FDEM}
\end{definition}

\noindent
Like in the Euler-Poisson case, (\ref{1F_fn_m+1_fd2_mod}) is obtained from impliciting the central discretization part of the classical flux (\ref{1FEM_hydro_flux}), while keeping the numerical viscosity term  explicit. Then, using the momentum balance equation (\ref{1F_qx_fd}) and the same kind of lumping of the electric field average as for (\ref{1FEP_fn_m+1_fd2_mod}), it is easy to derive (\ref{1F_fn_m+1_fd2_mod}). The details are left to the reader. 

The current $(nu)_x$ in the $x$-component of the Ampere equation (\ref{1F_amperex_fd}) is evaluated by using the mass flux $\tilde f_n|_{k+1/2}^{m+1}$. At the level of the continuous problem, these two quantities are identical. Therefore, this approximation is consistent. However, using the mass flux rather than the current allows us to guarantee a perfect consistency with the Gauss equation, as shown below: 

\begin{proposition}
The Gauss equation 
\begin{eqnarray}
& & \hspace{-1cm}  \lambda^2  h^{-1} (E_x|_{k+1/2}^{m} - E_x|_{k-1/2}^{m}) = 1 - n|_{k}^{m} ,  \label{1F_gauss_fd_2}
\end{eqnarray}
is satisfied for all $m \in {\mathbb N}^*$ provided that it is satisfied by the initial condition (i.e. with $m=0$). 
\label{prop_gauss_EM}
\end{proposition}

\noindent
{\bf Proof: } Taking the difference of (\ref{1F_amperex_fd}) evaluated at $x_{k+1/2}$ and $x_{k-1/2}$ and using (\ref{1F_n_fd}), we easily check that:
\begin{eqnarray}
& & \hspace{-1cm}  \lambda^2  h^{-1} (E_x|_{k+1/2}^{m+1} - E_x|_{k-1/2}^{m+1}) + n|_{k}^{m+1} = \lambda^2  h^{-1} (E_x|_{k+1/2}^{m} - E_x|_{k-1/2}^{m}) + n|_{k}^{m}.  \label{1F_gauss_fd}
\end{eqnarray}
Then, proposition \ref{prop_gauss_EM} follows. \endproof

\noindent
We note that this proof would apply to the classical scheme as well. In the $y$-component of the Ampere equation (\ref{1F_amperex_fd}), the current is evaluated using the usual approximation $(n u_y)|_k^{m+a}$ because, in a one-dimensional problem, the $y$-component of the mass flux is independent of $x$ and does not enter the mass balance. In a 2 or 3-dimensional problem, one should evaluate all components of the current using the corresponding components of the mass flux, to ensure consistency with the Gauss equation. 

We now reformulate the FD-EM scheme in order to study the limit $\lambda \to 0$.

%%%%%%%%%%%%%%%%%%%%%%%%%%%%%%%%%%%%%%%%%%%%%%%%%%%%%%%%%%
\subsubsection{Reformulation and AP property}
\label{subsubsec_1F_EM_spatial_AP_reform}

\begin{definition}
The Reformulated Fully-Discrete Euler-Maxwell scheme (RFD-EM \\
scheme) is
\begin{eqnarray}
& & \hspace{-1cm}  \delta^{-1} (n|_k^{m+1} - n|_k^m) + h^{-1} (\tilde f_n|_{k+1/2}^{m+1} - \tilde f_n|_{k-1/2}^{m+1})  = 0, \label{1F_n_fd_ref} \\
& & \hspace{-1cm} \delta^{-1}  ((nu_x)|_k^{m+1} - (nu_x)|_k^m) + h^{-1} (f_{u_x}|_{k+1/2}^{m} - f_{u_x}|_{k-1/2}^{m})  = \nonumber \\
& & \hspace{5cm}  =  - n|_k^{m} \tilde E_x|_k^{m+1}  - (nu_y)|_k^m \,  \tilde B_z|_k^m, \label{1F_qx_fd_ref} \\
& & \hspace{-1cm} \delta^{-1}  ((nu_y)|_k^{m+1} - (nu_y)|_k^m) + h^{-1} (f_{u_y}|_{k+1/2}^{m} - f_{u_y}|_{k-1/2}^{m})  = \nonumber \\
& & \hspace{5cm}  =  - n|_k^{m} E_y|_k^{m+1}  + (nu_x)|_k^m \, \tilde  B_z|_k^m, \label{1F_qy_fd_ref} \\
& & \hspace{-1cm}  \delta^{-1} (B_z|_{k+1/2}^{m+1} - B_z|_{k+1/2}^m) + h^{-1} (E_y|_{k+1}^{m+1} - E_y|_{k}^{m+1}) = 0, \label{1F_faraday_fd_ref} \\
& & \hspace{-1cm}  (\lambda^2  +  \frac{\delta^2}{2} (n|_{k+1}^m + n|_{k}^m) )  \, E_x|_{k+1/2}^{m+1}  = \lambda^2 E_x|_{k+1/2}^m + \delta f_{n}|_{k+1/2}^{m}  \nonumber \\
& & \hspace{0cm} - \frac{\delta^2 h^{-1}}{2} (f_{u_x}|_{k+3/2}^{m} - f_{u_x}|_{k-1/2}^{m}) - \frac{\delta^2}{2}( (nu_y)|_k^m  + (nu_y)|_{k+1}^m ) \, B_z|_{k+1/2}^m , 
\label{1F_amperex_fd_mod}  \\
& & \hspace{-1cm} (\lambda^2  + \delta^{2} n|_k^m) E_y|_k^{m+1} - \delta^{2} h^{-2} (E_y|_{k+1}^{m+1} - 2 E_y|_k^{m+1} + E_y|_{k-1}^{m+1}) = \lambda^2 E_y|_k^m  + \delta (nu_y)|_k^m \nonumber \\
& & \hspace{-0.5cm}   - \delta h^{-1} (B_z|_{k+1/2}^{m} - B_z|_{k-1/2}^{m})  - \delta^{2} h^{-1}  (f_{u_y}|_{k+1/2}^{m} - f_{u_y}|_{k-1/2}^{m}) + \delta^{2} (nu_x)|_k^m \, \tilde  B_z|_k^m
, \label{1F_amperey_ref_fd}     
\end{eqnarray}
with $\tilde E_x|_k^{m}$ and $\tilde  B_z|_k^m$ given by (\ref{interp_Bz_Ex}).
\label{def_RFDEM}
\end{definition}

\begin{proposition}
The FD-EM and RFD-EM schemes are equivalent. 
\label{prop_equiv_FDEM-RFDEM}
\end{proposition}

\noindent
{\bf Proof:} We first consider $E_y^{m+1}$: inserting (\ref{1F_faraday_fd}) and (\ref{1F_qy_fd}) into (\ref{1F_amperey_fd}) to eliminate $B_z|_{k+1/2}^{m+1}$, and $(nu_y)|_k^{m+1}$ respectively, we find that (\ref{1F_amperey_fd}) is equivalent to (\ref{1F_amperey_ref_fd}). 

We now examine $E_x^{m+1}$. To get (\ref{1F_amperex_fd_mod}), we simply insert the expression (\ref{1F_fn_m+1_fd2_mod}) of the mass flux into (\ref{1F_amperex_fd}). \endproof

\noindent 
Eq. (\ref{1F_amperey_ref_fd}) is a well-posed discrete elliptic equation provided that suitable boundary conditions are defined. It allows to compute $E_y|_k^{m+1}$ from known values. Eq. (\ref{1F_amperex_fd_mod}) provides a direct explicit evaluation of $E_x|_{k+1/2}^{m+1}$.

We now investigate the $\lambda \to 0$ limit. We have:

\begin{proposition}
(i) The formal $\lambda \to 0$ limit of the RFD-EM scheme (with 'well-prepared' initial data satisfying the discrete Gauss eq. (\ref{1F_gauss_fd_2}) at $m=0$ for all $\lambda$ and such that their limits satisfy $u_y|_k^0 - h^{-1} (B_z|_{k+1/2}^{0} - B_z|_{k-1/2}^{0}) = 0$) is
\begin{eqnarray}
& & \hspace{-1cm}  n|_k^{m+1} = 1, \label{FDQN_n_fd_ref} \\
& & \hspace{-1cm} \delta^{-1}  (u_x|_k^{m+1} - u_x|_k^m) + h^{-1} (f_{u_x}|_{k+1/2}^{m} - f_{u_x}|_{k-1/2}^{m})  =   -  \tilde E_x|_k^{m+1}  - u_y|_k^m \,  \tilde B_z|_k^m, \label{FDQN_qx_fd_ref} \\
& & \hspace{-1cm} \delta^{-1}  (u_y|_k^{m+1} - u_y|_k^m) + h^{-1} (f_{u_y}|_{k+1/2}^{m} - f_{u_y}|_{k-1/2}^{m})  =  - E_y|_k^{m+1}  + u_x|_k^m \, \tilde  B_z|_k^m, \label{FDQN_qy_fd_ref} \\
& & \hspace{-1cm}  \delta^{-1} (B_z|_{k+1/2}^{m+1} - B_z|_{k+1/2}^m) + h^{-1} (E_y|_{k+1}^{m+1} - E_y|_{k}^{m+1}) = 0, \label{FDQN_faraday_fd_ref} \\
& & \hspace{-1cm}   \, E_x|_{k+1/2}^{m+1}  =    - \frac{h^{-1}}{2} (f_{u_x}|_{k+3/2}^{m} - f_{u_x}|_{k-1/2}^{m}) - \frac{1}{2}( u_y|_k^m  + u_y|_{k+1}^m ) \, B_z|_{k+1/2}^m + \nonumber \\
& & \hspace{9.5cm}  + \delta^{-1}  f_{n}|_{k+1/2}^{m} , 
\label{FDQN_amperex_fd_mod}  \\
& & \hspace{-1cm}   E_y|_k^{m+1} - h^{-2} (E_y|_{k+1}^{m+1} - 2 E_y|_k^{m+1} + E_y|_{k-1}^{m+1}) =  \nonumber \\
& & \hspace{3cm} = -  h^{-1}  (f_{u_y}|_{k+1/2}^{m} - f_{u_y}|_{k-1/2}^{m}) +  u_x|_k^m \, \tilde  B_z|_k^m . \label{FDQN_amperey_ref_fd_2}     
\end{eqnarray}
with $\tilde E_x|_k^{m}$ and $\tilde  B_z|_k^m$ given by (\ref{interp_Bz_Ex}) and with the numerical fluxes: 
\begin{eqnarray}
& & \hspace{-1.5cm}  \left( \begin{array}{c} f_n|_{k+1/2}^{m} \\ f_{u_x}|_{k+1/2}^{m} \\ f_{u_y}|_{k+1/2}^{m} \end{array} \right) = \frac{1}{2}  \left( \begin{array}{c}   u_x|_k^m + u_x|_{k+1}^m \\ u_x^2|_k^m + u_x^2|_{k+1}^m\\ \left(u_x u_y \right)|_k^m + \left( u_x u_y \right)|_{k+1}^m 
\end{array} \right) 
+ \frac{\mu_{k+1/2}^m}{2} \, \left( \begin{array}{c}  0  \\ u_x|_k^m - u_x|_{k+1}^m  \\ u_y|_k^m - u_y|_{k+1}^m  \end{array} \right) , 
\label{1FEM_hydro_flux_n=1}
\end{eqnarray}

(ii)  We denote by:
\begin{eqnarray}
& & \hspace{-1cm}
 \left( \begin{array}{c} D_n|_{k+1/2}^m \\ D_{u_x}|_{k+1/2}^m  \\ D_{u_y}|_{k+1/2}^m  \end{array} \right) =  \mu_{k+1/2}^m \, \left( \begin{array}{c}  0  \\ u_x|_k^m - u_x|_{k+1}^m \\ u_y|_k^m - u_y|_{k+1}^m \end{array} \right).
\label{1FEM_hydro_num_visc}
\end{eqnarray}
Then, for $m \geq 1$, (\ref{FDQN_amperex_fd_mod}) is equivalent to: 
\begin{eqnarray}
& & \hspace{-1cm}   \, E_x|_{k+1/2}^{m+1}  =    - \frac{h^{-1}}{2} (f_{u_x}|_{k+3/2}^{m} - f_{u_x}|_{k-1/2}^{m}) - \frac{1}{2}( u_y|_k^m  + u_y|_{k+1}^m ) \, B_z|_{k+1/2}^m  \nonumber \\
& & \hspace{1cm}  -\frac{1}{4} (E_x|_{k+3/2}^{m} -2 E_x|_{k+1/2}^{m} + E_x|_{k-1/2}^{m}) \nonumber \\
& & \hspace{1cm} 
-\frac{1}{4} ( u_y|_{k+1}^{m-1} (B_z|_{k+3/2}^{m-1} - B_z|_{k+1/2}^{m-1}) - u_y|_k^{m-1}  (B_z|_{k+1/2}^{m-1} -  B_z|_{k-1/2}^{m-1}))  \nonumber \\
& & \hspace{1cm} + \frac{\delta^{-1}}{2} ( D_n|_{k+1/2}^m - D_n|_{k+1/2}^{m-1}) . 
\label{FDQN_amperex_fd_mod_2}  
\end{eqnarray}
for all $m \geq 1$. Then, this scheme (the Reformulated Fully-Discrete Quasi-Neutral Euler-Maxwell (RFD-QN) scheme) is consistent with the RQN-EM model. 
\label{prop_RFDEM_lto0}
\end{proposition}

\medskip
\noindent
{\bf Proof:} (i) Taking the limit $\lambda \to 0$ in the RFD-EM scheme only modifies (\ref{1F_amperex_fd_mod}), (\ref{1F_amperey_ref_fd}) into 
\begin{eqnarray}
& & \hspace{-1cm}   \frac{\delta}{2} (n|_{k+1}^m + n|_{k}^m)  \, E_x|_{k+1/2}^{m+1}  =  f_{n}|_{k+1/2}^{m}  \nonumber \\
& & \hspace{0cm} - \frac{\delta h^{-1}}{2} (f_{u_x}|_{k+3/2}^{m} - f_{u_x}|_{k-1/2}^{m}) - \frac{\delta}{2}( (nu_y)|_k^m  + (nu_y)|_{k+1}^m ) \, B_z|_{k+1/2}^m , 
\label{QN_amperex_fd_mod}  \\
& & \hspace{-1cm}  \delta n|_k^m E_y|_k^{m+1} - \delta h^{-2} (E_y|_{k+1}^{m+1} - 2 E_y|_k^{m+1} + E_y|_{k-1}^{m+1}) =  (nu_y)|_k^m \nonumber \\
& & \hspace{-1cm}   -  h^{-1} (B_z|_{k+1/2}^{m} - B_z|_{k-1/2}^{m})  - \delta h^{-1}  (f_{u_y}|_{k+1/2}^{m} - f_{u_y}|_{k-1/2}^{m}) + \delta (nu_x)|_k^m \, \tilde  B_z|_k^m
. \label{QN_amperey_ref_fd}     
\end{eqnarray}
Comparing (\ref{QN_amperex_fd_mod}) and (\ref{1F_fn_m+1_fd2_mod}) shows that $\tilde f_{n}|_{k+1/2}^{m+1} = 0$, and, inserting it into (\ref{1F_n_fd_ref}), that (\ref{FDQN_n_fd_ref}) is satisfied for all $m \geq 1$. With the assumption that the initial condition satisfies (\ref{1F_gauss_fd_2}) for all $\lambda$, we deduce that $n|_k^{m} = 1$ for all integer $m \geq 0$. Inserting this into (\ref{1F_qx_fd_ref}), (\ref{1F_qy_fd_ref}), (\ref{1F_faraday_fd_ref}), (\ref{QN_amperex_fd_mod}) and (\ref{1FEM_hydro_flux}), we are led to (\ref{FDQN_qx_fd_ref}), (\ref{FDQN_qy_fd_ref}), (\ref{FDQN_faraday_fd_ref}), (\ref{FDQN_amperex_fd_mod}) and (\ref{1FEM_hydro_flux_n=1}).  

Concerning (\ref{QN_amperey_ref_fd}), at this point, we get: 
\begin{eqnarray}
& & \hspace{-1cm}   E_y|_k^{m+1} - h^{-2} (E_y|_{k+1}^{m+1} - 2 E_y|_k^{m+1} + E_y|_{k-1}^{m+1}) =  \nonumber \\
& & \hspace{3cm} = -  h^{-1}  (f_{u_y}|_{k+1/2}^{m} - f_{u_y}|_{k-1/2}^{m}) +  u_x|_k^m \, \tilde  B_z|_k^m \nonumber \\
& & \hspace{5cm}  + \delta^{-1} (u_y|_k^m  - h^{-1} (B_z|_{k+1/2}^{m} - B_z|_{k-1/2}^{m}) ) 
. \label{FDQN_amperey_ref_fd}     
\end{eqnarray}
However, taking the differences of (\ref{FDQN_faraday_fd_ref}) for $k+1/2$ and $k-1/2$, multiplying the result by $h^{-1}$ and subtracting the result to (\ref{FDQN_qy_fd_ref}) leads to 
\begin{eqnarray}
& & \hspace{-1cm}   u_y|_k^{m+1}  - h^{-1} (B_z|_{k+1/2}^{m+1} - B_z|_{k-1/2}^{m+1}) = 0  
, \label{FDQN_amperey_ref_fd_5}     
\end{eqnarray}
for all $m \geq 0$. But with the well-prepared initial data assumption, (\ref{FDQN_amperey_ref_fd_5}) is valid with $m+1$ repaced by $m$, for all $m \geq 0$. Inserting this identity into (\ref{FDQN_amperey_ref_fd}) leads to (\ref{FDQN_amperey_ref_fd_2}). This concludes point (i) of the proposition.

(ii) All equations in the resulting scheme are consistent with the RQN-EM model, except (\ref{FDQN_amperex_fd_mod}),  where an $O(\delta^{-1})$ terms appear: $\delta^{-1} f_{n}|_{k+1/2}^{m}$. In this part of the proof, we show that $\delta^{-1} f_{n}|_{k+1/2}^{m}$ can be substituted with the last three lines of (\ref{FDQN_amperex_fd_mod_2}). Indeed, taking half the sum of (\ref{FDQN_qx_fd_ref}) for $k$ and $k+1$, we can write: 
\begin{eqnarray}
& & \hspace{-1cm} \delta^{-1}  ( (f_{n}|_{k+1/2}^{m+1} - \frac{1}{2} D_n|_{k+1/2}^{m+1} ) -  ( f_{n}|_{k+1/2}^{m} - \frac{1}{2} D_n|_{k+1/2}^{m} ) ) = \nonumber \\
& & \hspace{1.6cm} = - \frac{h^{-1}}{2} (f_{u_x}|_{k+3/2}^{m} - f_{u_x}|_{k-1/2}^{m}) \nonumber \\
& & \hspace{2cm} -\frac{1}{4} (E_x|_{k+3/2}^{m+1} +2 E_x|_{k+1/2}^{m+1} + E_x|_{k-1/2}^{m+1}) \nonumber \\
& & \hspace{2cm} 
-\frac{1}{4} ( u_y|_{k+1}^m (B_z|_{k+3/2}^{m} + B_z|_{k+1/2}^{m}) + u_y|_k^m  (B_z|_{k+1/2}^{m} +  B_z|_{k-1/2}^{m}))  . 
\label{FDQN_amperex_fd_mod_3}  
\end{eqnarray}
The, subtracting (\ref{FDQN_amperex_fd_mod}) to (\ref{FDQN_amperex_fd_mod_3}) leads to
\begin{eqnarray}
& & \hspace{-1cm} \delta^{-1} f_{n}|_{k+1/2}^{m+1} = -\frac{1}{4} (E_x|_{k+3/2}^{m+1} -2 E_x|_{k+1/2}^{m+1} + E_x|_{k-1/2}^{m+1}) \nonumber \\
& & \hspace{2cm} 
-\frac{1}{4} ( u_y|_{k+1}^m (B_z|_{k+3/2}^{m} - B_z|_{k+1/2}^{m}) - u_y|_k^m  (B_z|_{k+1/2}^{m} -  B_z|_{k-1/2}^{m})) \nonumber \\
& & \hspace{2cm} + \frac{\delta^{-1}}{2} (  D_n|_{k+1/2}^{m+1} - D_n|_{k+1/2}^{m} ) . 
\label{FDQN_amperex_fd_mod_4}  
\end{eqnarray}
Substituting $\delta^{-1} f_{n}|_{k+1/2}^{m}$ in (\ref{FDQN_amperex_fd_mod})  by its expression deduced from (\ref{FDQN_amperex_fd_mod_4}) (with $m$ replaced by $m-1$) leads to (\ref{FDQN_amperex_fd_mod_2}) for all $m \geq 1$. 

Now, we see that the second and third lines of (\ref{FDQN_amperex_fd_mod_2}) are at leading order, equal to $-(h^2/4)$ $((\partial_x^2 E_x)|_{k+1/2}^m + (\partial_x (u_y \partial_x B_z))|_{k+1/2}^{m-1} )$ and tend to zero with $h$. 

Then, the fourth line can be expanded using Taylor's expansion and is of the order of $(\partial_t D_n)(x_{k+1/2},t^{m-1/2})$ (supposing that the artificial viscosity term can be interpolated by a smooth function $D_n(x,t)$). But the artificial viscosity is $O(h)$ (see (\ref{1FEM_hydro_num_visc})). Therefore, this term is $O(h)$ and tends to zero with $h$ as well.

Therefore, (\ref{FDQN_amperex_fd_mod_2}) is consistent with (\ref{QN1F_E_ref}) (having in mind that, in this 1-D geometry, the $x$ component of $\nabla \times (\nabla \times E)$ is identically zero. Then, the RFD-QN scheme is clearly consistent with the RQN-EM model. This concludes the second point of the proposition. \endproof

\noindent
This proposition shows that the RFD-EM scheme is AP. The assumption of well-prepared initial conditions can be removed. In this case, the scheme takes its form as stated in the proposition for $m \geq 1$. We note that, in the Euler-Maxwell case, there is no need for an inverse CFL condition when the viscosity terms $D_n|_{k+1/2}^{m}$ are not identically zero (see discussion at the end of section \ref{subsubsec_QN_EP_1F_space_AP_lim_l_to_0} for comparison with the Euler-Poisson case).

%%%%%%%%%%%%%%%%%%%%%%%%%%%%%%%%%%%%%%%%%%%%%%%%%%%%%%%%%%
\subsection{Euler-Maxwell problem: conclusion}
\label{subsec_QN_EM_1F_conclusion}

In this section, following the same strategy as for the Euler-Poisson problem, we have provided an Asymptotic-Preserving (AP) scheme for the one-fluid Euler-Maxwell problem in the quasineutral limit. However, the quasi-neutral limit of the Euler-Maxwell model is more complicated than that of the Euler-Poisson problem and gives rise to infinite propagation speed of electromagnetic waves. Again, following the previous strategy, we provide a reformulation of the Euler-Maxwell model which is not singular when $\lambda \to 0$. The AP scheme is therefore designed to mimic this reformulated Euler-Maxwell system. Again, it can be based on any classical shock capturing scheme and consists in perturbing the standard hydrodynamic fluxes by corrective terms which are of the order of the time step or the mesh step. However, additionally, a fully implicit treatment of the Maxwell equations must also be implemented. A linearized stability analysis has been reviewed. It confirms that the stability condition of the AP-scheme is independent of $\lambda$ when $\lambda \to 0$. Again, non-linear stability results are still open.

%%%%%%%%%%%%%%%%%%%%%%%%%%%%%%%%%%%%%%%%%%%%%%%%%%%%%%%%%%%%%%%%%%%%%%%%%%%%%%%%%%%%%%%%%%%%%%%%
%%%%%%%%%%%%%%%%%%%%%%%%%%%%%%%%%%%%%%%%%%%%%%%%%%%%%%%%%%%%%%%%%%%%%%%%%%%%%%%%%%%%%%%%%%%%%%%%
\setcounter{equation}{0}
\section{Extensions}
\label{sec_QN_extension}

\medskip
\noindent
{\bf Full Euler, Navier-Stokes, etc.} The AP-methodology extends straightforwardly when the isothermal or isentropic gas dynamics model is replaced by a full Euler system including an energy equation. This extension is straightforward in both the Euler-Poisson and Euler-Maxwell cases. Similarly, viscosity or heat conductivity terms can be considered without altering the principles of the method.

\medskip
\noindent
{\bf Two fluids or more.} The case of two-fluid models, where each of the ion and electron species is modeled by its own Euler system of equations, and are coupled to the electric potential or electro-magnetic field by the electrical sources (charges and currents), has been considered. References \cite{Cri_Deg_Vig_CRAS05} and \cite{Cri_Deg_Vig_07} for the Euler-Poisson case, and \cite{Deg_Del_Sav_prep} for the Euler-Maxwell case show that the approach extends easily to this case (and would also apply to the multiple ion species case as well). The numerical results have been obtained in this case, with the physical electron to ion mass ratio. We refer to the refereces for more details.

\medskip
\noindent
{\bf Euler-Poisson-Boltzmann.} A commonly used approximation in plasma physics is to suppose that the electrons follow a Boltzmann law. The Boltzmann law provides a linear relationship between the electric potential and the logarithm of the electron density (or chemical potential). It is obtained from the electron momentum conservation equation in the limit of vanishing inertia. Then, the Euler-Poisson-Boltzmann (EPB) system consists of a pressureless gas dynamics model for the ions with an electrical forcing. The electric potential is obtained by solving the Poisson equation where the electric charge takes into account the ion and electron densities, the latter through the Boltzmann law. The resulting Poisson equation is nonlinear. In the quasineutral limit, the EPB model reduces to the compressible gas dynamics equations, the electrical force acting as a pressure term for the ions. The AP methodology has been applied to the EPB model in \cite{Deg_Del_Liu_Sav_Vig_prep}.

%% file: drift.tex
%%%%%%%%%%%%%%%%%%%%%%%%%%%%%%%%%%%%%%%%%%%%%%%%%%%%%%%%%%%%%%%%%%%%%%%%%%%%%%%%%%%%%%%%%%%%%%%%
%%%%%%%%%%%%%%%%%%%%%%%%%%%%%%%%%%%%%%%%%%%%%%%%%%%%%%%%%%%%%%%%%%%%%%%%%%%%%%%%%%%%%%%%%%%%%%%%
\setcounter{equation}{0}
\section{The isentropic Euler-Lorentz model}
\label{sec_DF_EL}

%%%%%%%%%%%%%%%%%%%%%%%%%%%%%%%%%%%%%%%%%%%%%%%%%%%%%%%%%%%%%%%%%%%%%%%%%%%%%%%%%%%%%%%%%%%%%%%%
\subsection{Introduction}
\label{subsec_DF_EL_intro}

This section and the following ones are concerned with the numerical approximation of the  
Euler equations for charged particles subject to the
Lorentz force (the 'Euler-Lorentz' system), when the magnetic field is large, or equivalently, 
when the parameter $\tau$ representing the reciprocal of the non-dimensional 
cyclotron frequency tends to zero. In this regime, the so-called drift-fluid (or 
gyro-fluid) approximation is obtained. In this limit, 
the parallel motion relative to the magnetic field 
direction splits from the perpendicular motion. The latter is given by an algebraic relation which describes the various drifts of the fluid across the magnetic field lines. The parallel motion is given 
implicitly by the constraint of zero total force along the
magnetic  field lines. In these sections, we construct Asymptotic-Preserving (AP) schemes which give rise to both a consistent approximation
of the Euler-Lorentz model when $\tau$ is finite and a consistent
approximation of the drift limit when $\tau \to 0$. Above all,
they do not require any constraint on the space and time steps related
to the small value of $\tau$.

%%%%%%%%%%%%%%%%%%%%%%%%%%%%%%%%%%%%%%%%%%%%%%%%%%%%%%%%%%%%%%%%%%%%%%%%%%%%%%%%%%%%%%%%%%%%%%%%
\subsection{Setting of the problem}
\label{subsec_DF_EL_setting}

The Isentropic Euler-Lorentz (IEL) model consists of the system of isentropic Euler equations subject to the Lorentz force. In this study, the electro-magnetic field is supposed given with an arbitrary spatio-temporal dependence. Of course, we have in mind that it satisfies the Maxwell system or any system derived from it, but the only information that we shall use from it is that the magnetic field is divergence-free. For the same reason, a single plasma species is considered (the ions) but again, all considerations below would extend to the electron species and any kind of coupling between these two species (either through Poisson's equations or through Maxwell's equations, or again, through quasi-neutrality). 

In this framework, the IEL model is written: 
\begin{eqnarray}
& & \hspace{-1cm} \partial_t n + \nabla \cdot (nu) = 0, \label{DFEL_n} \\
& & \hspace{-1cm} m ( \partial_t (nu) + \nabla \cdot (nu \otimes u)) + \nabla p(n) = e n (E + u \times B), \label{DFEL_u} 
\end{eqnarray}
where, $n(x,t) \geq 0$ and $u(x,t) \in {\mathbb R}^3 $ stand for the ion density and ion velocity and depend on the space-variable $x \in {\mathbb R}^3$ and on the time $t\geq0$. The electro-magnetic field $(E(x,t), B(x,t))  \in {\mathbb R}^3 \times {\mathbb R}^3$ is supposed given and satisfies $\nabla \cdot B = 0$. The positive elementary charge is denoted by $e$ and the ion mass, by $m$. The ion pressure $p$ is supposed to be a given function of the density (e.g. $p(n) = k_B T n$ in the isothermal case, where $T$ is the ion temperature and $k_B$, the Boltzmann constant, or $p(n) = C n^\gamma$, where $\gamma >1$ and $C>0$ are given constants, in the isentropic case).

The following scaling allows us to highlight the large magnetic field regime. We denote the scaling units for length, time, velocity, density, pressure, electric field and magnetic field by $x_0$, $t_0$, $u_0$, $n_0$, $p_0$, $E_0$, $B_0$. As usual, we relate the velocity scale to the time and space scales by $x_0=u_0t_0$. We also relate the electric and magnetic field scales by $E_0 = u_0 B_0$. This relation indicates that the typical electric field in the plasma is of the same order as the electric field induced by the motion of the plasma accross the magnetic field lines. We introduce the ion sound speed $c_s = (p_0/(m n_0))^{1/2}$. 
The ion gyro-frequency, i.e. the angular velocity of the gyration motion about the magnetic field lines is given by $\omega_0 = eB_0/m = eE_0/(mu_0)$. Two dimensionless parameters
appear: the Mach number $M$ and the scaled gyro-period $\tau$ given by 
\begin{eqnarray}
& & \hspace{-1cm} M = \frac{u_0}{c_s}, \quad \quad \tau = \frac{1}{\omega_0 t_0}  = \frac{m}{e B_0 t_0}
, \label{def_M_tau} 
\end{eqnarray}
This leads to the following scaled IEL model: 
\begin{eqnarray*}
& & \hspace{-1cm} \partial_t n + \nabla \cdot (nu) = 0, \\
& & \hspace{-1cm}  \partial_t (nu) + \nabla \cdot (nu \otimes u) + \frac{1}{M^2} \nabla p(n) = \frac{1}{\tau} n (E + u \times B).
\end{eqnarray*}
Now, there are two interesting scales:
\begin{enumerate}
\item Case: $M^2 = \tau \ll 1$. In this case, the pressure force is of the same order of magnitude as the Lorentz force, but much larger than the inertia terms. Without the Lorentz force, this scaling corresponds to the low Mach-number regime. In the presence of a Lorentz force, the low Mach-number regime still applies to the parallel motion, in a modified form, as we will see below. 
\item Case: $M^2 = O(1)$, $\tau \ll 1$. In this case, the Lorentz force is smaller than the Lorentz force. This scaling requires that $E_\parallel = 0(\tau) E_\bot$ and the parallel dynamics remains that of a compressible fluid. 
\end{enumerate}
From the viewpoint of AP schemes, the second case is simpler to treat than the first one and, for this reason, we will focus on the first case. The developed schemes will obviously be suitable for the second case as well. Therefore, from now on, we assume: 
\begin{eqnarray}
& & \hspace{-1cm} M^2 = \tau , \label{M2=tau} 
\end{eqnarray}
which leads to the final scaled form of the IEL model:

\begin{definition}
The scaled Incompressible Euler-Lorentz model (IEL) is:  
\begin{eqnarray}
& & \hspace{-1cm} \partial_t n^\tau + \nabla \cdot (n^\tau u^\tau) = 0, \label{SDFEL_n} \\
& & \hspace{-1cm}  \tau \left\{ \partial_t (n^\tau u^\tau) + \nabla \cdot (n^\tau u^\tau \otimes u^\tau) \right\} + \nabla p(n^\tau) = n^\tau (E + u^\tau \times B) .  \label{SDFEL_u} 
\end{eqnarray}
\label{def_IEL}
\end{definition}

\noindent
The following notations will be useful: the direction of the magnetic
field is denoted by $ b = B/|B|$ wherever $B \not = 0$. Any vector quantity $ v $ can be split into its
parallel and perpendicular parts as follows:      
\begin{eqnarray} 
& & \hspace{-1cm}
   v =  v_{\parallel}  b +
   v_{\perp}\,, \quad v_{\parallel} =  v \cdot  b \,, \quad
   v_{\perp} =  b \times \left( v \times  b\right) = (\mbox{Id} - b \otimes b) v \,, \label{decomp}
\end{eqnarray}
where the matrix $(\mbox{Id} - b \otimes b)$ is nothing but the projection matrix onto the perpendicular plane to $b$. Next, we introduce the parallel and perpendicular gradients of a scalar function $\phi$ by 
\begin{eqnarray} 
& & \hspace{-1cm}
 \nabla_\parallel \phi = (\nabla \phi)_\parallel = b \cdot \nabla \phi, \quad \quad \nabla_\bot \phi = (\nabla \phi)_\bot = b \times \left( \nabla \phi \times  b\right) = (\mbox{Id} - b \otimes b) \phi. 
\label{decomp_grad}
\end{eqnarray}
Similarly, the parallel and perpendicular divergence of a vector field $v$ are defined by: 
\begin{eqnarray} 
& & \hspace{-1cm}
 \nabla_\parallel \cdot v  = \nabla \cdot (v_\parallel b), \quad \nabla_\bot \cdot v = \nabla \cdot v_\bot, \quad \nabla \cdot v = \nabla_\parallel \cdot v + \nabla_\bot \cdot v. 
\label{decomp_div}
\end{eqnarray}
We also note that, since $|b|=1$, any derivative of $b$ is orthogonal to $b$, i.e. 
\begin{eqnarray} 
& & \hspace{-1cm}
b \cdot \partial_t b = 0, \quad b \cdot \partial_{x_k} b = 0, \, \, \forall k \in \{1,2,3\}. 
\label{bdotpab=0}
\end{eqnarray}

%%%%%%%%%%%%%%%%%%%%%%%%%%%%%%%%%%%%%%%%%%%%%%%%%%%%%%%%%%%%%%%%
%%%%%%%%%%%%%%%%%%%%%%%%%%%%%%%%%%%%%%%%%%%%%%%%%%%%%%%%%%%%%%%%
\setcounter{equation}{0}
\section{The Drift-fluid Limit $\tau \to 0$}
\label{section_drift_limit}

In this section, we investigate various formulations of the model obtained by letting $\tau \to 0$ in the IEL model (\ref{SDFEL_n}), (\ref{SDFEL_u}).

%%%%%%%%%%%%%%%%%%%%%%%%%%%%%%%%%%%%%%%%%%%%%%%%%%%%%%%%%%%%%%%%
\subsection{The Isentropic Drift-Fluid (IDF) model; a first reformulation}
\label{subsection_drift_limit}

The formal limit $\tau \to 0$ in the IEL model (\ref{SDFEL_n}), (\ref{SDFEL_u}) leads to the so-called Isentropic Drift-Fluid (IDF) model:

\begin{definition}
The Isentropic Drift-Fluid model (IDF) is:
\begin{eqnarray}
& & \hspace{-1cm} \partial_t n^0 + \nabla \cdot (n^0 u^0) = 0, \label{LDFEL_n_0} \\
& & \hspace{-1cm} \nabla p(n^0) =  n^0 (E + u^0 \times B) .  \label{LDFEL_u} 
\end{eqnarray}
\label{def_IDF}
\end{definition}

\noindent
We obviously have: 

\begin{proposition}
The formal limit of the IEL model is the IDF model. 
\label{prop_IEL-IDF}
\end{proposition}

\noindent
We now propose a first reformulation of the IDF model: 

\begin{proposition}
A solution of the IDF model is a solution to the following reformulation (first Refomulated IDF model or (RIDF-1) model): 
\begin{eqnarray}
& & \hspace{-1cm} \partial_t n^0 + \nabla \cdot (n^0 u^0) = 0, \label{LDFEL_n_3} \\
& & \hspace{-1cm} n^0 u^0_{\perp} = \frac1{B} b \times  (\nabla p(n^0) - n^0 E ) \,, \label{LDFEL_u_perp_3}\\
&&  \hspace{-1cm} - \nabla_\parallel \left( p'(n^0) \nabla_\parallel \cdot 
  (n^0  u^0_\parallel ) \right) = 
  \partial_t ( n^0 E_\parallel ) -  \partial_t b
  \cdot \nabla p(n^0) + \nabla_\parallel \left( p'(n^0) \nabla_\bot \cdot (n^0
  u^0_\bot ) \right)\,. 
  \label{LDFEL_u_par_3} 
\end{eqnarray}
The converse is true provided that the initial data satisfy the constraint $(\nabla_\parallel p(n^0) - n^0 E_\parallel)|_{t=0} = 0$.
\label{prop_IDF_equiv_RIDF1}
\end{proposition}

\noindent
{\bf Proof:} 
We split (\ref{LDFEL_u}) into its parallel and transverse components with respect to the magnetic field direction $b$. First, concerning the transverse component, we take the vector product of (\ref{LDFEL_u}) with $b$. The resulting equation can easily be resolved for $u^0_{\bot}$ in the form (\ref{LDFEL_u_perp_3}). 

Taking now the scalar product of (\ref{LDFEL_u}) with $b$, we find: 
\begin{eqnarray}
& & \hspace{-1cm} \nabla_\parallel p(n^0) - n^0 E_\parallel = 0 \, .
\label{LDFEL_u_par}
\end{eqnarray} 
This equation does not explicitly contain $u^0_{\parallel}$ but defines a constraint which indirectly determines it. The resolution of this constraint leads to the elliptic equation (\ref{LDFEL_u_par_3}). To show it, we first, multiply  (\ref{LDFEL_n_0}) by $p'(n^0)$ and we get: 
\begin{eqnarray}
& & \hspace{-1cm} \partial_t p(n^0) + p'(n^0) \left( \nabla_\bot \cdot (n^0 u^0_\bot )
  + \nabla_\parallel \cdot (n^0  u^0_\parallel ) \right)  =
  0\,.
  \label{LDFEL_n_2} 
\end{eqnarray}   
Applying $\nabla_\parallel$ to (\ref{LDFEL_n_2}), noting that
$[\nabla_\parallel, \partial_t] = - \partial_t b \cdot \nabla$ (where
$[\cdot,\cdot]$ denotes the commutator) and inserting (\ref{LDFEL_u_par})
leads to (\ref{LDFEL_u_par_3}).   
Reciprocally, it is straightforward to see that system (\ref{LDFEL_n_3}), (\ref{LDFEL_u_perp_3}), (\ref{LDFEL_u_par_3}), implies system (\ref{LDFEL_n_0}), (\ref{LDFEL_u}) provided that eq. (\ref{LDFEL_u_par}) is satisfied at time $t=0$. This ends the proof of proposition \ref{prop_IDF_equiv_RIDF1}. \endproof

\noindent
After dividing by $n^0$, we find that the first term at the right-hand side of (\ref{LDFEL_u_perp_3}) is the diamagnetic drift velocity while the second one is the $ E \times B$ drift velocity. 
Eq. (\ref{LDFEL_u_par_3}) is a well-posed one-dimensional elliptic equation for $u^0_\parallel$ posed along
the magnetic field lines, provided that adequate boundary conditions are given.  The determination of the boundary conditions for (\ref{LDFEL_u_par_3}) will be discussed in more details below.

%%%%%%%%%%%%%%%%%%%%%%%%%%%%%%%%%%%%%%%%%%%%%%%%%%%%%%%%%%%%%%%%
\subsection{Analogy with the low Mach-number limit of compressible fluids}
\label{subsection_analogy_low_Mach}

In this section, we depart from the drift-fluid limit and consider the low Mach-number limit of ordinary isentropic compressible fluids. We show that the procedure which leads to (\ref{LDFEL_u_par_3}) is specific to the Euler-Lorentz problem and cannot be reproduced in the case of ordinary fluids. In the next section, we will use the analogy with the low Mach-number limit of ordinary fluids and devise an alternate expression of the limit problem (\ref{LDFEL_n_3})-(\ref{LDFEL_u_par_3}). 

The scaled Isentropic Compressible Euler (ICE) system with the low Mach-number scaling is written as follows
\begin{eqnarray}
& & \hspace{-1cm} \partial_t n^\tau + \nabla \cdot (n^\tau u^\tau) = 0, \label{SLMIE_n} \\
& & \hspace{-1cm}  \tau \left\{ \partial_t (n^\tau u^\tau) + \nabla \cdot (n^\tau u^\tau \otimes u^\tau) \right\} + \nabla p(n^\tau) = 0 .  \label{SLMIE_u} 
\end{eqnarray}
Then, in the limit $\tau \to 0$, we formally get:  
\begin{eqnarray}
& & \hspace{-1cm} \partial_t n^0 + \nabla \cdot (n^0 u^0) = 0, \label{ILMIE_n_0} \\
& & \hspace{-1cm} \nabla p(n^0) = 0 .  \label{ILMIE_u_0} 
\end{eqnarray}
We suppose that the boundary conditions are such that (\ref{ILMIE_u_0}) is well-posed and gives $p(n^0) = $ Constant. This occurs for instance if the boundary conditions for $n^\tau$ are mixed Dirichlet or Neumann conditions and if the values along the Dirichlet boundary are uniform. More general conditions are of course possible but will not be detailed here. Here, we additionally impose that these conditions lead to the fact that $p(n^0)$  and therefore $n^0$ are independent of both space and time. As a consequence, (\ref{ILMIE_n_0}) leads to 
\begin{eqnarray}
& & \hspace{-1cm} \nabla \cdot  u^0 = 0, \label{ILMIE_n_01} 
\end{eqnarray}
However, this is not enough to determine $u^0$. But, dividing (\ref{SLMIE_u}) by $\tau$ and letting $\tau \to 0$ leads to the existence of a scalar function $p^1$ such that 
\begin{eqnarray}
& & \hspace{-1cm} \partial_t u^0 + \nabla \cdot (u^0 \otimes u^0) + \nabla p^1 = 0 .  \label{ILMIE_u_01} 
\end{eqnarray}
The hydrostatic pressure $p^1$ is the first order (in $\tau$) perturbation pressure, i.e.
\begin{eqnarray}
& & \hspace{-1cm} p^1 = \left( \lim_{\tau \to 0} \frac{p(n^\tau)}{\tau} \right) ,  \label{ILMIE_pi} 
\end{eqnarray}
and can be determined from the incompressibility constraint (\ref{ILMIE_n_01}). Indeed, taking the divergence of (\ref{ILMIE_u_01}) and using (\ref{ILMIE_n_01}), we find 
\begin{eqnarray}
& & \hspace{-1cm} \Delta p^1 + \nabla^2 : (u^0 \otimes u^0) = 0 ,  \label{ILMIE_pi_2} 
\end{eqnarray}
where $\nabla^2$ denotes the Hessian matrix (matrix of second order derivatives) and $:$ the contracted product of tensors. Eq. (\ref{ILMIE_pi_2}) is an elliptic equation which determines $p^1$ provided appropriate boundary conditions are given. These conditions can be deduced from those for $n^\tau$.  

As a summary, the low Mach-number limit of the ICE eqs. (\ref{SLMIE_n}), (\ref{SLMIE_u}) is the Incompressible Euler (IE) eqs.: 
\begin{eqnarray}
& & \hspace{-1cm} n^0 = \mbox{Constant}, \label{ILMIE_n_11} \\
& & \hspace{-1cm} \nabla \cdot  u^0 = 0, \label{ILMIE_n_12} \\
& & \hspace{-1cm} \partial_t u^0 + \nabla \cdot (u^0 \otimes u^0) + \nabla p^1 = 0 .  \label{ILMIE_u_1} 
\end{eqnarray}

To identify the limit problem, the strategy of section \ref{subsection_drift_limit} would not be not adequate. Indeed, if we take the gradient of (\ref{ILMIE_n_0}) and use (\ref{ILMIE_u_0}), we are led to 
\begin{eqnarray}
& & \hspace{-1cm} \nabla( \nabla \cdot u^0) = 0, \label{ILMIE_n_02} 
\end{eqnarray}
which is {\bf not} a well-posed problem for $u^0$. Indeed, the operator $- \nabla( \nabla \cdot u) = - \Delta u + \nabla \times (\nabla \times u)$ is not elliptic  because of the term $\nabla \times (\nabla \times u)$, except in dimension 1 where this term does not appear. By contrast, in the case of the IDF model, the problem is well-posed, thanks to the presence of the Lorentz force. On the one hand, the Lorentz force provides the explicit algebraic relation (\ref{LDFEL_u_perp_3}) for $u_\bot$. On the other hand, $u_\parallel$ is determined by inverting the operator $- \nabla_\parallel( \nabla_\parallel \cdot u_\parallel)$ (see \ref{LDFEL_u_par_3}), which now leads to a well-posed elliptic equation because $u_\parallel$ is a scalar and the equation is posed on a one-dimensional manifold (the magnetic field line).

%%%%%%%%%%%%%%%%%%%%%%%%%%%%%%%%%%%%%%%%%%%%%%%%%%%%%%%%%%%%%%%%
\subsection{A second formulation of the drift-fluid limit $\tau \to 0$ using the analogy with the low Mach-number limit}
\label{subsection_drift_limit_2}

In the present section, we use the analogy with the low Mach-number limit of ordinary fluids developed in the previous section to devise an alternate expression of the limit problem (\ref{LDFEL_n_3})-(\ref{LDFEL_u_par_3}), which in turn will be useful for the numerical developments below. Specifically, we want to exploit the analogy of the constraint (\ref{LDFEL_u_par}) with (\ref{ILMIE_u_0}), and of the continuity eq. (\ref{LDFEL_n_0}) with (\ref{ILMIE_n_0}) (we leave the algebraic relation (\ref{LDFEL_u_perp_3}) apart because it will not play any role in the discussion). In the case of the low Mach-number limit, these two equations respectively lead to (\ref{ILMIE_n_11}) and to (\ref{ILMIE_n_12}), while in the case of the drift-fluid limit, they are unchanged, meaning that they cannot be further simplified. However, in the low Mach-number limit, we know that these two relations are not sufficient to characterize the limit solution  and an additional relation was sought by dividing the momentum conservation equation by $\tau$ and taking the limit. We explore a similar strategy here with the parallel component of the momentum equation. 

We first introduce 
\begin{eqnarray}
& & \hspace{-1cm} n^1 = \lim_{\tau \to 0} \frac{n^\tau - n^0}{\tau}. 
\label{LDFEL_n1} 
\end{eqnarray}
We have 
\begin{eqnarray}
& & \hspace{-1cm}  p^1 := \lim_{\tau \to 0} \frac{p(n^\tau) - p(n^0)}{\tau} = p'(n^0) n^1 
\label{LDFEL_p1} 
\end{eqnarray}
Upon dividing (\ref{SDFEL_u}) by $\tau$ and taking the limit $\tau \to 0$, we deduce that 
\begin{eqnarray}
 & \hspace{-1cm}  b \cdot  \left( \partial_t (n^0 u^0) + \nabla \cdot (n^0 u^0 \otimes u^0) \right)  = -  \nabla_\parallel (p'(n^0) n^1 ) + n^1 E_\parallel  := F_\parallel^1.  \label{SDFEL_u_par_4} 
\end{eqnarray}

Then, we supplement the IDF model (\ref{LDFEL_n_0})-(\ref{LDFEL_u}) with the additional eq. (\ref{SDFEL_u_par_4}) and introduce the following augmented model: 

\begin{definition}
The 'augmented IDF' model (AIDF model) is : 
\begin{eqnarray}
& & \hspace{-1cm} \nabla_\parallel p(n^0) - n^0 E_\parallel = 0
  \, , \label{LDFEL_u_par_51} \\
& & \hspace{-1cm} \partial_t n^0 + \nabla \cdot (n^0 u^0) = 0, \label{LDFEL_n_5} \\
& & \hspace{-1cm}  b \cdot  \left( \partial_t (n^0 u^0) + \nabla \cdot (n^0 u^0 \otimes u^0) \right)  = -  \nabla_\parallel (p'(n^0) n^1 ) + n^1 E_\parallel ,  \label{SDFEL_u_par_52} \\
& & \hspace{-1cm} n^0 u^0_{\perp} = \frac1{B} b \times  \left(\nabla p(n^0) - n^0 E\right) \,. \label{LDFEL_u_perp_5}
\end{eqnarray}
\label{def_AIDF}
\end{definition}

\noindent
We have the obvious: 

\begin{proposition}
Any solution of the AIDF model is a solution of the IDF model. Reciprocally, a solution of the IDF model gives rise to a solution of the AIDF by solving (\ref{SDFEL_u_par_52}) for $n^1$. 
\label{prop_equiv_IDF_AIDF}
\end{proposition}

\noindent
The AIDF model allows us to change the viewpoint and instead of seeing (\ref{LDFEL_u_par_51}) as a constraint which implicitly determines $u^0_\parallel$, we can see it as a constraint which determines $n^1$ while $u^0_\parallel$ is found by solving the evolution eq. (\ref{SDFEL_u_par_52}). This viewpoint is highlighted in

\begin{proposition}
Any solution of the AIDF model provides a solution of the following model (the second reformulation of the Isentropic Drift-Fluid model or RIDF2 model): 
\begin{eqnarray}
& & \hspace{-1cm} \nabla_\parallel p(n^0) - n^0 E_\parallel = 0
  \, , \label{LDFEL_u_par_61} \\
& & \hspace{-1cm} - \nabla_\parallel \cdot (\nabla_\parallel (p'(n^0) n^1 ) - n^1 E_\parallel ) = - \partial^2_t n^0 +  \nabla \cdot ((b \otimes b)  \nabla \cdot (n^0 u^0 \otimes u^0)) \nonumber \\
& & \hspace{2.8cm} - \nabla_\bot \cdot (n^0 u^0_\parallel \partial_t b) - \nabla_\parallel \cdot (n^0 u^0_\bot \cdot \partial_t b) - \partial_t \nabla_\bot \cdot (n^0 u^0_\bot). \label{LDFEL_n_6} \\
& & \hspace{-1cm}  b \cdot  \left( \partial_t (n^0 u^0) + \nabla \cdot (n^0 u^0 \otimes u^0) \right)  = -  \nabla_\parallel (p'(n^0) n^1 ) + n^1 E_\parallel ,  \label{SDFEL_u_par_62} \\
& & \hspace{-1cm} n^0 u^0_{\perp} = \frac1{B} b \times  \left(\nabla p(n^0) - n^0 E\right) \,. \label{LDFEL_u_perp_6}
\end{eqnarray}
\label{prop_equiv_AIDF_RIDF2}
Conversely, any solution of the RIDF-2 model such that the mass conservation eq. (\ref{LDFEL_n_5}) is satisfied at time $t=0$, is a solution of the AIDF model. 
\end{proposition}

\noindent
{\bf Proof:} We proceed like in section \ref{subsection_analogy_low_Mach}. Taking the time derivative of (\ref{LDFEL_n_5}) and using (\ref{decomp_div}), we get: 
\begin{eqnarray}
& & \hspace{-1cm} \partial^2_t n^0 + \partial_t \nabla_\parallel \cdot (n^0 u^0_\parallel) + \partial_t \nabla_\bot \cdot (n^0 u^0_\bot) = 0. \label{LDFEL_n_6_1} 
\end{eqnarray}
We note the identity: 
\begin{eqnarray}
\partial_t \nabla_\parallel \cdot (n^0 u^0_\parallel) &  = & \partial_t \nabla \cdot (n^0 u^0_\parallel b) = \partial_t \nabla \cdot ((b \otimes b) (n^0 u^0)) = \nonumber  \\
& = & \nabla \cdot ((b \otimes b) \partial_t (n^0 u^0)) + \nabla \cdot ((\partial_t b \otimes b + b \otimes \partial_t b) (n^0 u^0)) \nonumber \\
& = & \nabla \cdot \left( b ( b \cdot \partial_t (n^0 u^0)) \right) + \nabla_\bot \cdot (n^0 u^0_\parallel \partial_t b) + \nabla_\parallel \cdot (n^0 u^0_\bot \cdot \partial_t b).   \label{pat_napar_nupar}
\end{eqnarray}
Multiplying (\ref{SDFEL_u_par_52}) by the vector $b$, inserting it into (\ref{pat_napar_nupar}), and using definition (\ref{SDFEL_u_par_4}) of $F^1_\parallel$, we find:
\begin{eqnarray}
& & \hspace{-1cm} \partial_t \nabla_\parallel \cdot (n^0 u^0_\parallel) = 
-  \nabla \cdot ((b \otimes b)  \nabla \cdot (n^0 u^0 \otimes u^0))  + \nabla_\parallel \cdot F^1_\parallel + \nonumber \\
& & \hspace{5cm} + \nabla_\bot \cdot (n^0 u^0_\parallel \partial_t b) + \nabla_\parallel \cdot (n^0 u^0_\bot \cdot \partial_t b).  \label{pat_napar_nupar_2}
\end{eqnarray}
Inserting (\ref{pat_napar_nupar_2}) into (\ref{LDFEL_n_6_1}) leads to 
\begin{eqnarray}
& & \hspace{-1cm} \partial^2_t n^0 -  \nabla \cdot ((b \otimes b)  \nabla \cdot (n^0 u^0 \otimes u^0))  + \nabla_\parallel \cdot F^1_\parallel + \nonumber \\
& & \hspace{2cm} + \nabla_\bot \cdot (n^0 u^0_\parallel \partial_t b) + \nabla_\parallel \cdot (n^0 u^0_\bot \cdot \partial_t b) + \partial_t \nabla_\bot \cdot (n^0 u^0_\bot) = 0. \label{LDFEL_n_6_2} 
\end{eqnarray}
In view of definition (\ref{SDFEL_u_par_4}), (\ref{LDFEL_n_6_2}) leads to (\ref{LDFEL_n_6}). Reciprocally, if $(n^0,u^0, n^1)$ is a solution of the RIDF-2 model, it is an easy matter to see that it satisfies the AIDF model, provided that the mass conservation eq. (\ref{LDFEL_n_5}) is satisfied at time $t=0$. This end the proof of the proposition. \endproof

\noindent
(\ref{LDFEL_n_6}) is a one-dimensional non-linear elliptic equation for $n^1$ posed along the magnetic field lines. 

Like in the quasi-neutral limit case, the key for designing AP schemes in the drift-fluid limit case is to reformulate the Euler-Lorentz model in the form of a singular perturbation of the Drift-Fluid model. In the next section, we provide two different such reformulation, which are based on the two reformulations of the Drift-fluid model established above.

%%%%%%%%%%%%%%%%%%%%%%%%%%%%%%%%%%%%%%%%%%%%%%%%%%%%%%%%%%%%%%%%
%%%%%%%%%%%%%%%%%%%%%%%%%%%%%%%%%%%%%%%%%%%%%%%%%%%%%%%%%%%%%%%%
\setcounter{equation}{0}
\section{Reformulations of the Euler-Lorentz model}
\label{section_EL_ref}

In this section, we construct two equivalent reformulations of the Euler-Lorentz model. These reformulations will be the bases for two different AP-schemes.

%%%%%%%%%%%%%%%%%%%%%%%%%%%%%%%%%%%%%%%%%%%%%%%%%%%%%%%%%%%%%%%%
\subsection{First reformulation of the Euler-Lorentz model: wave equation formulation for $n u_\parallel$}
\label{subsection_EL_ref1}

The RIDF-1 reformulation of the IDF model has a counterpart at the level of the IEL model as shown in the following:

\begin{proposition}
Any solution of the IEL model is a solution of the following first Reformulated IEL model (RIEL-1): 
\begin{eqnarray}
& & \hspace{-1cm} \partial_t n^\tau + \nabla \cdot (n^\tau u^\tau) = 0, \label{RIEL1_n} \\
& & \hspace{-1cm}  n^\tau u_\bot^\tau = \frac{1}{|B|} b \times \left\{ \nabla p(n^\tau) - n^\tau E + \tau \left[ \partial_t (n^\tau u^\tau) + \nabla \cdot (n^\tau u^\tau \otimes u^\tau) \right] \right\} ,  \label{RIEL1_u_perp} \\
& & \hspace{-1cm}  \tau \partial_t \left( \left\{ \partial_t (n^\tau u^\tau) + \nabla \cdot (n^\tau u^\tau \otimes u^\tau) \right\}_\parallel \right) - \nabla_\parallel( p'(n^\tau) \nabla_\parallel \cdot 
  \left(n^\tau  u^\tau_\parallel \right) ) = \nonumber \\
& & \hspace{2cm}   =  \partial_t \left( n^\tau E_\parallel \right) -  \partial_t b
 \cdot \nabla p(n^\tau) + \nabla_\parallel ( p'(n^\tau) \nabla_\bot \cdot \left(n^\tau  u^\tau_\bot \right) ).   \label{RIEL1_u_par} 
\end{eqnarray}
Conversely, any solution ot the RIEL-1 model such that (\ref{SDFEL_u_par}) is satisfied at $t=0$ is a solution of the IEL model. 
\label{prop_equiv_IEL-RIEL1}
\end{proposition} 

\noindent
{\bf Proof:} We apply the same algebraic manipulations to the IEL model (\ref{SDFEL_n}), (\ref{SDFEL_u}), as we did to the IDF one (\ref{LDFEL_n_0}), (\ref{LDFEL_u}) in section \ref{subsection_drift_limit}. We first split (\ref{DFEL_u}) into its parallel and perpedicular components. This leads to: 
\begin{eqnarray}
& & \hspace{-1cm}  \tau \left\{ \partial_t (n^\tau u^\tau) + \nabla \cdot (n^\tau u^\tau \otimes u^\tau) \right\}_\bot + \nabla_\bot p(n^\tau) = n^\tau (E_\bot + u_\bot^\tau \times B) ,  \label{SDFEL_u_perp} \\
& & \hspace{-1cm}  \tau \left\{ \partial_t (n^\tau u^\tau) + \nabla \cdot (n^\tau u^\tau \otimes u^\tau) \right\}_\parallel + \nabla_\parallel p(n^\tau) = n^\tau E_\parallel .  \label{SDFEL_u_par} 
\end{eqnarray}
The transverse component eq. (\ref{SDFEL_u_perp}) can be recast in the form:
\begin{eqnarray}
& & \hspace{-1cm}  n^\tau u_\bot^\tau = \frac{1}{|B|} b \times \left\{ \nabla p(n^\tau) - n^\tau E + \tau \left[ \partial_t (n^\tau u^\tau) + \nabla \cdot (n^\tau u^\tau \otimes u^\tau) \right] \right\} ,  \label{SDFEL_u_perp_1} 
\end{eqnarray}
and clearly appears as a singular perturbation of (\ref{LDFEL_u_perp_3}). 
For the parallel component eq. (\ref{SDFEL_u_par}), we use (\ref{LDFEL_n_2}), which is also valid for finite $\tau$. Like in section \ref{subsection_drift_limit}, we apply $\nabla_\parallel$ to (\ref{LDFEL_n_2}), commute $\partial_t p(n^\tau)$ and $\nabla_\parallel$ and use (\ref{SDFEL_u_par}) to eliminate $\nabla_\parallel p(n^\tau)$. This leads to (\ref{RIEL1_u_par}).   
Conversely, it is an easy matter to see that any solution of the RIEL-1 model provided that (\ref{SDFEL_u_par}) is satisfied at $t=0$. This ends the proof.  \endproof

\noindent
Eq. (\ref{RIEL1_u_par}) is a wave equation for $n^\tau  u^\tau_\parallel$. Indeed, we have, using (\ref{bdotpab=0}): 
\begin{eqnarray}
& & \hspace{-1cm}  \left\{ \partial_t (n u) + \nabla \cdot (n u \otimes u) \right\}_\parallel =  \partial_t (n u_\parallel) + \nabla \cdot ( n u u_\parallel) -  (\partial_t + u \cdot \nabla ) b  \cdot n u_\bot .   \label{transp_mom_par} 
\end{eqnarray}
Then, (\ref{RIEL1_u_par}) can be rewritten:
\begin{eqnarray}
& & \hspace{-1cm} \tau  \partial_{tt} (n^\tau u^\tau_\parallel) 
- \nabla_\parallel( p'(n^\tau) \nabla_\parallel \cdot 
  \left(n^\tau  u^\tau_\parallel \right) ) + \tau \partial_t \nabla \cdot ( n^\tau u^\tau u^\tau_\parallel)  = \nonumber \\
& & \hspace{4cm} = \tau \partial_t  ( (\partial_t + u^\tau \cdot \nabla ) b \cdot n^\tau u^\tau_\bot)  +   \partial_t \left( n^\tau E_\parallel \right)  \nonumber \\
& & \hspace{4.5cm}  -  \partial_t b
 \cdot \nabla_\bot p(n^\tau) + \nabla_\parallel ( p'(n^\tau) \nabla_\bot \cdot \left(n^\tau  u^\tau_\bot \right) ).   \label{SDFEL_u_par_2} 
\end{eqnarray}
For small $u$, the third term of the left-hand side of (\ref{SDFEL_u_par_2}) can be neglected to leading order. Then, the principal symbol of the differential operator acting on \, $n^\tau u^\tau_\parallel$ \,  is \, $\tau \partial_{tt} - p'(n^\tau) (\nabla_\parallel)^2$, \, which is a wave operator associated to the acoustic wave speed $c_s = (p'(n^\tau)/\tau)^{1/2}$. For the sake of simplicity, let us take an isothermal equation-of-state $p(n) = nT$ where $T$ is a fixed temperature. Then, $c_s = (T/\tau)^{1/2}$ is the typical velocity of acoustic waves in the low Mach-number scaling.

The requirement that (\ref{SDFEL_u_par}) must be satisfied at $t=0$ sets up the additional initial condition that is needed for the time second order differential equation (\ref{RIEL1_u_par}). 

We note the obvious

\begin{proposition}
In the limit $\tau \to 0$, the RIEL-1 formulation of the Euler-Lorentz model formally converges to the RIDF-1 formulation of the Drift-Fluid model. 
\label{prop_RIEL1-RIDF1}
\end{proposition}

\noindent
A first class of AP-schemes for the IEL model will rely on an implicit discretization of the wave eq. (\ref{RIEL1_u_par}).

%%%%%%%%%%%%%%%%%%%%%%%%%%%%%%%%%%%%%%%%%%%%%%%%%%%%%%%%%%%%%%%%
\subsection{Second reformulation of the Euler-Lorentz model: wave equation formulation for $n$}
\label{subsection_EL_ref2}

Now, we show that the RIDF-2 reformulation of the IDF model has also a counterpart at the level of the IEL model. By contrast to the previous one, this reformulation involves a wave equation for $n$.  To this aim, we apply the methodology of section \ref{subsection_drift_limit_2}. 

\begin{proposition} 
Any solution of the IEL model is a solution of the following second Reformulated IEL model (RIEL-2): 
\begin{eqnarray}
& & \hspace{-1cm} \tau \partial_{tt} n^\tau - \nabla_\parallel \cdot (\nabla_\parallel p(n^\tau) - n^\tau E_\parallel ) = \tau \left\{ \nabla \cdot ((b \otimes b)  \nabla \cdot (n^\tau u^\tau \otimes u^\tau)) \right. \nonumber \\
& & \hspace{2.8cm} \left. - \nabla_\bot \cdot (n^\tau u^\tau_\parallel \partial_t b) - \nabla_\parallel \cdot (n^\tau u^\tau_\bot \cdot \partial_t b) - \partial_t \nabla_\bot \cdot (n^\tau u^\tau_\bot) \right\}, \label{RIEL2_n} \\
& & \hspace{-1cm}  n^\tau u_\bot^\tau = \frac{1}{|B|} b \times \left\{ \nabla p(n^\tau) - n^\tau E + \tau \left[ \partial_t (n^\tau u^\tau) + \nabla \cdot (n^\tau u^\tau \otimes u^\tau) \right] \right\} ,  \label{RIEL2_u_perp} \\
& & \hspace{-1cm}  \left\{ \partial_t (n^\tau u^\tau) + \nabla \cdot (n^\tau u^\tau \otimes u^\tau) \right\}_\parallel + \frac{1}{\tau} (\nabla_\parallel p(n^\tau) - n^\tau E_\parallel ) = 0.     \label{RIEL2_u_par} 
\end{eqnarray}
Conversely, any solution ot the IREL-2 model such that (\ref{SDFEL_n}) is satisfied at $t=0$ is a solution of the IEL model. 
\label{prop_equiv_IEL-RIEL2}
\end{proposition} 

\noindent
{\bf Proof:} 
We leave (\ref{SDFEL_u_perp_1}) unchanged. We use (\ref{LDFEL_n_6_1}) and (\ref{pat_napar_nupar}), which are obviously valid for finite $\tau$. To eliminate $\partial_t (n^\tau u^\tau)$ in (\ref{pat_napar_nupar}), we use (\ref{SDFEL_u_par}) in the form: 
\begin{eqnarray}
& & \hspace{-1cm}  (b \otimes b ) \left\{ \partial_t (n^\tau u^\tau) + \nabla \cdot (n^\tau u^\tau \otimes u^\tau) \right\} + \frac{1}{\tau} (b \otimes b) (\nabla p(n^\tau) - n^\tau E ) = 0 .   
\label{SDFEL_u_par_5} 
\end{eqnarray}
This leads to (\ref{RIEL2_n}). 

Reciprocally, it is an easy matter to see that any solution of the RIEL-2 model satisfies the original IEL model provided that (\ref{SDFEL_n}) is satisfied at $t=0$. This ends the proof. \endproof

\noindent
Eq. (\ref{RIEL2_n}) is a wave equation for $n$. If $u$ is small, at the leading order in $u$, the principal symbol of the differential operator acting on $n$ is again $\tau \partial_{tt} - p'(n^\tau) (\nabla_\parallel)^2$, and is again a wave operator associated to the acoustic wave speed $c_s = (p'(n^\tau)/\tau)^{1/2}$.

The requirement that (\ref{SDFEL_n}) should be satisfied at $t=0$ sets up the additional initial condition that is needed for the time second order differential equation (\ref{RIEL2_n}). 

Again, we have the obvious:

\begin{proposition}
In the limit $\tau \to 0$, the RIEL-2 formulation of the Euler-Lorentz model formally converges to the RIDF-2 formulation of the Drift-Fluid model. 
\label{prop_RIEL2-RIDF2}
\end{proposition}

\noindent
A second class of AP-schemes for the IEL model will rely on an implicit discretization of the wave eq. (\ref{RIEL2_n}).

%%%%%%%%%%%%%%%%%%%%%%%%%%%%%%%%%%%%%%%%%%%%%%%%%%%%%%%%%%%%%%%%
\subsection{Boundary conditions}
\label{subsection_EL_BC}

The question of boundary conditions is complex for several reasons. First, the theory of boundary value problems for hyperbolic systems of conservation laws is still in its infancy (see e.g. \cite{Bar_Ler_Ned_79, Dub_Lef_88, Serre_00}\footnote{For the last reference, see chapters 14 and 15}). Unfortunately, the cases that can be rigorously treated seldom apply to practical situations. A practical rule is that the number of boundary conditions that can be imposed corresponds to the number of entering characteristics. However, this number depends on the solution itself. In practice, the state variables are supposed known outside the domain and a boundary Riemann problem is solved between the prescribed outer values and the current inner values of the state variables. This permits the computation of the entering fluxes and the advancement of the solution. Neumann boundary conditions can be prescribed by supposing that the value of the corresponding state variable in the outer cell is equal to the value in the inner cell and again solving a boundary Riemann problem. Here the problem is complexified by the fact that the limit $\tau \to 0$ leads to a change of type, from hyperbolic to elliptic, of some of the equations. This situation is similar as in the low Mach-number limit, with the additional feature that the elliptic equations are one-dimensional, posed along a magnetic field line. 

Here, we propose some boundary conditions which are operational and, in particular, which avoid the appearance of boundary layers. However, a rigorous mathematical theory is not available yet. The geometry of the magnetic field lines plays an important role. We assume that the problem is posed in a bounded domain $\Omega$ with boundary $\Gamma$ decomposed into 
\begin{eqnarray}
& & \hspace{-1cm} \Gamma = \Gamma_+ \cup \Gamma_- \cup \Gamma_0, \label{def_Gamma} \\
& & \hspace{-1cm} \Gamma_\pm = \{ x \in \Gamma \, | \, \pm b(x) \cdot \nu >0 \}, \quad \Gamma_0 = \{ x \in \Gamma \, | \,  b(x) \cdot \nu = 0 \} 
 .   \label{def_Gamma2}
\label{Gamma} 
\end{eqnarray}
Relative to the magnetic field lines, $\Gamma_-$ is the incoming boundary, $\Gamma_+$ the outgoing one and $\Gamma_0$, the tangential one. Our proposal of boundary conditions is guided by two principles in hierarchical order: first, avoid boundary layers and second, find artificial boundary conditions that are as close as possible to the free-space situation (transparent boundary conditions). 

The prescription of boundary conditions for $n^\tau$ on $\Gamma_- \cup \Gamma_+$ is guided by the first principle. $n^\tau$ satisfies the elliptic equation (\ref{RIEL2_n}). The elliptic operator is, at leading order in $\tau$, equal to $- \nabla ((b \otimes b)  (\nabla p(n^\tau) - n^\tau E ))$. We propose homogeneous Neumann boundary conditions associated to the conormal derivative of this elliptic operator on $\Gamma_- \cup \Gamma_+$, namely: 
\begin{eqnarray}
& & \hspace{-1cm}   \nabla_\parallel p(n^\tau) - n^\tau E_\parallel = 0, \quad \mbox{on} \quad \Gamma_- \cup \Gamma_+. 
\label{IEL_BC_n_pm} 
\end{eqnarray}
In this way, the boundary condition is compatible with the leading order operator inside the domain. We can prescribe non-homogeneous Neumann boundary conditions, provided that the right-hand side in (\ref{IEL_BC_n_pm}) is $O(\tau)$. On $\Gamma_0$, we apply the second principle and propose a homogeneous Neumann boundary condition:
\begin{eqnarray}
& & \hspace{-1cm}  \nu \cdot \nabla n^\tau  = 0, \quad \mbox{on} \quad \Gamma_0, 
\label{IEL_BC_n_0} 
\end{eqnarray}
which models the flatness of the density profile near the boundary. This is intended to be an approximation of the case where the domain is the entire space. 

For $n^\tau u^\tau_\parallel$, we proceed in a similar fashion, considering the elliptic equation (\ref{RIEL1_u_par}). At the leading order in $\tau$, the principal part of the associated elliptic operator is $(b \cdot \nabla) (p'(n^\tau) \nabla \cdot ( b n^\tau u^\tau_\parallel ))$. The homogeneous Neumann boundary conditions associated to the conormal derivative of this elliptic operator on $\Gamma_- \cup \Gamma_+$, are given by (assuming that $p'(n^\tau)$ never vanishes): 
\begin{eqnarray}
& & \hspace{-1cm} \nabla \cdot ( b n^\tau u^\tau_\parallel ) = 0, \quad \mbox{on} \quad \Gamma_- \cup \Gamma_+. 
\label{IEL_BC_u_par_pm} 
\end{eqnarray}
The prescription of this boundary condition follows the first principle. Again, inhomogeneous boundary conditions can be used provided that the right-hand side of (\ref{IEL_BC_u_par_pm}) is $O(\tau)$. The second principle leads us to prescribe homogeneous Neumann boundary conditions for the parallel momentum on $\Gamma_0$: 
\begin{eqnarray}
& & \hspace{-1cm}   \nu \cdot \nabla ( n^\tau  u^\tau_\parallel )  = 0, \quad \mbox{on} \quad \Gamma_0. 
\label{IEL_BC_u_par_0} 
\end{eqnarray}

Now, we consider the transverse momentum and eq. (\ref{RIEL2_u_perp}). The first principle (avoidance of boundary layers) leads us to propose Dirichlet boundary conditions for $n^\tau u^\tau_\bot$: 
\begin{eqnarray}
& & \hspace{-1cm}  u_\bot^\tau|_{\Gamma} (y) =  u_{\bot B}^\tau (y), \quad \forall y \in \Gamma,       
\label{IEL_BC_u_perp} 
\end{eqnarray}
with $u_{\bot B}^\tau$ satisfying:
\begin{eqnarray}
& & \hspace{-1cm}  n^\tau u_{\bot B}^\tau = \frac{1}{|B|} b \times \left\{ \nabla p(n^\tau) -  n^\tau E  \right\} + O(\tau) , \quad \mbox{on} \quad \Gamma.  \label{IEL_BC_u_perp_2} 
\end{eqnarray}

We comment on the conditions (\ref{IEL_BC_n_pm}) and (\ref{IEL_BC_u_par_pm}). Introducing functions $h(n)$ such that $h'(n) = p'(n)/n$, and $\phi(x)$ such that locally: 
\begin{eqnarray}
& & \hspace{-1cm}   E_\parallel  = - \nabla_\parallel \phi,       
\label{def_pot_par} 
\end{eqnarray}
(\ref{IEL_BC_n_pm}) can be rewritten as 
\begin{eqnarray}
& & \hspace{-1cm}   b \cdot \nabla (h(n^\tau) + \phi) = 0, \quad \mbox{on} \quad \Gamma_- \cup \Gamma_+. 
\label{IEL_BC_n_pm_1} 
\end{eqnarray}
It expresses that $h(n^\tau) + \phi$ is locally constant in the direction of the field line on $\Gamma_- \cup \Gamma_+$. Now, using that $\nabla \cdot B = 0$, we have 
$$ \nabla \cdot b = - b \cdot \nabla \ln |B|. $$
Then, (\ref{IEL_BC_u_par_pm}) can be recast as 
\begin{eqnarray}
& & \hspace{-1cm}   b \cdot \nabla \ln \left( \frac{n^\tau  u^\tau_\parallel}{|B|} \right) = 0, \quad \mbox{on} \quad \Gamma_- \cup \Gamma_+, 
\label{IEL_BC_u_par_pm_1} 
\end{eqnarray}
which expresses that the quantity $n^\tau  u^\tau_\parallel / |B|$ is locally constant in the direction of the field line on $\Gamma_- \cup \Gamma_+$.

A last comment is that eq. (\ref{RIEL2_n}) with boundary conditions (\ref{IEL_BC_n_pm}), (\ref{IEL_BC_n_0}) or eq. (\ref{RIEL1_u_par}) with boundary conditions (\ref{IEL_BC_u_par_pm}), (\ref{IEL_BC_u_par_0}) are well posed for $\tau >0$ but ill-posed in the limit $\tau \to 0$ because the solution is then defined up to a constant (per field line). This induces a bad conditioning of these equations when $\tau$ is small which will require some special treatment. We also remark the strong anisotropy of the problems in the direction of the field lines. Since $b$ may be time-dependent, we wish to develop solution strategies which do not rely on a special set of coordinates related to $b$. The resolution of these elliptic problem will be considered in detail in a forthcoming section.

%%%%%%%%%%%%%%%%%%%%%%%%%%%%%%%%%%%%%%%%%%%%%%%%%%%%%%%%%%%%%%%%
%%%%%%%%%%%%%%%%%%%%%%%%%%%%%%%%%%%%%%%%%%%%%%%%%%%%%%%%%%%%%%%%
\setcounter{equation}{0}
\section{Time semi-discrete AP scheme}
\label{section_EL_semi_discrete}

%%%%%%%%%%%%%%%%%%%%%%%%%%%%%%%%%%%%%%%%%%%%%%%%%%%%%%%%%%%%%%%%
\subsection{Time semi-discrete schemes: general setting}
\label{subsection_EL_semidis_gene}

We rewrite the IEL system using the conservative variables $n^\tau$ and $q^\tau = n^\tau u^\tau$: 
\begin{eqnarray}
& & \hspace{-1cm} \partial_t n^\tau + \nabla \cdot q^\tau = 0, \label{SDFEL_n_10} \\
& & \hspace{-1cm}  \tau \left\{ \partial_t q^\tau + \nabla \cdot \left( \frac{q^\tau \otimes q^\tau}{n^\tau} \right) \right\}+ \nabla p(n^\tau) = n^\tau E + q^\tau \times B ,  \label{SDFEL_u_10} 
\end{eqnarray}
We devise two Asymptotic-Preserving (AP) schemes corresponding to the RIEL-1 and RIEL-2 (respectively) reformulations of the Euler-Lorentz system. We start from a discretization of system (\ref{SDFEL_n_10}), (\ref{SDFEL_u_10}) and design the schemes in such a way that the manipulations which have led to the RIEL-1 and RIEL-2 reformulations can be performed at the discrete level. 

There are several reasons for not using the RIEL-1 or RIEL-2 formulations directly. First, these formulations are quite complicated and involve many terms. It is not clear how to discretize them in a good way. Second, the scheme must provide consistent solutions in both the regimes $\tau = O(1)$ and $\tau \ll 1$. The RIEL-1 and RIEL-2 forms are adequate for the regime $\tau \ll 1$ but not for the regime $\tau = O(1)$. In this regime, the problem is a standard system of conservation laws with source terms, for which a huge literature is available (see e.g. \cite{God_Rav_96, Leveque_92, Leveque_02, Toro_99}). This literature can be directly adapted to the form (\ref{SDFEL_n_10}), (\ref{SDFEL_u_10}) but much less obviously to the the RIEL-1 or RIEL-2 forms. For these reasons, we develop our schemes starting from (\ref{SDFEL_n_10}), (\ref{SDFEL_u_10}).

We first consider the time semi-discretization, because, in the present example, like in many other instances, the design of an AP scheme is primarily a question of time-discretization. in a forthcoming section, we will discuss the full discretization of these equations by AP methods. The time semi-discretization serves as a preparation for this last step. Surprisingly, the algebra is slightly simpler in the discrete than in the continuous case. We successively discuss the two reformulations.

%%%%%%%%%%%%%%%%%%%%%%%%%%%%%%%%%%%%%%%%%%%%%%%%%%%%%%%%%%%%%%%%
\subsection{AP scheme based on the first reformulation}
\label{subsection_EL_semidis_ref1}

We start with some notations and preliminaries. We denote by $n^{\tau,m}$, $q^{\tau,m}$ approximations of $n^\tau$ and $q^\tau$ at time $t^m = m \delta$. Since $b$ may depend on time, we index the 'perpendicular' and 'parallel' components of a vector field $v$ by the time index $m$, i.e. $b^m = b(t^m)$ and
\begin{eqnarray} 
& & \hspace{-1cm}
   v =  v_{\parallel^m}  b^m +
   v_{\perp^m}\,, \quad v_{\parallel^m} =  v \cdot  b^m \,, \quad
   v_{\perp^m} =  b^m \times \left( v \times  b^m \right) = (\mbox{Id} - b^m \otimes b^m) v \,, \label{decomp_m}
\end{eqnarray}
and similarly for the parallel gradient and divergence operators. We suppose that $E$ and $B$ are known in the course of time and that these projections are available at all times without any approximation. The coupling of the Euler equations with a time evolution of $E$ and $B$ (e.g. through Maxwell's equations) will not be discussed here, but is of course an important and interesting question for future works. 

We now introduce the:

\begin{definition}
The First Semi-Discrete AP scheme (SDAP-1 scheme) is defined by: 
\begin{eqnarray}
& & \hspace{-1cm} \delta^{-1} ( n^{\tau,m+1} - n^{\tau,m} ) + \nabla_{\parallel^{m+1}} \cdot q^{\tau,m+1}_{\parallel^{m+1}} + \nabla_{\bot^{m+1}} \cdot q^{\tau,m}_{\bot^{m+1}} = 0, \label{TSDEL_n_2} \\
& & \hspace{-1cm}  \tau \left\{ \delta^{-1} ( q^{\tau,m+1} - q^{\tau,m} ) + \nabla \cdot \left( \frac{q^{\tau,m} \otimes q^{\tau,m}}{n^{\tau,m}} \right) \right\} + \nonumber \\
& & \hspace{1cm} + \nabla \left[ p(n^{\tau,m}) - \delta \,  p'(n^{\tau,m}) \, (\nabla_{\parallel^{m+1}} \cdot q^{\tau,m+1}_{\parallel^{m+1}} + \nabla_{\bot^{m+1}} \cdot q^{\tau,m}_{\bot^{m+1}}) \right] = \nonumber \\
& & \hspace{6cm}  = n^{\tau,m+1} E^{m+1} + q^{\tau,m+1} \times B^{m+1} .  \label{TSDEL_u_2} 
\end{eqnarray}
\label{def_SDAP1}
\end{definition}

\noindent
The rationale for this approximation is as follows. We start from the following implicit scheme: 
\begin{eqnarray}
& & \hspace{-1cm} \delta^{-1} ( n^{\tau,m+1} - n^{\tau,m} ) + \nabla \cdot q^{\tau,m+1} = 0, \label{TSDEL_n_1} \\
& & \hspace{-1cm}  \tau \left\{ \delta^{-1} ( q^{\tau,m+1} - q^{\tau,m} ) + \nabla \cdot \left( \frac{q^{\tau,m} \otimes q^{\tau,m}}{n^{\tau,m}} \right) \right\} + \nabla p(n^{\tau,m+1}) = \nonumber \\
& & \hspace{6cm}  = n^{\tau,m+1} E^{m+1} + q^{\tau,m+1} \times B^{m+1} .  \label{TSDEL_u_1} 
\end{eqnarray}
We then split $\nabla \cdot q^{\tau,m+1}$ into its parallel and transverse components and evaluate the transverse component explicitly: 
\begin{eqnarray}
\nabla \cdot q^{\tau,m+1}  &=& \nabla_{\parallel^{m+1}} \cdot q^{\tau,m+1}_{\parallel^{m+1}} + \nabla_{\bot^{m+1}} \cdot q^{\tau,m+1}_{\bot^{m+1}}  \nonumber \\
&  = &\nabla_{\parallel^{m+1}} \cdot q^{\tau,m+1}_{\parallel^{m+1}} + \nabla_{\bot^{m+1}} \cdot q^{\tau,m}_{\bot^{m+1}} + O(\delta) .   
\label{TSDEL_div_q_2} 
\end{eqnarray}
Inserting (\ref{TSDEL_div_q_2}) into (\ref{TSDEL_n_1}) leads to (\ref{TSDEL_n_2}). Then, we Taylor expand $p(n^{\tau,m+1})$, insert (\ref{TSDEL_n_1}) and use (\ref{TSDEL_div_q_2}) again:
\begin{eqnarray} 
p(n^{\tau,m+1}) &=& p(n^{\tau,m}) + p'(n^{\tau,m}) ( n^{\tau,m+1} - n^{\tau,m} ) + o(\delta) \nonumber \\
&=& p(n^{\tau,m}) - \delta \,  p'(n^{\tau,m}) \, \nabla \cdot q^{\tau,m+1} + o(\delta)   \nonumber \\
& = & p(n^{\tau,m}) - \delta \,  p'(n^{\tau,m}) \, (\nabla_{\parallel^{m+1}} \cdot q^{\tau,m+1}_{\parallel^{m+1}} + \nabla_{\bot^{m+1}} \cdot q^{\tau,m}_{\bot^{m+1}})  + o(\delta).   \label{TSDEL_p_expan_3} 
\end{eqnarray}
Now, we replace $p(n^{\tau,m+1})$ in (\ref{TSDEL_u_1} ) by the right-hand side of (\ref{TSDEL_p_expan_3}). This leads to (\ref{TSDEL_u_2}). 

None of the above listed manipulations has altered the conservative character of the scheme, nor its consistency. Thanks to these modifications, this scheme is consistent with the RIEL-1 formulation of the Euler-Lorentz model, and in the limit $\tau \to 0$, with the RIDF-1 formulation of the Drift-Fluid model. This is precisely stated in the two following propositions:

\begin{proposition}
The SDAP-1 scheme can be equivalently formulated as follows: 
\begin{eqnarray}
& & \hspace{-1cm} \delta^{-1} ( n^{\tau,m+1} - n^{\tau,m} ) + \nabla_{\parallel^{m+1}} \cdot q^{\tau,m+1}_{\parallel^{m+1}} + \nabla_{\bot^{m+1}} \cdot q^{\tau,m}_{\bot^{m+1}} = 0, \label{SD1_n} \\
& & \hspace{-1cm}  
q_{\bot^{m+1}}^{\tau,m+1} = \frac{b^{m+1}}{|B^{m+1}|}  \times \left(  \nabla_{\bot^{m+1}} \left[ p(n^{\tau,m}) - \delta \,  p'(n^{\tau,m}) \, (\nabla_{\parallel^{m+1}} \cdot q^{\tau,m+1}_{\parallel^{m+1}} + \nabla_{\bot^{m+1}} \cdot q^{\tau,m}_{\bot^{m+1}}) \right] \right. \nonumber \\
& & \hspace{1cm}  
\left. - n^{\tau,m+1} E^{m+1}_{\bot^{m+1}} + \tau \left\{ \delta^{-1} ( q_{\bot^{m+1}}^{\tau,m+1} - q_{\bot^{m+1}}^{\tau,m} ) + ( \nabla \cdot ( \frac{q^{\tau,m} \otimes q^{\tau,m}}{n^{\tau,m}} ) )_{\bot^{m+1}} \right\} \right) .  
\label{SD1_u_perp} \\
& & \hspace{-1cm}  \tau  \delta^{-1}  q_{\parallel^{m+1}}^{\tau,m+1}  - \delta \, \nabla_{\parallel^{m+1}} ( p'(n^{\tau,m}) \, (\nabla_{\parallel^{m+1}} \cdot q^{\tau,m+1}_{\parallel^{m+1}}) ) + \delta\,  E^{m+1}_{\parallel^{m+1}} \, \nabla_{\parallel^{m+1}} \cdot q^{\tau,m+1}_{\parallel^{m+1}}  =    \nonumber \\
& & \hspace{4cm} = \tau  \delta^{-1} 
q_{\parallel^{m+1}}^{\tau,m} - \tau \, \,  ( \nabla \cdot ( \frac{q^{\tau,m} \otimes q^{\tau,m}}{n^{\tau,m}} ) )_{\parallel^{m+1}} \nonumber \\
& & \hspace{4.5cm} - \nabla_{\parallel^{m+1}} \left[ p(n^{\tau,m}) - \delta \,  p'(n^{\tau,m}) \,  \nabla_{\bot^{m+1}} \cdot q^{\tau,m}_{\bot^{m+1}} \right] + \nonumber \\
& & \hspace{4.5cm}  +  ( n^{\tau,m} - \delta \nabla_{\bot^{m+1}} \cdot q^{\tau,m}_{\bot^{m+1}})  E^{m+1}_{\parallel^{m+1}}  ,  \label{TSDEL_u_par_5} 
\end{eqnarray}
The SDAP-1 scheme is consistent with the RIEL-1 formulation of the Euler-Lorentz model
\label{prop_SDAP1_reform}
\end{proposition}

{\bf Proof:}  We first take the parallel component of (\ref{TSDEL_u_2}) and get: 
\begin{eqnarray}
& & \hspace{-1cm}  \tau \left\{ \delta^{-1} ( q_{\parallel^{m+1}}^{\tau,m+1} - q_{\parallel^{m+1}}^{\tau,m} ) + ( \nabla \cdot ( \frac{q^{\tau,m} \otimes q^{\tau,m}}{n^{\tau,m}} ) )_{\parallel^{m+1}} \right\} + \nonumber \\
& & \hspace{-0.5cm} + \nabla_{\parallel^{m+1}} \left[ p(n^{\tau,m}) - \delta \,  p'(n^{\tau,m}) \, (\nabla_{\parallel^{m+1}} \cdot q^{\tau,m+1}_{\parallel^{m+1}} + \nabla_{\bot^{m+1}} \cdot q^{\tau,m}_{\bot^{m+1}}) \right] = \nonumber \\
& & \hspace{10cm}  =  n^{\tau,m+1} E^{m+1}_{\parallel^{m+1}}  .  \label{TSDEL_u_par_3} 
\end{eqnarray}
Using (\ref{TSDEL_n_2}) to eliminate $n^{\tau,m+1}$ at the right-hand side of (\ref{TSDEL_u_par_3}), and bringing all terms involving $q_{\parallel^{m+1}}^{\tau,m+1}$ to the left-hand side and all other terms to the right-hand side, we find (\ref{TSDEL_u_par_5}). Now, taking the cross product of (\ref{TSDEL_u_2}) with $b^{m+1}$, we get (\ref{SD1_u_perp}). Eq. (\ref{TSDEL_u_par_5}) is a time-integrated version of the wave equation (\ref{RIEL1_u_par}). Therefore, the whole SDAP-1 scheme is consistent with the RIEL-1 model. This ends the proof. \endproof

\noindent
Eq. (\ref{TSDEL_u_par_5}) takes the form of an elliptic problem for $q_{\parallel^{m+1}}^{\tau,m+1}$ where the right-hand side is known. This elliptic equation, supplemented with the Neumann boundary conditions (\ref{IEL_BC_u_par_pm}), is well posed and provides the updated value $q_{\parallel^{m+1}}^{\tau,m+1}$. Once $q_{\parallel^{m+1}}^{\tau,m+1}$ is known, $n^{\tau,m+1}$ can be computed using (\ref{SD1_n}). Finally, eq. (\ref{SD1_u_perp}), which is clearly a time discretization of (\ref{RIEL1_u_perp}), can be solved. Alternately (\ref{SD1_u_perp}) can be written: 
\begin{eqnarray}
& & \hspace{-1cm}  
\left( \mbox{Id} - \frac{\tau}{\delta |B^{m+1}|} b^{m+1} \times \right) q_{\bot^{m+1}}^{\tau,m+1} = b^{m+1} \times Y^{m+1/2}, \label{TSDEL_u_per_5} \\
& & \hspace{-1cm}  
Y^{m+1/2} =  \frac{1}{|B^{m+1}|} \left(  \nabla_{\bot^{m+1}} \left[ p(n^{\tau,m}) - \delta \,  p'(n^{\tau,m}) \, (\nabla_{\parallel^{m+1}} \cdot q^{\tau,m+1}_{\parallel^{m+1}} + \nabla_{\bot^{m+1}} \cdot q^{\tau,m}_{\bot^{m+1}}) \right] \right. \nonumber \\
& & \hspace{1cm}  
\left. - n^{\tau,m+1} E^{m+1}_{\bot^{m+1}} + \tau \left\{ - \delta^{-1} q_{\bot^{m+1}}^{\tau,m} + ( \nabla \cdot ( \frac{q^{\tau,m} \otimes q^{\tau,m}}{n^{\tau,m}} ) )_{\bot^{m+1}} \right\} \right) ,  
\label{TSDEL_u_per_6} 
\end{eqnarray}
where \, Id \, denotes the identity matrix and $b^{m+1} \times$ denote the matrix of the vector product by $b^{m+1}$. The vector $Y^{m+1/2}$ is constructed with known quantities and (\ref{TSDEL_u_per_5}) can be easily solved for $q_{\bot^{m+1}}^{\tau,m+1}$ as follows: 
\begin{eqnarray}
& & \hspace{-1cm}  
q_{\bot^{m+1}}^{\tau,m+1} = \frac{\delta |B^{m+1}|}{\tau^2 + \delta^2 |B^{m+1}|^2} \, ( - \tau Y^{m+1/2} \, + \,  \delta |B^{m+1}| \, \, 
b^{m+1} \times Y^{m+1/2}). 
\label{TSDEL_u_per_7} 
\end{eqnarray}
This algebraic relation provides the update $q_{\bot^{m+1}}^{\tau,m+1}$. 

\medskip
We have the obvious:

\begin{proposition}
The limit $\tau \to 0$ in the SDAP-1 scheme leads to the following First Semi-Discrete Drift-Fluid scheme or SDDF-1 scheme:  
\begin{eqnarray}
& & \hspace{-1cm} \delta^{-1} ( n^{0,m+1} - n^{0,m} ) + \nabla_{\parallel^{m+1}} \cdot q^{0,m+1}_{\parallel^{m+1}} + \nabla_{\bot^{m+1}} \cdot q^{0,m}_{\bot^{m+1}} = 0, \label{SD1_n_tau=0} \\
& & \hspace{-1cm}  
q_{\bot^{m+1}}^{0,m+1} = \frac{b^{m+1}}{|B^{m+1}|}  \times \left(  \nabla_{\bot^{m+1}} \left[ p(n^{0,m}) - \delta \,  p'(n^{0,m}) \, (\nabla_{\parallel^{m+1}} \cdot q^{0,m+1}_{\parallel^{m+1}} + \nabla_{\bot^{m+1}} \cdot q^{0,m}_{\bot^{m+1}}) \right] \right. \nonumber \\
& & \hspace{10.5cm}  
\left. - n^{0,m+1} E^{m+1}_{\bot^{m+1}} \right) .  
\label{SD1_u_perp_tau=0} \\
& & \hspace{-1cm}  - \delta \, \nabla_{\parallel^{m+1}} ( p'(n^{0,m}) \, (\nabla_{\parallel^{m+1}} \cdot q^{0,m+1}_{\parallel^{m+1}}) ) =    - \nabla_{\parallel^{m+1}} \left[ p(n^{0,m}) - \delta \,  p'(n^{0,m}) \,  \nabla_{\bot^{m+1}} \cdot q^{0,m}_{\bot^{m+1}} \right] + \nonumber \\
& & \hspace{10.5cm}  +  n^{0,m+1} E^{m+1}_{\parallel^{m+1}}  .  \label{SD1_u_par_tau=0} 
\end{eqnarray}
The SDDF-1 scheme is consistent with the RIDF-1 reformulation of the Drift-Fluid model. 
\label{prop_SDAP1-SDDF1}
\end{proposition}

\noindent
{\bf Proof:} Eqs. (\ref{SD1_n_tau=0}) and (\ref{SD1_u_perp_tau=0}) are clearly consistent with (\ref{LDFEL_n_3}), (\ref{LDFEL_u_perp_3}), while (\ref{SD1_u_par_tau=0}) is a time-integrated version of (\ref{LDFEL_u_par_3}). Therefore, the SDDF-1 scheme is consistent with the RIDF-1 model. \endproof 

\noindent
This last proposition shows that the SDAP-1 scheme si AP.

%%%%%%%%%%%%%%%%%%%%%%%%%%%%%%%%%%%%%%%%%%%%%%%%%%%%%%%%%%%%%%%%
\subsection{AP scheme based on the second reformulation}
\label{subsection_EL_semidis_ref2}

\begin{definition}
The Second Semi-Discrete AP scheme (SDAP-2 scheme) is defined by: 
\begin{eqnarray}
& & \hspace{-1cm} \delta^{-1} ( n^{\tau,m+1} - n^{\tau,m} ) + \nabla \cdot ( q_{\parallel^{m+1}}^{\tau,m+1} b^{m+1} + q_{\bot^{m+1}}^{\tau,m} ) = 0, \label{TSDEL_n_12} \\
& & \hspace{-1cm}  \tau \left\{ \delta^{-1} ( q^{\tau,m+1} - q^{\tau,m} ) + \nabla \cdot \left( \frac{q^{\tau,m} \otimes q^{\tau,m}}{n^{\tau,m}} \right) \right\} + \nabla p(n^{\tau,m+1}) = \nonumber \\
& & \hspace{6cm}  = n^{\tau,m+1} E^{m+1} + q^{\tau,m+1} \times B^{m+1} .  \label{TSDEL_u_12} 
\end{eqnarray}
\label{def_SDAP2}
\end{definition}

\noindent
The rationale for this scheme is the same as for the SDAP-1 scheme. We start from 
(\ref{TSDEL_n_1}), (\ref{TSDEL_u_1}) and transform the mass flux using (\ref{TSDEL_div_q_2}). However, here, we do not transform the pressure term in the momentum balance eq. This leads to the SDAP-2 scheme.

Again, none of these manipulations has altered the conservative character of the scheme, nor its consistency. We show that this scheme is consistent with the RIEL-2 formulation of the Euler-Lorentz model, and in the limit $\tau \to 0$, with the RIDF-2 formulation of the Drift-Fluid model.

\begin{proposition}
The SDAP-2 scheme can be equivalently formulated as follows: 
\begin{eqnarray}
& & \hspace{-1cm} \tau \delta^{-1} n^{\tau,m+1} - \delta \, \nabla_{\parallel^{m+1}} \cdot  (  \nabla_{\parallel^{m+1}} p(n^{\tau,m+1}) - n^{\tau,m+1} E_{\parallel^{m+1}}^{m+1} ) = \nonumber \\
& & \hspace{1cm} = \tau \delta^{-1} n^{\tau,m} + \tau \left\{ - \nabla \cdot  
q^{\tau,m} + \delta  ( \nabla \cdot ( \frac{q^{\tau,m} \otimes q^{\tau,m}}{n^{\tau,m}} ) )_{\parallel^{m+1}}  \right\} , \label{SD2L_n} \\
& & \hspace{-1cm}  
q_{\bot^{m+1}}^{\tau,m+1} = \frac{b^{m+1}}{|B^{m+1}|}  \times (  \nabla_{\bot^{m+1}}  p(n^{\tau,m+1}) - n^{\tau,m+1} E^{m+1}_{\bot^{m+1}}  \nonumber \\
& & \hspace{2cm}  + \tau \{ \delta^{-1} ( q_{\bot^{m+1}}^{\tau,m+1} - q_{\bot^{m+1}}^{\tau,m} ) + ( \nabla \cdot ( \frac{q^{\tau,m} \otimes q^{\tau,m}}{n^{\tau,m}} ) )_{\bot^{m+1}} \} ) ,  
\label{SD2_u_perp}  \\
& & \hspace{-1cm} \delta^{-1} ( q_{\parallel^{m+1}}^{\tau,m+1} - q_{\parallel^{m+1}}^{\tau,m} ) + (\nabla \cdot ( \frac{q^{\tau,m} \otimes q^{\tau,m}}{n^{\tau,m}} ) )_{\parallel^{m+1}} + \nonumber \\
& & \hspace{4cm} + \tau^{-1} \left( \nabla_{\parallel^{m+1}} p(n^{\tau,m+1}) - n^{\tau,m+1} E_{\parallel^{m+1}}^{m+1} \right) =0 .  \label{SD2_u_par} 
\end{eqnarray}
It is consistent with the RIEL-2 formulation of the Euler-Lorentz model
\label{prop_SDAP2_reform}
\end{proposition}

\noindent
{\bf Proof:} Taking the parallel component of (\ref{TSDEL_u_12}), we find (\ref{SD2_u_par}).  Inserting the value of $q_{\parallel^{m+1}}^{\tau,m+1}$ found from (\ref{SD2_u_par}) in (\ref{TSDEL_n_12}) leads to (\ref{SD2L_n}). Taking the cross product of $b^{m+1}$ with (\ref{TSDEL_u_12}) leads to (\ref{SD2_u_perp}). 
Eq. (\ref{SD2L_n}) is a time-integrated version of the wave equation (\ref{RIEL2_n}). Therefore, the whole SDAP-2 scheme is consistent with the RIEL-2 formulation of the Euler-Lorentz model. \endproof 

\noindent
Eq. (\ref{SD2L_n}) is a nonlinear elliptic equation for $n^{\tau,m+1}$, the right-hand side of which is known from previous time steps. It has a unique solution thanks to the boundary conditions (\ref{IEL_BC_n_pm}). Furthermore, these boundary conditions guarantee that 
\begin{eqnarray}
& & \hspace{-1cm} \nabla_{\parallel^{m+1}} p(n^{\tau,m+1}) - n^{\tau,m+1} E_{\parallel^{m+1}}^{m+1} = O(\tau). \label{TSDEL_n_17} 
\end{eqnarray}
Eq. (\ref{SD2_u_par}) looks singular as $\tau \to 0$. However, with (\ref{TSDEL_n_17}), the seemingly singular term at the right-hand side of (\ref{SD2_u_par}) is of order unity and the equation is not singular as $\tau \to 0$. 
Finally, (\ref{SD2_u_perp}) can be alternately written: 
\begin{eqnarray}
& & \hspace{-1cm}  
\left( \mbox{Id} - \frac{\tau}{\delta |B^{m+1}|} b^{m+1} \times \right) q_{\bot^{m+1}}^{\tau,m+1} = b^{m+1} \times Z^{m+1/2}, \label{TSDEL_u_per_19} \\
& & \hspace{-1cm}  
Z^{m+1/2} =  \frac{1}{|B^{m+1}|} (  \nabla_{\bot^{m+1}} p(n^{\tau,m+1}) - n^{\tau,m+1} E^{m+1}_{\bot^{m+1}} + \nonumber \\
& & \hspace{3cm}  
+ \tau \{ - \delta^{-1} q_{\bot^{m+1}}^{\tau,m} + ( \nabla \cdot ( \frac{q^{\tau,m} \otimes q^{\tau,m}}{n^{\tau,m}} ) )_{\bot^{m+1}} \} ) ,  
\label{TSDEL_u_per_20} 
\end{eqnarray}
and has the solution 
\begin{eqnarray}
& & \hspace{-1cm}  
q_{\bot^{m+1}}^{\tau,m+1} = \frac{\delta |B^{m+1}|}{\tau^2 + \delta^2 |B^{m+1}|^2} \, ( - \tau Z^{m+1/2} \, + \,  \delta |B^{m+1}| \, \, 
b^{m+1} \times Z^{m+1/2}). 
\label{TSDEL_u_per_21} 
\end{eqnarray}

\medskip
We investigate the limit $\tau \to 0$ in the following proposition whose proof is easy and left to the reader:

\begin{proposition}
Taking the limit $\tau \to 0$ in the SDAP-2 scheme, expanding $n^{\tau,m+1} = n^{0,m+1} + \tau n^{1,m+1} + o(\tau)$ and using the boundary condition (\ref{IEL_BC_n_pm}) leads to the following Second Semi-Discrete Drift-Fluid scheme or SDDF-2 scheme:  
\begin{eqnarray}
& & \hspace{-1cm}  \nabla_{\parallel^{m+1}} p(n^{0,m+1}) - n^{0,m+1} E_{\parallel^{m+1}}^{m+1}  = 0 , \label{SD2L_n_tau=0} \\
& & \hspace{-1cm} \delta^{-1} n^{0,m+1} - \delta \, \nabla_{\parallel^{m+1}} \cdot  (  \nabla_{\parallel^{m+1}} (p'(n^{0,m+1}) n^{1,m+1} ) - n^{1,m+1} E_{\parallel^{m+1}}^{m+1} ) = \nonumber \\
& & \hspace{1cm} = \delta^{-1} n^{0,m} + \left\{ - \nabla \cdot  
q^{0,m} + \delta  ( \nabla \cdot ( \frac{q^{0,m} \otimes q^{0,m}}{n^{0,m}} ) )_{\parallel^{m+1}}  \right\} , \label{SD2L_n1} \\
& & \hspace{-1cm} \delta^{-1} ( q_{\parallel^{m+1}}^{0,m+1} - q_{\parallel^{m+1}}^{0,m} ) + (\nabla \cdot ( \frac{q^{0,m} \otimes q^{0,m}}{n^{0,m}} ) )_{\parallel^{m+1}} + \nonumber \\
& & \hspace{4cm} + \left( \nabla_{\parallel^{m+1}} (p'(n^{0,m+1}) n^{1,m+1} ) - n^{1,m+1} E_{\parallel^{m+1}}^{m+1} \right) =0 .  \label{SD2_u_par_tau=0} \\
& & \hspace{-1cm}  
q_{\bot^{m+1}}^{0,m+1} = \frac{b^{m+1}}{|B^{m+1}|}  \times (  \nabla_{\bot^{m+1}}  p(n^{0,m+1}) - n^{0,m+1} E^{m+1}_{\bot^{m+1}} )   ,   
\label{SD2_u_perp_tau=0} 
\end{eqnarray}
The SDDF-2 scheme is consistent with the RIDF-2 reformulation of the Drift-Fluid model. 
\label{prop_SDAP2-SDDF2}
\end{proposition}

\noindent
This last proposition shows that the SDAP-2 scheme si AP.

%%%%%%%%%%%%%%%%%%%%%%%%%%%%%%%%%%%%%%%%%%%%%%%%%%%%%%%%%%%%%%%%
\subsection{Time semi-discrete schemes: conclusions}
\label{subsection_EL_semidis_conclu}

In the previous sections, we have derived two different semi-discrete AP schemes (the SDAP-1 and SDAP-2 schemes). In the limit $\tau \to 0$, these two schemes are respectively consistent with the two previously established reformulations of the Euler-Lorentz problem (the RIEL-1 and RIEL-2 reformulations). To derive these schemes, we have started from the time continuous problem in its original form (the IEL form) instead of using the reformulated forms. Both schemes are derived from an implicit scheme where the flux in the mass conservation equation, the pressure flux in the momentum conservation equation, and the Lorentz force are evaluated implicitly. Then, the two schemes are transformed somehow similarly: in the SDAP-1 scheme, the implicit pressure gives rise to an elliptic equation for the parallel momentum through the use of the mass conservation equation. A symmetric operation is performed on the SDAP-2 scheme, where the implicit mass flux gives rise to an elliptic equation on the pressure through the use of the momentum balance equation. Both elliptic equations are degenerate: they are one-dimensional elliptic equations posed in the direction of the magnetic field lines. 

The derivation of time semi-discrete schemes is a preparation for the development of the fully discrete ones which will be performed in the next section. Time semi-discrete schemes are also convenient for  linearized stability analyses in the spirit of section \ref{subsec_QN_EP_1F_stability} or \ref{subsec_1F_time}. The stability analysis of these schemes will be considered in future work.

%%%%%%%%%%%%%%%%%%%%%%%%%%%%%%%%%%%%%%%%%%%%%%%%%%%%%%%%%%%%%%%%
%%%%%%%%%%%%%%%%%%%%%%%%%%%%%%%%%%%%%%%%%%%%%%%%%%%%%%%%%%%%%%%%
\setcounter{equation}{0}
\section{Fully discrete AP scheme}
\label{section_EL_fully_discrete}

%%%%%%%%%%%%%%%%%%%%%%%%%%%%%%%%%%%%%%%%%%%%%%%%%%%%%%%%%%%%%%%%
\subsection{Fully discrete schemes: general setting}
\label{subsection_EL_fully_discrete_general}

%%%%%%%%%%%%%%%%%%%%%%%%%%%%%%%%%%%%%%%%%%%%%%%%%%%%%%%%%%%%%%%%
\subsubsection{Classical schemes}
\label{subsubsection_EL_fully_class}

Before considering AP schemes, we recall the framework of the classical explicit shock-capturing schemes. We start from the original IEL formulation of the problem (\ref{SDFEL_n_10}), (\ref{SDFEL_u_10}),  which we write as follows: 
\begin{eqnarray}
& & \hspace{-1cm} \partial_t U + \sum_{j=1}^3 \partial_{x_j} f_j(U)  = \tau^{-1} g(t,U),
\label{gene_cons_law}
\end{eqnarray}
where $U=U(x,t)$ is the vector of conservative variables, $f_j(U)$, the flux and $g$, the source term: 
\begin{eqnarray}
& & \hspace{-1cm} U = \left( \begin{array}{c} n \\ q \end{array} \right), \quad f_j(U) = \left( \begin{array}{c} q_j \\ n^{-1} q_j q + \tau^{-1} p(n) e_j \end{array} \right), 
\label{gene_cons_law_2} \\
& & \hspace{-1cm}  g(t,U) = \left( \begin{array}{c} 0 \\ n E(t) + q \times B(t) \end{array} \right).
\label{gene_cons_law_3} 
\end{eqnarray}
We denote by $e_j$ the  unit vector in the $j$-th direction (for instance, $e_2 = (0,1,0)$). Here, we always assume that the space dimension is $3$.  

We develop a finite-volume formulation on a structured, cartesian mesh. However, the concepts would easily be generalized to a finite volume method on an unstructured mesh. Let $K = (k_1, k_2, k_3) \in {\mathbb Z}^3$ be a multi-index, and $C_K = h_1 [k_1 - 1/2, k_1 + 1/2] \times h_2 [k_2 - 1/2, k_2 + 1/2] \times h_3 [k_3 - 1/2, k_3 + 1/2]$ be the associated finite-volume cell, with space steps $h_j$ in the $j$-th direction. We denote by $U|_K^m$ an approximation of $U(x_K, t^m)$, where $x_K = (k_1 h_1, k_2 h_2, k_3 h_3)$ is the center of cell $c_K$. Similarly, $f_j|_{K+e_j/2}^m$ denotes an approximation of $f_j(U(x_{K+e_j/2}, t^m))$, with $x_{K+e_1/2} =  ((k_1+1/2) h_1, k_2 h_2, k_3 h_3)$, and similarly for $j=2,3$. Finally, $g^m(U) \approx g(t^m,U)$. 

The classical schemes are written as follows

\begin{definition}
Classical explicit shock-capturing explicit schemes are given by:
\begin{eqnarray}
& & \hspace{-1cm} \delta^{-1} (U|_K^{m+1} - U|_K^m) + 
\sum_{j=1}^3 h_j^{-1} ( f_j|_{K+e_j/2}^{m} - f_j|_{K-e_j/2}^{m} ) = \tau^{-1} 
g^{m} (U|_K^{m}) , 
\label{gene_cons_law_disc}
\end{eqnarray}
The fluxes are the sum of a central discretization term and a numerical viscosity term, according to:
\begin{eqnarray}
& & \hspace{-1cm} f_j|_{K+e_j/2}^{m} = \frac{1}{2} \left[ f_j(U|_K^m) + f_j(U|_{K+e_j}^m)  + \mu_j|_{K+e_j/2}^{m} (U|_K^m - U|_{K+e_j}^m) \right]
, 
\label{gene_cons_law_flux}
\end{eqnarray}
with a suitable viscosity matrix $\mu_j|_{K+e_j/2}^{m}$. 
\label{def_fd_class_schemes}
\end{definition}

\medskip
\noindent
We denote by 
\begin{eqnarray}
& & \hspace{-1cm}  D_j|_{K+e_j/2}^{m} := \mu_j|_{K+e_j/2}^{m} (U|_K^m - U|_{K+e_j}^m) , 
\label{gene_cons_law_num_visc} 
\end{eqnarray}
the numerical viscosity. In explicit shock capturing methods, $\mu_j|_{K+e_j/2}^{m}$ is derived from the Jacobian matrix of the flux functions. For instance, the Roe scheme corredponds to 
\begin{eqnarray}
& & \hspace{-1cm}  \mu_j|_{K+e_j/2}^{m} = \left| \frac{\partial f_j}{\partial U} (U|_{K+e_j/2}^m) \right|, 
\label{gene_cons_law_Roe} 
\end{eqnarray}
where $U|_{K+e_j/2}^m$ is a conveniently chosen average state between $U|_K^m$ and $U|_{K+e_j}^m$. The Rusanov scheme would correspond to $\mu_j|_{K+e_j/2}^{m}$ begin a scalar such that 
\begin{eqnarray}
& & \hspace{-1cm}  \mu_j|_{K+e_j/2}^{m} = \max \{ \, \max \{ \, |\lambda_j(U|_{K}^{m})|\, ,\,  |\lambda_j(U|_{K+e_j/2}^m)|\, ,\,  |\lambda_j(U|_{K+e_j}^{m})| \, \} ,    \nonumber \\
& & \hspace{3cm}   \mbox{ such that }   \lambda_j(U) \, \mbox{is an  eigenvalue of } \frac{\partial f_j}{\partial U} (U) \}
, 
\label{gene_cons_law_Rusanov} 
\end{eqnarray}
The CFL stability condition, which guarantees the stability of the scheme, is as follows 
\begin{eqnarray}
& & \hspace{-1cm}  \mu  \delta \leq h = \min_j h_j, 
\label{gene_cons_law_CFL} 
\end{eqnarray}
with 
\begin{eqnarray}
& & \hspace{-1cm} \mu = \max_{K \in {\mathbb Z}^3, \, m \in {\mathbb N}, \, j \in \{1,2,3 \} } \mu_j|_{K+e_j/2}^{m} 
, 
\label{gene_cons_law_CFL2} 
\end{eqnarray}
Any other shock capturing methods can be considered as well. We refer the reader to \cite{God_Rav_96, Leveque_92, Leveque_02, Toro_99} and references therein.

%%%%%%%%%%%%%%%%%%%%%%%%%%%%%%%%%%%%%%%%%%%%%%%%%%%%%%%%%%%%%%%%
\subsubsection{AP schemes}
\label{subsubsection_EL_fully_AP}

By contrast to our presentation of the Semi-Discrete schemes in section \ref{section_EL_semi_discrete}, we gradually derive the final expression of the fully-Discrete AP schemes from a general semi-implicit shock capturing scheme framework. In this section, we present the common starting point for the two AP schemes which we consider in this work.  Then, in two forthcoming sections, we will develop the specificities of each of these schemes which are intended to be consistent discretizations of the Reformulated Euler-Lorentz models RIEL-1 and RIEL-2. 

We plan to design AP-schemes from minor modifications from classical shock capturing schemes in order to ensure that the discretization of the left-hand side of (\ref{SDFEL_n_10}), (\ref{SDFEL_u_10}) is in conservative form. The conservativity property guarantees correct shock speeds at the discrete level. It cannot not be guaranteed if the scheme is developed from the RIEL-1 or RIEL-2 reformulated systems because of the presence of many and sometimes complex terms in these formulations. Another reason for dealing with the original system (\ref{SDFEL_n_10}), (\ref{SDFEL_u_10}) is the need for numerical viscosity to stabilize the discretization. While it is easy to adapt the literature to (\ref{SDFEL_n_10}), (\ref{SDFEL_u_10}), it is uneasy to decide where and how numerical viscosity should be added to the RIEL-1 or RIEL-2 formulations. 

The common framework for both AP schemes is a modification of (\ref{gene_cons_law_disc}) where some kind of implicitness in the flux and source terms is introduced: 
\begin{eqnarray}
& & \hspace{-1cm} \delta^{-1} (U|_K^{m+1} - U|_K^m) + 
\sum_{j=1}^3 h_j^{-1} ( \tilde f_j|_{K+e_j/2}^{m+1} - \tilde f_j|_{K-e_j/2}^{m+1} ) = \tau^{-1} 
g^{m+1} (U|_K^{m+1}) . 
\label{gene_cons_law_disc_impl}
\end{eqnarray}
where the implicit fluxes are denoted $\tilde f_j|_{K+e_j/2}^{m+1}$ to distinguish them from the explicit ones (\ref{gene_cons_law_flux}). The AP schemes will be such that the implicit fluxes $\tilde f_j|_{K+e_j/2}^{m+1}$ are fairly simple modifications of the explicit fluxes (\ref{gene_cons_law_flux}). 
Like in the time semi-discrete case, we construct them by making the mass and pressure fluxes implicit. Additionally, the implicitness only applies to the central part of the flux, because impliciting the viscosity is not needed to make the scheme AP. The implicit fluxes are thus given by: 
\begin{eqnarray}
& & \hspace{-1cm} \tilde f_j|_{K+e_j/2}^{m+1} = \frac{1}{2} \left[ \bar f_j|_K^{m+1} + \bar f_j|_{K+e_j}^{m+1}  + \mu_j|_{K+e_j/2}^{m} (U|_K^m - U|_{K+e_j}^m) \right] .
\label{gene_cons_law_flux_imp}
\end{eqnarray}
The bars denote implicit central fluxes $f_j|_K^{m+1}$ given by: 
\begin{eqnarray}
& & \hspace{-1cm}  \bar f_j|_K^{m+1} = \left( \begin{array}{c} q_j|_K^{m+1} \\ (n^{-1} q_j q)|_K^m + \tau^{-1} p(n|_K^{m+1}) e_j \end{array} \right),  
\label{gene_cons_law_flux_imp_cent} 
\end{eqnarray}
and $(n^{-1} q_j q)|_K^m$ is a short-hand writing for $(n^{-1})|_K^m \,  \,q_j|_K^m \,  \,q|_K^m$. Note that this part of the momentum flux is kept explicit, because impliciting it is not needed to make the scheme AP. The viscosity matrix and the CFL conditions are constructed only from the explicit part of the system, i.e. they are associated to the flux:
\begin{eqnarray}
& & \hspace{-1cm} f_j(U) = \left( \begin{array}{c} 0 \\ n^{-1} q_j q  \end{array} \right). 
\label{gene_cons_law_flux_expl} 
\end{eqnarray}
This systems has eigenvalues $0$ and $u_j = q_j/n$. For instance, for the Rusanov scheme, $ \mu_j|_{K+e_j/2}^{m}$ is a scalar equal to the maximal value of $|u|$ in the adjacent cells: \begin{eqnarray}
& & \hspace{-1cm}  \mu_j|_{K+e_j/2}^{m} = \max \{ \, |u|_{K}^m| \, , \, |u|_{K + e_j/2}^m| \, , \, |u|_{K + e_j}^m|\, \}
.
\label{gene_cons_law_Rus_expl} 
\end{eqnarray}
In doing so, neither the numerical viscosity nor the CFL condition depend on $\tau$, a condition for the scheme to be AP. Any other shock capturing methods can be considered as well. It may improve the stability of the scheme to keep some small explicit part in the mass and pressure fluxes, in the spirit of the method proposed in \cite{Deg_Tan}. We will defer the development of this idea to future work. 

Inserting (\ref{gene_cons_law_num_visc}) and (\ref{gene_cons_law_flux_imp_cent}) into (\ref{gene_cons_law_flux_imp}) and the resulting expression into (\ref{gene_cons_law_disc}), we get the following expression for the scheme: 
\begin{eqnarray}
& & \hspace{-1cm} \delta^{-1} (n|_K^{m+1} - n|_K^m) + 
\sum_{j=1}^3 h_j^{-1} ( \tilde f_{nj}|_{K+e_j/2}^{m+1} - \tilde f_{nj}|_{K-e_j/2}^{m+1} ) = 0, 
\label{gene_AP_n} \\
& & \hspace{-1cm}\delta^{-1} (q_i|_K^{m+1} - q_i|_K^m) + 
\sum_{j=1}^3 h_j^{-1} ( \tilde f_{q_ij}|_{K+e_j/2}^{m+1} - \tilde f_{q_ij}|_{K-e_j/2}^{m+1} ) =  \nonumber \\
& & \hspace{1cm} = \tau^{-1} 
( n|_K^{m+1} E_i|_K^{m+1} + (q|_K^{m+1} \times B|_K^{m+1})_i ) , \quad i = 1,2,3,
\label{gene_AP_q}
\end{eqnarray}
Here, we have supposed that the electric and magnetic fields are appxomitated by cell-centered discretizations $E|_K^{m}$, $B|_K^{m}$. The fluxes are given by: 
\begin{eqnarray}
& & \hspace{-1cm} \tilde f_{nj}|_{K+e_j/2}^{m+1} = \frac{1}{2} \left[ q_j|_K^{m+1} + q_j|_{K+e_j}^{m+1} + D_{nj}|_{K+e_j/2}^{m} \right] , 
\label{gene_AP_fn} \\
& & \hspace{-1cm} \tilde f_{q_ij}|_{K+e_j/2}^{m+1} = \frac{\tau^{-1}}{2} \left[ p(n_j|_K^{m+1}) + p(n_j|_{K+e_j}^{m+1}) \right] \delta_{ij} + \nonumber \\
& & \hspace{1cm} + \frac{1}{2} \left[  (n^{-1} q_i q_j)|_K^m +  (n^{-1} q_i q_j)|_{K+e_j}^m + D_{q_ij}|_{K+e_j/2}^{m} \right],  \,  \, i,j = 1,2,3,
\label{gene_AP_fq}
\end{eqnarray}
where $D_{nj}|_{K+e_j/2}^{m}$ and $D_{q_ij}|_{K+e_j/2}^{m}$ are the entries of the numerical viscosity vector $D_j|_{K+e_j/2}^{m}$: 
\begin{eqnarray}
& & \hspace{-1cm}  D_j|_{K+e_j/2}^{m} := \left( D_{nj}|_{K+e_j/2}^{m}, D_{q_1j}|_{K+e_j/2}^{m}, D_{q_2j}|_{K+e_j/2}^{m}, D_{q_3j}|_{K+e_j/2}^{m} \right) . 
\label{gene_cons_law_num_visc_vect} 
\end{eqnarray}
and $D_j|_{K+e_j/2}^{m}$ is given by (\ref{gene_cons_law_num_visc}) (but with the viscosity matrix associated to the explicit part of the flux, as pointed out in the previous paragraph). For later usage, we define: 
\begin{eqnarray}
& & \hspace{-1cm}  F_{q_ij}|_{K+e_j/2}^{m} = \frac{1}{2} \left[  (n^{-1} q_i q_j)|_K^m +  (n^{-1} q_i q_j)|_{K+e_j}^m + D_{q_ij}|_{K+e_j/2}^{m} \right],  \, \, i,j = 1,2,3. 
\label{gene_AP_Fdef}
\end{eqnarray}
Alternately, inserting (\ref{gene_AP_fn}), (\ref{gene_AP_fq}) into (\ref{gene_AP_n}), (\ref{gene_AP_q}), we can write: 
\begin{eqnarray}
& & \hspace{-1cm} \delta^{-1} (n|_K^{m+1} - n|_K^m) + 
\sum_{j=1}^3 \frac{h_j^{-1}}{2}  \left[ q_j|_{K+e_j}^{m+1} - q_j|_{K-e_j}^{m+1} \right] + \Delta_n|_K^m = 0, 
\label{gene_AP_n_2} \\
& & \hspace{-1cm}\delta^{-1} (q_i|_K^{m+1} - q_i|_K^m) + \frac{\tau^{-1} h_i^{-1}}{2}  \left[ p(n_i|_{K+e_i}^{m+1}) - p(n_i|_{K-e_i}^{m+1}) \right] + \Delta_{q_i}|_K^m =  \nonumber \\
& & \hspace{1cm} = \tau^{-1} 
( n|_K^{m+1} E_i|_K^{m+1} + (q|_K^{m+1} \times B|_K^{m+1})_i ) , \quad i = 1,2,3,
\label{gene_AP_q_2}
\end{eqnarray}
with 
\begin{eqnarray}
& & \hspace{-1cm} \Delta_n|_K^m = \sum_{j=1}^3   \frac{h_j^{-1}}{2} \left[ D_{nj}|_{K+e_j/2}^{m} - D_{nj}|_{K-e_j/2}^{m} \right], 
\label{gene_AP_Delta_n} \\
& & \hspace{-1cm} \Delta_{q_i}|_K^m = \sum_{j=1}^3   h_j^{-1}  \left[ F_{q_ij}|_{K+e_j/2}^{m} - F_{q_ij}|_{K-e_j/2}^{m} \right]
 , \quad i = 1,2,3. 
\label{gene_AP_Delta_qi}
\end{eqnarray}

We are now going to modify this scheme in two different ways in order to make each of the resulting scheme consistent with either the RIEL-1 or the RIEL-2 formulation of the Euler-Lorentz model, and a fully discrete counterpart of the time semi-discrete SDAP-1 and SDAP-2 schemes respectively.

%%%%%%%%%%%%%%%%%%%%%%%%%%%%%%%%%%%%%%%%%%%%%%%%%%%%%%%%%%%%%%%%
\subsection{Fully discrete AP scheme based on the first reformulation}
\label{subsection_EL_fully_ref1}

First, we decompose the momentum into its parallel and perpendicular parts. Specifically, we denote by 
\begin{eqnarray}
& & \hspace{-1cm} (\mbox{Id} - b \otimes b)|_K^{m} = \mbox{Id} - b|_K^{m} \otimes b|_K^{m}, \quad q_\bot|_K^{m} = (\mbox{Id} - b \otimes b)|_K^{m} \, q|_K^{m} \label{FDAP_I-bb} \\
& & \hspace{-1cm}q_\parallel|_K^{m} = q|_K^{m} \cdot b|_K^{m}, \quad q|_K^{m} = q_\parallel|_K^{m} \, b|_K^{m}  \, +  \, q_\bot|_K^{m}
. \label{AP1_qpar_qper}
\end{eqnarray}
Therefore, we can write
\begin{eqnarray}
q|_K^{m+1} &=& q_\parallel|_K^{m+1} \, b|_K^{m+1}  \, +  \, (\mbox{Id} - b \otimes b)|_K^{m+1} \, q|_K^{m+1}, \nonumber \\
&=& q_\parallel|_K^{m+1} \, b|_K^{m+1}  \, +  \, (\mbox{Id} - b \otimes b)|_K^{m+1} \, q|_K^{m} + O(\delta) \label{AP1_qdecom}
\end{eqnarray}
and insert this approximation into (\ref{gene_AP_fn}). This leads to a modified expression of the discrete mass flux: 
\begin{eqnarray}
& & \hspace{-1cm}
\tilde f_{nj}|_{K+e_j/2}^{m+1} = \frac{1}{2} \left[ q_\parallel|_K^{m+1} \, b_j|_K^{m+1}  \, +  q_\parallel|_{K+e_j}^{m+1} \, b_j|_{K+e_j}^{m+1}  \right] \,  + \, \tilde D_{nj}|_{K+e_j/2}^{m}, \label{AP1_fn} 
\end{eqnarray}
with 
\begin{eqnarray}
& & \hspace{-1cm}
\tilde D_{nj}|_{K+e_j/2}^{m} = \frac{1}{2} \left[  ((\mbox{Id} - b \otimes b)|_K^{m+1} \, q|_K^{m})_j \, +   \, ((\mbox{Id} - b \otimes b)|_{K+e_j}^{m+1} \, q|_{K+e_j}^{m})_j \right] + \nonumber \\
& & \hspace{10cm} + D_{nj}|_{K+e_j/2}^{m}. 
\label{AP1_def_tilde_D} 
\end{eqnarray}
The discrete mass balance eq. (\ref{gene_AP_n}) with the modified flux (\ref{AP1_fn}) is now written: 
\begin{eqnarray}
& & \hspace{-1cm} \delta^{-1} (n|_K^{m+1} - n|_K^m) + 
\sum_{j=1}^3 \frac{h_j^{-1}}{2}  \left[  q_\parallel|_{K+e_j}^{m+1} \, b_j|_{K+e_j}^{m+1} -  q_\parallel|_{K-e_j}^{m+1} \, b_j|_{K-e_j}^{m+1} \right] + \tilde \Delta_n|_K^m = 0, 
\label{AP1_n_2} 
\end{eqnarray}
with 
\begin{eqnarray}
& & \hspace{-1cm} \tilde \Delta_n|_K^m = \sum_{j=1}^3  h_j^{-1} \left[ \tilde D_{nj}|_{K+e_j/2}^{m} - \tilde D_{nj}|_{K-e_j/2}^{m} \right]. 
\label{AP1_tilde_Delta_n} 
\end{eqnarray}

Proceeding like in section \ref{subsection_EL_semidis_ref1}. We now expand $p(n_j|_K^{m+1})$, using (\ref{gene_AP_n_2}) and (\ref{AP1_n_2}): 
\begin{eqnarray}
& & \hspace{-1cm} p(n|_K^{m+1}) = p(n|_K^{m}) + p'(n|_K^{m}) (n|_K^{m+1} - n|_K^m) + O(\delta), 
\label{AP1_p_expan} \\
& & \hspace{0cm} = p(n|_K^{m}) - \delta \, p'(n|_K^{m}) \, (\sum_{j=1}^3 \frac{h_j^{-1}}{2}  \left[  q_\parallel|_{K+e_j}^{m+1} \, b_j|_{K+e_j}^{m+1} -  q_\parallel|_{K-e_j}^{m+1} \, b_j|_{K-e_j}^{m+1} \right] + \nonumber \\
& & \hspace{9cm} + \tilde \Delta_n|_K^m + O(\delta) \, ), \label{AP1_p_expan_2} \\
& & \hspace{0cm} = P|_K^{m} - \delta \, p'(n|_K^{m}) \, \sum_{j=1}^3 \frac{h_j^{-1}}{2}  \left[  q_\parallel|_{K+e_j}^{m+1} \, b_j|_{K+e_j}^{m+1} -  q_\parallel|_{K-e_j}^{m+1} \, b_j|_{K-e_j}^{m+1} \right] + O(\delta), \label{AP1_p_expan_3} 
\end{eqnarray}
with 
\begin{eqnarray}
& & \hspace{-1cm} P|_K^{m} = p(n|_K^{m}) - \delta \, p'(n|_K^{m}) \, \tilde \Delta_n|_K^m . 
\label{AP1_def_Pkm} 
\end{eqnarray}
Then, replacing $p(n|_K^{m+1})$ in (\ref{gene_AP_fq}) by its approximation (\ref{AP1_p_expan_3}) leads to a modified momentum flux: 
\begin{eqnarray}
& & \hspace{-1cm} \tilde f_{q_ij}|_{K+e_j/2}^{m+1}  = \tilde F_{q_ij}|_{K+e_j/2}^{m} + 
\nonumber \\
& & \hspace{-0.7cm} 
- \frac{\delta \tau^{-1}}{2} \, \left\{ p'(n|_K^{m}) \, \sum_{r=1}^3 \frac{h_r^{-1}}{2}  \left[  q_\parallel|_{K+e_r}^{m+1} \, b_r|_{K+e_r}^{m+1} -  q_\parallel|_{K-e_r}^{m+1} \, b_r|_{K-e_r}^{m+1} \right] \right.  \nonumber \\
& & \hspace{-0.2cm} 
\left. + p'(n|_{K+e_j}^{m}) \, \sum_{r=1}^3 \frac{h_r^{-1}}{2}  \left[  q_\parallel|_{K+e_j+e_r}^{m+1} \, b_r|_{K+e_j+e_r}^{m+1} -  q_\parallel|_{K+e_j-e_r}^{m+1} \, b_r|_{K+e_j-e_r}^{m+1} \right] \right\} \delta_{ij} ,  
\label{AP1_fq}
\end{eqnarray}
with 
\begin{eqnarray}
& & \hspace{-1cm} \tilde F_{q_ij}|_{K+e_j/2}^{m} = \frac{\tau^{-1}}{2} \left[ P|_K^{m} + P|_{K+e_j}^{m} \right] \delta_{ij} + F_{q_ij}|_{K+e_j/2}^{m} , \quad i,j = 1,2,3.   
\label{AP1_def_tilde_F}
\end{eqnarray}
The momentum balance eq. (\ref{gene_AP_q}) with the modified flux (\ref{AP1_fq}) is now written: 
\begin{eqnarray}
& & \hspace{-1cm} \delta^{-1} (q_i|_K^{m+1} - q_i|_K^m) 
+ \tilde \Delta_{q_i}|_K^m \nonumber \\
& & \hspace{-0.5cm} - \frac{\delta \tau^{-1} h_i^{-1}}{4} \, \left\{ p'(n|_{K+e_i}^{m}) \, \sum_{j=1}^3 h_j^{-1} \left[  q_\parallel|_{K+e_i+e_j}^{m+1} \, b_j|_{K+e_i+e_j}^{m+1} -  q_\parallel|_{K+e_i-e_j}^{m+1} \, b_j|_{K+e_i-e_j}^{m+1} \right] \right.  \nonumber \\
& & \hspace{-0.cm} 
\left. - p'(n|_{K-e_i}^{m}) \, \sum_{j=1}^3 h_j^{-1} \left[  q_\parallel|_{K-e_i+e_j}^{m+1} \, b_j|_{K-e_i+e_j}^{m+1} -  q_\parallel|_{K-e_i-e_j}^{m+1} \, b_j|_{K-e_i-e_j}^{m+1} \right] \right\}  =  \nonumber \\
& & \hspace{2cm} = \tau^{-1} 
( n|_K^{m+1} E_i|_K^{m+1} + (q|_K^{m+1} \times B|_K^{m+1})_i ) , \quad i = 1,2,3,
\label{AP1_q_2}
\end{eqnarray}
with 
\begin{eqnarray}
& & \hspace{-1cm} \tilde \Delta_{q_i}|_K^m = \sum_{j=1}^3   h_j^{-1}  \left[ \tilde F_{q_ij}|_{K+e_j/2}^{m} - \tilde F_{q_ij}|_{K-e_j/2}^{m} \right]
 , \quad i = 1,2,3. 
\label{AP1_tilde_Delta_qi}
\end{eqnarray}

Thus, we are led to the following definition:

\begin{definition}
The first Fully-Discrete AP scheme (or FDAP-1 scheme) is given by 
\begin{eqnarray}
& & \hspace{-1cm} \delta^{-1} (n|_K^{m+1} - n|_K^m) + 
\sum_{j=1}^3 \frac{h_j^{-1}}{2}  \left[  q_\parallel|_{K+e_j}^{m+1} \, b_j|_{K+e_j}^{m+1} -  q_\parallel|_{K-e_j}^{m+1} \, b_j|_{K-e_j}^{m+1} \right] + \tilde \Delta_n|_K^m = 0, 
\label{AP1_n_3} \\
& & \hspace{-1cm} \delta^{-1} (q_i|_K^{m+1} - q_i|_K^m) 
+ \tilde \Delta_{q_i}|_K^m \nonumber \\
& & \hspace{-0.5cm} - \frac{\delta \tau^{-1} h_i^{-1}}{4} \, \left\{ p'(n|_{K+e_i}^{m}) \, \sum_{j=1}^3 h_j^{-1} \left[  q_\parallel|_{K+e_i+e_j}^{m+1} \, b_j|_{K+e_i+e_j}^{m+1} -  q_\parallel|_{K+e_i-e_j}^{m+1} \, b_j|_{K+e_i-e_j}^{m+1} \right] \right.  \nonumber \\
& & \hspace{-0.cm} 
\left. - p'(n|_{K-e_i}^{m}) \, \sum_{j=1}^3 h_j^{-1} \left[  q_\parallel|_{K-e_i+e_j}^{m+1} \, b_j|_{K-e_i+e_j}^{m+1} -  q_\parallel|_{K-e_i-e_j}^{m+1} \, b_j|_{K-e_i-e_j}^{m+1} \right] \right\}  =  \nonumber \\
& & \hspace{2cm} = \tau^{-1} 
( n|_K^{m+1} E_i|_K^{m+1} + (q|_K^{m+1} \times B|_K^{m+1})_i ) , \quad i = 1,2,3,
\label{AP1_q_3}
\end{eqnarray}
with $\tilde \Delta_n|_K^m$ given by (\ref{AP1_tilde_Delta_n}), (\ref{AP1_def_tilde_D}), (\ref{gene_cons_law_num_visc}) on the one hand and $\tilde \Delta_{q_i}|_K^m$ by (\ref{AP1_tilde_Delta_qi}), (\ref{AP1_def_tilde_F}), (\ref{AP1_def_Pkm}), (\ref{gene_AP_Fdef}), (\ref{gene_cons_law_num_visc}) on the other hand. 
\label{def_FDAP1}
\end{definition}

\medskip
\noindent
The quantities $\tilde \Delta_n|_K^m$ and $\tilde \Delta_{q_i}|_K^m$ only depend on known quantities at the previous time steps (with the exception of the magnetic field which is supposed to be know but taken at the current time step). $\tilde \Delta_n|_K^m$ collects a consistent approximation of $\nabla_\bot \cdot q_\bot$ and the contribution of the numerical viscosity (see  (\ref{AP1_def_tilde_D})). The quantity $\tilde \Delta_{q_i}|_K^m$ collects a consistent approximation of $\nabla \cdot (n^{-1} q \otimes q) + \tau^{-1} \nabla p(n)$ and the contribution of the numerical viscosity (see (\ref{AP1_def_tilde_F}) and (\ref{gene_AP_Fdef})). 

We now show that this scheme can be recast into a consistent approximation of the RIEL-1 reformulation and that it is actually AP.

\begin{proposition}
The FDAP-1 scheme can be equivalently formulated: 
\begin{eqnarray}
& & \hspace{-1cm} \delta^{-1} (n|_K^{m+1} - n|_K^m) + 
\sum_{j=1}^3 \frac{h_j^{-1}}{2}  \left[  q_\parallel|_{K+e_j}^{m+1} \, b_j|_{K+e_j}^{m+1} -  q_\parallel|_{K-e_j}^{m+1} \, b_j|_{K-e_j}^{m+1} \right] + \tilde \Delta_n|_K^m = 0, 
\label{AP1_n_8} \\
& & \hspace{-1cm} \tau q_\parallel|_K^{m+1}  - \sum_{i,j = 1}^3 \frac{\delta^2  h_i^{-1} h_j^{-1}}{4} \, \left\{ p'(n|_{K+e_i}^{m}) \,  b_i|_K^{m+1}\, \left[  q_\parallel|_{K+e_i+e_j}^{m+1} \, b_j|_{K+e_i+e_j}^{m+1} -  q_\parallel|_{K+e_i-e_j}^{m+1} \, b_j|_{K+e_i-e_j}^{m+1} \right] \right.  \nonumber \\
& & \hspace{1cm} 
\left. - p'(n|_{K-e_i}^{m}) \,  b_i|_K^{m+1} \, \left[  q_\parallel|_{K-e_i+e_j}^{m+1} \, b_j|_{K-e_i+e_j}^{m+1} -  q_\parallel|_{K-e_i-e_j}^{m+1} \, b_j|_{K-e_i-e_j}^{m+1} \right] \right\}  +  \nonumber \\
& & \hspace{1cm} 
+ \delta^2 E_\parallel|_K^{m+1} \sum_{j=1}^3 \frac{h_j^{-1}}{2}  \left[  q_\parallel|_{K+e_j}^{m+1} \, b_j|_{K+e_j}^{m+1} -  q_\parallel|_{K-e_j}^{m+1} \, b_j|_{K-e_j}^{m+1} \right] = \tilde R|_K^m,
\label{AP1_q_parallel} \\
& & \hspace{-1cm} q_\bot|_K^{m+1}  - \frac{\tau}{\delta \, \left|  B|_K^{m+1}  \right| } \, b|_K^{m+1} \times q_\bot|_K^{m+1} = \nonumber \\
& & \hspace{3cm} = \frac{b|_K^{m+1}}{\left|  B|_K^{m+1}  \right| } \times \left \{ G|_K^m - n|_K^{m+1} E|_K^{m+1} - \tau \delta^{-1} q|_K^m \right\}. 
\label{AP1_q_perp} 
\end{eqnarray}
with 
\begin{eqnarray}
& & \hspace{-1cm} \tilde R|_K^m = R|_K^m + \delta E_\parallel|_K^{m+1} (n|_K^{m} - \delta \tilde \Delta_n|_K^m ) , \quad E_\parallel|_K^{m+1} = \sum_{i=1}^3 b_i|_K^{m+1} E_i|_K^{m+1} . 
\label{AP1_tilde_RKm} \\
& & \hspace{-1cm} R|_K^m = \tau \sum_{i=1}^3 b_i|_K^{m+1} \, q_i|_K^m 
- \tau \delta \sum_{i=1}^3 b_i|_K^{m+1} \tilde \Delta_{q_i}|_K^m .
\label{AP1_RKm} \\
& & \hspace{-1cm} \tau^{-1} G_i|_K^m = \tilde \Delta_{q_i}|_K^m \nonumber \\
& & \hspace{-0.5cm} - \frac{\delta \tau^{-1} h_i^{-1}}{4} \, \left\{ p'(n|_{K+e_i}^{m}) \, \sum_{j=1}^3 h_j^{-1} \left[  q_\parallel|_{K+e_i+e_j}^{m+1} \, b_j|_{K+e_i+e_j}^{m+1} -  q_\parallel|_{K+e_i-e_j}^{m+1} \, b_j|_{K+e_i-e_j}^{m+1} \right] \right.  \nonumber \\
& & \hspace{-0.cm} 
\left. - p'(n|_{K-e_i}^{m}) \, \sum_{j=1}^3 h_j^{-1} \left[  q_\parallel|_{K-e_i+e_j}^{m+1} \, b_j|_{K-e_i+e_j}^{m+1} -  q_\parallel|_{K-e_i-e_j}^{m+1} \, b_j|_{K-e_i-e_j}^{m+1} \right] \right\}  .
\label{AP1_def_G} 
\end{eqnarray}
This scheme is consistent with the RIEL-1 formulation of the Euler-Lorentz model and is a full discretization of the SDAP-1 time semi-discrete scheme. 
\label{prop-FDAP1_ref}
\end{proposition}

\noindent
{\bf Proof:} We first consider the parallel component of the momentum and take the dot product of (\ref{AP1_q_3}) by $b|_K^{m+1}$ (i.e. we multiply (\ref{AP1_q_3}) by $b_i|_K^{m+1}$ and sum over $i$). We multiply by $\tau \delta$ and we find: 
\begin{eqnarray}
& & \hspace{-1cm} \tau q_\parallel|_K^{m+1}  - \sum_{i,j = 1}^3 \frac{\delta^2  h_i^{-1} h_j^{-1}}{4} \, \left\{ p'(n|_{K+e_i}^{m}) \,  b_i|_K^{m+1}\, \left[  q_\parallel|_{K+e_i+e_j}^{m+1} \, b_j|_{K+e_i+e_j}^{m+1} -  q_\parallel|_{K+e_i-e_j}^{m+1} \, b_j|_{K+e_i-e_j}^{m+1} \right] \right.  \nonumber \\
& & \hspace{1cm} 
\left. - p'(n|_{K-e_i}^{m}) \,  b_i|_K^{m+1} \, \left[  q_\parallel|_{K-e_i+e_j}^{m+1} \, b_j|_{K-e_i+e_j}^{m+1} -  q_\parallel|_{K-e_i-e_j}^{m+1} \, b_j|_{K-e_i-e_j}^{m+1} \right] \right\}  =  \nonumber \\
& & \hspace{7cm} 
= \delta \sum_{i=1}^3 n|_K^{m+1} b_i|_K^{m+1} E_i|_K^{m+1} + R|_K^m,
\label{AP1_q_parallel_0}
\end{eqnarray}
with $R|_K^m$ given by (\ref{AP1_RKm}). Inserting (\ref{AP1_n_3}) into (\ref{AP1_q_parallel_0}) allows us to replace $n|_K^{m+1}$ by an expression involving $q_\parallel|_K^{m+1}$ and known values at time $n$. This leads to (\ref{AP1_q_parallel}). We now consider the transverse component of the momentum. Taking the vector product of (\ref{AP1_q_3}) with $b|_K^{m+1}$ and dividing by $\left|  B|_K^{m+1}  \right|$, we find (\ref{AP1_q_perp}). 

We now show that the FDAP-1 scheme is consistent with the SDAP-1 time semi-discrete scheme (and consequently, with the RIEL-1 reformulation of the Euler-Lorentz model). Eq. (\ref{AP1_q_parallel}) is a discrete version of the elliptic equation: 
\begin{eqnarray}
& & \hspace{-1cm} \tau q_\parallel  - \delta^2 \, b \cdot  (\nabla (p'(n) \nabla \cdot (q_\parallel b)) 
- E \nabla \cdot (q_\parallel b)) = \tilde R.  \label{AP1_q_parallel_consistent}
\end{eqnarray}
Developing the term $\tilde R|_K^m$ using  (\ref{AP1_tilde_Delta_qi}), (\ref{AP1_def_tilde_F}), (\ref{AP1_def_Pkm}), (\ref{gene_AP_Fdef}), we find: 
\begin{eqnarray}
& & \hspace{-1cm} \tilde  R|_K^m = - \frac{\delta}{2} \sum_{i=1}^3 b_i|_K^{m+1}  \left[ P|_{K+e_i}^{m}  - P|_{K-e_i}^{m} \right] + \delta E_\parallel|_K^{m+1} (n|_K^{m} - \delta \tilde \Delta_n|_K^m )  \nonumber \\
& & \hspace{0.2cm} + \tau \left\{ - \frac{\delta}{2}  \sum_{i=1}^3 b_i|_K^{m+1} \sum_{j=1}^3 h_j^{-1}  \left[  (n^{-1} q_i q_j)|_{K+e_j}^m -  (n^{-1} q_i q_j)|_{K-e_j}^m \right] \right.  \nonumber \\
& & \hspace{0.2cm} \left.   - \frac{\delta}{2} \sum_{i=1}^3 b_i|_K^{m+1} \, \sum_{j=1}^3 h_j^{-1} \left[ D_{q_ij}|_{K+e_j/2}^{m} - D_{q_ij}|_{K-e_j/2}^{m} \right] + \sum_{i=1}^3 b_i|_K^{m+1} \, q_i|_K^m  \right\} . \label{AP1_RKm_2} 
\end{eqnarray}
We easily convince ourselves that $\delta^{-1} R|_K^m $ is a consistent approximation of the right-hand side of (\ref{TSDEL_u_par_5}), and consequently, that (\ref{AP1_q_parallel}) is consistent with (\ref{TSDEL_u_par_5}). Eq. (\ref{AP1_n_8}) is clearly a consistent approximation of (\ref{SD1_n}). Finally, $G|_K^m$ appears as a consistent approximation of $\nabla p(n) + \tau \nabla \cdot ( n^{-1} q \otimes q)$ and therefore, (\ref{AP1_q_perp}) is a consistent approximation of  (\ref{SD1_u_perp}). This ends the proof. \endproof

\medskip
\noindent
Supplemented with the Neumann boundary conditions (\ref{IEL_BC_u_par_pm}), the elliptic equation (\ref{AP1_q_parallel}) is well-posed. Its inversion allows us to compute the values of $q_\parallel|_K^{m+1}$ from the right-hand side $\tilde R|_K^m$ which only involves known values. From the knowledge of $q_\parallel|_K^{m+1}$, we can use (\ref{AP1_n_8}) to find the new values $n|_K^{m+1}$ of the density. Finally, the quantities lying at the right-hand side of (\ref{AP1_q_perp}) are known at this level of the recursion. Therefore, $q_\bot|_K^{m+1}$ can be computed thanks to the formula: 
\begin{eqnarray}
& & \hspace{-1cm}  q_\bot|_K^{m+1} = \frac{\delta \, \left|  B|_K^{m+1}  \right|}{\tau^2 + \delta^2 \, \left|  B|_K^{m+1}  \right|^2} \left(- \tau (\mbox{Id} - b \otimes b)|_K^{m+1} +  \right. 
\nonumber \\
& & \hspace{2cm} \left. + \delta \, \left|  B|_K^{m+1}  \right| \,  b|_K^{m+1} \times  \right) \,  \left\{ G|_K^m - n|_K^{m+1} E|_K^{m+1} - \tau \delta^{-1} q|_K^m \right\},  
\label{AP1_q_perp_2} 
\end{eqnarray}
where the right-hand side is known. 

\medskip
We now investigate the limit $\tau \to 0$.

\begin{proposition}
The limit $\tau \to 0$ of the FDAP-1 scheme is the following First Fully-Discrete Drift-Fluid scheme (FDDF-1 scheme): 
\begin{eqnarray}
& & \hspace{-1cm} \delta^{-1} (n|_K^{m+1} - n|_K^m) + 
\sum_{j=1}^3 \frac{h_j^{-1}}{2}  \left[  q_\parallel|_{K+e_j}^{m+1} \, b_j|_{K+e_j}^{m+1} -  q_\parallel|_{K-e_j}^{m+1} \, b_j|_{K-e_j}^{m+1} \right] + \tilde \Delta_n|_K^m = 0, 
\label{AP1_n_8_tau=0} \\
& & \hspace{-1cm} - \sum_{i,j = 1}^3 \frac{\delta^2  h_i^{-1} h_j^{-1}}{4} \, \left\{ p'(n|_{K+e_i}^{m}) \,  b_i|_K^{m+1}\, \left[  q_\parallel|_{K+e_i+e_j}^{m+1} \, b_j|_{K+e_i+e_j}^{m+1} -  q_\parallel|_{K+e_i-e_j}^{m+1} \, b_j|_{K+e_i-e_j}^{m+1} \right] \right.  \nonumber \\
& & \hspace{1cm} 
\left. - p'(n|_{K-e_i}^{m}) \,  b_i|_K^{m+1} \, \left[  q_\parallel|_{K-e_i+e_j}^{m+1} \, b_j|_{K-e_i+e_j}^{m+1} -  q_\parallel|_{K-e_i-e_j}^{m+1} \, b_j|_{K-e_i-e_j}^{m+1} \right] \right\}  +  \nonumber \\
& & \hspace{1cm} 
+ \delta^2 E_\parallel|_K^{m+1} \sum_{j=1}^3 \frac{h_j^{-1}}{2}  \left[  q_\parallel|_{K+e_j}^{m+1} \, b_j|_{K+e_j}^{m+1} -  q_\parallel|_{K-e_j}^{m+1} \, b_j|_{K-e_j}^{m+1} \right] = \nonumber \\
& & \hspace{1cm} = - \frac{\delta}{2} \sum_{i=1}^3 b_i|_K^{m+1}  \left[ P|_{K+e_i}^{m}  - P|_{K-e_i}^{m} \right] + \delta E_\parallel|_K^{m+1} (n|_K^{m} - \delta \tilde \Delta_n|_K^m ) ,
\label{AP1_q_parallel_tau=0} \\
& & \hspace{-1cm} q_\bot|_K^{m+1}   = \frac{b|_K^{m+1}}{\left|  B|_K^{m+1}  \right| } \times \left \{ G^0|_K^m - n|_K^{m+1} E|_K^{m+1} \right\}. 
\label{AP1_q_perp_tau=0} 
\end{eqnarray}
with 
\begin{eqnarray}
& & \hspace{-1cm}  G^0_i|_K^m = \frac{h_i^{-1}}{2}  \left[ P|_{K+e_i}^{m}  - P|_{K-e_i}^{m} \right] \nonumber \\
& & \hspace{-0.5cm} - \frac{\delta h_i^{-1}}{4} \, \left\{ p'(n|_{K+e_i}^{m}) \, \sum_{j=1}^3 h_j^{-1} \left[  q_\parallel|_{K+e_i+e_j}^{m+1} \, b_j|_{K+e_i+e_j}^{m+1} -  q_\parallel|_{K+e_i-e_j}^{m+1} \, b_j|_{K+e_i-e_j}^{m+1} \right] \right.  \nonumber \\
& & \hspace{-0.cm} 
\left. - p'(n|_{K-e_i}^{m}) \, \sum_{j=1}^3 h_j^{-1} \left[  q_\parallel|_{K-e_i+e_j}^{m+1} \, b_j|_{K-e_i+e_j}^{m+1} -  q_\parallel|_{K-e_i-e_j}^{m+1} \, b_j|_{K-e_i-e_j}^{m+1} \right] \right\}  .
\label{AP1_def_G_tau=0} 
\end{eqnarray}
This scheme is consistent with the RIDF-1 formulation of the Drift-Fluid model and is a consistent  space-discretization of the SDDF-1 time semi-discrete scheme. 
\label{prop_DFAP1-FDDF1}
\end{proposition}

\noindent
The last two propositions show that the FDAP-1 scheme is AP. We now consider a scheme based on the second reformulation.

%%%%%%%%%%%%%%%%%%%%%%%%%%%%%%%%%%%%%%%%%%%%%%%%%%%%%%%%%%%%%%%%
\subsection{Fully discrete AP scheme based on the second reformulation}
\label{subsection_EL_fully_ref2}

Following the strategy developed in section \ref{subsection_EL_semidis_ref1}, the second Fully-Discrete AP method (or FDAP-2 scheme) consists in using the modified mass flux (\ref{AP1_fn}) but the unmodified momentum flux (\ref{gene_AP_fq}). Therefore, the FDAP-2 scheme is defined by:

\begin{definition}
The Second Fully-Discrete AP scheme (or FDAP-2 scheme) is given by 
\begin{eqnarray}
& & \hspace{-1cm} \delta^{-1} (n|_K^{m+1} - n|_K^m) + 
\sum_{j=1}^3 \frac{h_j^{-1}}{2}  \left[  q_\parallel|_{K+e_j}^{m+1} \, b_j|_{K+e_j}^{m+1} -  q_\parallel|_{K-e_j}^{m+1} \, b_j|_{K-e_j}^{m+1} \right] + \tilde \Delta_n|_K^m = 0, 
\label{AP2_n_1} \\
& & \hspace{-1cm}\delta^{-1} (q_i|_K^{m+1} - q_i|_K^m) + 
\frac{\tau^{-1} h_i^{-1}}{2}  \left[ p(n|_{K+e_i}^{m+1}) - p(n|_{K-e_i}^{m+1}) \right] + \Delta_{q_i}|_K^m =  \nonumber \\
& & \hspace{1cm} = \tau^{-1} 
( n|_K^{m+1} E_i|_K^{m+1} + (q|_K^{m+1} \times B|_K^{m+1})_i ) , \quad i = 1,2,3,
\label{AP2_q_1}
\end{eqnarray}
with $\tilde \Delta_n|_K^m$ given by (\ref{AP1_tilde_Delta_n}), (\ref{AP1_def_tilde_D}), (\ref{gene_cons_law_num_visc}) on the one hand and $\Delta_{q_i}|_K^m$ is given by (\ref{gene_AP_Delta_qi}), (\ref{gene_AP_Fdef}) on the other hand and we recall that the definition of the numerical viscosities is given by (\ref{gene_cons_law_num_visc}).  
\label{def_FDAP2}
\end{definition}

\medskip
\noindent
The quantities $\tilde \Delta_n|_K^m$ and $ \Delta_{q_i}|_K^m$ only depend on known quantities at the previous time steps. $\tilde \Delta_n|_K^m$ collects a consistent approximation of $\nabla_\bot \cdot q_\bot$ and the contribution of the numerical viscosity. The quantity $\Delta_{q_i}|_K^m$ collects a consistent approximation of $\nabla \cdot (n^{-1} q \otimes q)$ and the contribution of the numerical viscosity. We now show that this scheme can be recast into a consistent approximation of the RIEL-1 reformulation and that it is actually AP.

\begin{proposition}
The FDAP-2 scheme can be equivalently formulated: 
\begin{eqnarray}
& & \hspace{-1cm} 
-  \delta \sum_{i,j=1}^3 \frac{h_j^{-1}}{2} \left\{ b_j|_{K+e_j}^{m+1} \, b_i|_{K+e_j}^{m+1} \left( \frac{ h_i^{-1}}{2} \left[ p(n|_{K+e_j+e_i}^{m+1}) - p(n|_{K+e_j-e_i}^{m+1}) \right] - n|_{K+e_j}^{m+1} E_i|_{K+e_j}^{m+1} \right) \right.  \nonumber \\
& & \hspace{1cm} 
\left. - b_j|_{K-e_j}^{m+1} \,  b_i|_{K-e_j}^{m+1} \left( \frac{ h_i^{-1}}{2} \left[ p(n|_{K-e_j+e_i}^{m+1}) - p(n|_{K-e_j-e_i}^{m+1}) \right] - n|_{K-e_j}^{m+1} E_i|_{K-e_j}^{m+1} \right) \right\}
 \nonumber \\
& & \hspace{8cm} 
+ \tau \delta^{-1} n|_K^{m+1} + \tau \tilde {\tilde \Delta}_n|_K^m  = 0, 
\label{AP2_n_2} \\
& & \hspace{-1cm}  q_\parallel|_K^{m+1} = 
- \delta \tau^{-1} \sum_{i=1}^3  b_i|_K^{m+1} \left( \frac{h_i^{-1}}{2} \left[ p(n|_{K+e_i}^{m+1}) - p(n|_{K-e_i}^{m+1}) \right] - n|_K^{m+1} E_i|_K^{m+1} \right) + Q|_K^m ,\nonumber \\
& & \hspace{-1cm} 
\label{AP2_q_2} \\
& & \hspace{-1cm} q_\bot|_K^{m+1}  - \frac{\tau}{\delta \, \left|  B|_K^{m+1}  \right| } \, b|_K^{m+1} \times q_\bot|_K^{m+1} = \nonumber \\
& & \hspace{3cm} = \frac{b|_K^{m+1}}{\left|  B|_K^{m+1}  \right| } \times \left \{ \tilde G|_K^m - n|_K^{m+1} E|_K^{m+1} - \tau \delta^{-1} q|_K^m \right\}. 
\label{AP2_q_perp} 
\end{eqnarray}
with 
\begin{eqnarray}
& & \hspace{-1cm} \tilde {\tilde \Delta}_n|_K^m = \tilde \Delta_n|_K^m - \delta^{-1} n|_K^m +  
\sum_{j=1}^3 \frac{h_j^{-1}}{2}  \left[  Q|_{K+e_j}^{m} b_j|_{K+e_j}^{m+1} -  Q|_{K-e_j}^{m} b_j|_{K-e_j}^{m+1} \right].  
\label{AP2_def_delta_tilde_tilde} \\
& & \hspace{-1cm}  Q|_K^m = \sum_{i=1}^3 b_i|_K^{m+1} \left( q_i|_K^m - \delta \Delta_{q_i}|_K^m  \right) . 
\label{AP2_def_Q} \\
& & \hspace{-1cm} \tau^{-1} \tilde G_i|_K^m = \frac{\tau^{-1} h_i^{-1}}{2}  \left[ p(n|_{K+e_i}^{m+1}) - p(n|_{K-e_i}^{m+1}) \right] + \Delta_{q_i}|_K^m .
\label{AP1_def_tilde_G} 
\end{eqnarray}
This scheme is consistent with the RIEL-2 formulation of the Euler-Lorentz model and is a full discretization of the SDAP-2 time semi-discrete scheme. 
\label{prop-FDAP2_ref}
\end{proposition}

\noindent
{\bf Proof:}  Taking the parallel component of (\ref{AP2_q_1}), we get (\ref{AP2_q_2}). Inserting it into (\ref{AP2_n_1}) leads to (\ref{AP2_n_2}). We now consider the transverse momentum. Comparing (\ref{AP2_q_1}) with (\ref{AP1_q_3}) and having the expression (\ref{AP1_def_G}) of $G_i|_K^m$ in mind, we notice that relation (\ref{AP1_q_perp}), which was derived in the case of the FDAP-1 scheme, applies to the FDAP-2 scheme as well with $G_i|_K^m$ simply replaced by $\tilde G_i|_K^m$. This leads to (\ref{AP2_q_perp}). 

We now show that the FDAP-2 scheme is consistent with the SDAP-2 time semi-discrete scheme (and consequently, with the RIEL-2 reformulation of the Euler-Lorentz model). Eq. (\ref{AP2_n_2}) is a consistent approximation of 
$$- \delta \nabla \cdot ((b \otimes b) ( \nabla p - nE)) + \tau \delta^{-1} n + \tau \tilde {\tilde \Delta} = 0. $$
Furthermore, a more careful inspection of $\tilde {\tilde \Delta}_n|_K^m$ easily shows that it is a consistent approximation of the right-hand side of (\ref{SD2L_n}). Therefore, Eq. (\ref{AP2_n_2}) itself is consistent with (\ref{SD2L_n}). It is then clear that (\ref{AP2_q_2}) is consistent with (\ref{SD2_u_par}). Then, the quantity $\tilde G_i|_K^m $ is a consistent approximation of $\nabla p(n) + \tau \nabla \cdot ( n^{-1} q \otimes q)$ and therefore, (\ref{AP2_q_perp}) is consistent with (\ref{SD2_u_perp}). \endproof

\noindent
Eq. (\ref{AP2_n_2}) is a discrete elliptic equation. Supplemented with the Neumann boundary conditions (\ref{IEL_BC_n_pm}), it is well-posed. Its inversion allows us to compute $n|_K^{m+1}$ from the quantity $\tilde {\tilde \Delta}_n|_K^m$ which only involves known values. Once $n|_K^{m+1}$ has been found by the inversion of (\ref{AP2_n_2}), the parallel momentum eq. (\ref{AP2_q_2}) can be used to find $q_\parallel|_K^{m+1}$. Finally, the quantity $\tilde G_i|_K^m $ is known from the previous steps of the recursion. Eq. (\ref{AP1_q_perp_2}) applies with  $ G_i|_K^m$ replaced by $\tilde G_i|_K^m$ and determines $q_\bot|_K^{m+1}$ from the known values appearing at its right-hand side. 

\medskip
We now investigate the limit $\tau \to 0$. 

\begin{proposition}
Taking the limit $\tau \to 0$ in the FDAP-2 scheme, expanding $n^\tau|_{K}^{m+1}  = n^0|_{K}^{m+1} + \tau n^1|_{K}^{m+1} + o(\tau)$ (where the dependence of the solution upon $\tau$ has been restored) and using the boundary condition (\ref{IEL_BC_n_pm}) leads to the following Second Fully-Discrete Drift-Fluid scheme or FDDF-2 scheme:  
\begin{eqnarray}
& & \hspace{-1cm} 
\sum_{i}^3 \, b_i|_{K}^{m+1} \left( \frac{ h_i^{-1}}{2} \left[ p(n^0|_{K+e_i}^{m+1}) - p(n^0|_{K-e_i}^{m+1}) \right] - n^0|_{K}^{m+1} E_i|_{K}^{m+1} \right) = 0  . 
\label{AP2_n0_tau=0_2} \\
& & \hspace{-1cm} 
-  \delta \sum_{i,j=1}^3 \frac{h_j^{-1}}{2} \left\{ b_j|_{K+e_j}^{m+1} \, b_i|_{K+e_j}^{m+1} \left( \frac{ h_i^{-1}}{2} \left[ p'(n^0|_{K+e_j+e_i}^{m+1}) n^1|_{K+e_j+e_i}^{m+1} - p'(n^0|_{K+e_j-e_i}^{m+1} ) n^1|_{K+e_j-e_i}^{m+1} \right]  \right. \right. \nonumber \\
& & \hspace{11cm} \left. \left. - n^1|_{K+e_j}^{m+1} E_i|_{K+e_j}^{m+1} \right) \right.  \nonumber \\
& & \hspace{1cm} 
\left. - b_j|_{K-e_j}^{m+1} \,  b_i|_{K-e_j}^{m+1} \left( \frac{ h_i^{-1}}{2} \left[ p'(n^0|_{K-e_j+e_i}^{m+1}) n^1|_{K-e_j+e_i}^{m+1} - p'(n^0|_{K-e_j-e_i}^{m+1}) n^1|_{K-e_j-e_i}^{m+1}\right]
\right. \right. \nonumber \\
& & \hspace{11cm} \left. \left.  - n|_{K-e_j}^{m+1} E_i|_{K-e_j}^{m+1} \right) \right\}
 \nonumber \\
& & \hspace{8cm} 
+\delta^{-1} n^0|_K^{m+1} + \tilde {\tilde \Delta}_n|_K^m  = 0. 
\label{AP2_n1_tau=0} \\
& & \hspace{-1cm}  q_\parallel|_K^{m+1} = 
- \delta \sum_{i=1}^3  b_i|_K^{m+1} \left( \frac{h_i^{-1}}{2} \left[ p'(n^0|_{K+e_i}^{m+1}) n^1|_{K+e_i}^{m+1} - p'(n^0|_{K-e_i}^{m+1}) n^1|_{K-e_i}^{m+1} \right] \right. \nonumber \\
& & \hspace{8cm} \left. - n^1|_K^{m+1} E_i|_K^{m+1} \right) + Q|_K^m , 
\label{AP2_q_parallel_tau=0} \\
& & \hspace{-1cm} q_\bot|_K^{m+1}  = \frac{b|_K^{m+1}}{\left|  B|_K^{m+1}  \right| } \times \left \{ \frac{ h_i^{-1}}{2}  \left[ p(n|_{K+e_i}^{m+1}) - p(n|_{K-e_i}^{m+1}) \right]  - n|_K^{m+1} E|_K^{m+1} \right\}. 
\label{AP2_q_perp_tau=0} 
\end{eqnarray}
This scheme is consistent with the RIDF-2 formulation of the Drift-Fluid model and is a consistent  space-discretization of the SDDF-2 time semi-discrete scheme. 
\label{prop_FDAP2-FDDF2}
\end{proposition}

\noindent
{\bf Proof: } Expanding (\ref{AP2_n_2}) in powers of $\tau$, we find at the leading order: 
\begin{eqnarray}
& & \hspace{-1cm} 
-  \delta \sum_{i,j=1}^3 \frac{h_j^{-1}}{2} \left\{ b_j|_{K+e_j}^{m+1} \, b_i|_{K+e_j}^{m+1} \left( \frac{ h_i^{-1}}{2} \left[ p(n^0|_{K+e_j+e_i}^{m+1}) - p(n^0|_{K+e_j-e_i}^{m+1}) \right] - n^0|_{K+e_j}^{m+1} E_i|_{K+e_j}^{m+1} \right) \right.  \nonumber \\
& & \hspace{0cm} 
\left. - b_j|_{K-e_j}^{m+1} \,  b_i|_{K-e_j}^{m+1} \left( \frac{ h_i^{-1}}{2} \left[ p(n^0|_{K-e_j+e_i}^{m+1}) - p(n^0|_{K-e_j-e_i}^{m+1}) \right] - n^0|_{K-e_j}^{m+1} E_i|_{K-e_j}^{m+1} \right) \right\} = 0, \nonumber \\
& & \hspace{0cm} 
\label{AP2_n0_tau=0} 
\end{eqnarray}
and at the next order, (\ref{AP2_n1_tau=0}). With a discretized version of the homogeneous Neumann boundary conditions (\ref{IEL_BC_n_pm}), eq. (\ref{AP2_n0_tau=0}) can be integrated once and leads to (\ref{AP2_n0_tau=0_2}). Taking the $\tau \to 0$ limit in (\ref{AP2_q_2}) leads to (\ref{AP2_q_parallel_tau=0}), because the leading order (in $\tau^{-1}$) term cancels due to (\ref{AP2_n0_tau=0_2}). Finally taking the $\tau \to 0$ limit in (\ref{AP2_q_perp}) leads to (\ref{AP2_q_perp_tau=0}). This scheme is obviously consistent with the SDAP-2 scheme and consequently, with the RIDF-2 formulation of the Drift-Fluid model. \endproof

\noindent
The last two propositions show that the FDAP-2 scheme is AP.

%%%%%%%%%%%%%%%%%%%%%%%%%%%%%%%%%%%%%%%%%%%%%%%%%%%%%%%%%%%%%%%%
\subsection{Fully discrete AP scheme for the Euler-Lorentz model: conclusion}
\label{subsection_EL_fully_conclu}

In this section, we have derived two fully-discrete schemes for the Euler-Lorentz model, which are consistent with the Drift-Fluid limit when the parameter $\tau$ (representing the scaled cyclotron period and Mach number) tends to zero. They have been shown to be respectively discretizations of the two reformulations of the Euler-Lorentz model, the RIDf-1 and RIFD-2 models. These schemes are based on standard explicit shock-capturing schemes where implicit evaluations of the mass and pressure fluxes and of the source terms have been introduced. Then, some simple approximations have been performed. These approximations alter neither the conservative character of the scheme, nor its consistency but give rise to discrete elliptic equations for the parallel momentum and the density respectively. These elliptic equations are strongly anisotropic, with a diffusion which is concentrated along the magnetic field linres. Therefore, they lead  to AP schemes provided that they are uniformly well-posed as the parameter $\tau$ tends to $0$. The uniform invertibility of these elliptic equations as $\tau$ tends to $0$ as well as a practical methodology to solve them in coordinate systems which are independent of the magnetic field lines are investigated in the next section.

%% file: aniso.tex
%%%%%%%%%%%%%%%%%%%%%%%%%%%%%%%%%%%%%%%%%%%%%%%%%%%%%%%%%%%%%%%%%%%%%%%%%%%%%%%%%%%%%%%%%%%%%%%%
%%%%%%%%%%%%%%%%%%%%%%%%%%%%%%%%%%%%%%%%%%%%%%%%%%%%%%%%%%%%%%%%%%%%%%%%%%%%%%%%%%%%%%%%%%%%%%%%
%%%%%%%%%%%%%%%%%%%%%%%%%%%%%%%%%%%%%%%%%%%%%%%%%%%%%%%%%%%%%%%%%%%%%%%%%%%%%%%%%%%%%%%%%%%%%%%%
%%%%%%%%%%%%%%%%%%%%%%%%%%%%%%%%%%%%%%%%%%%%%%%%%%%%%%%%%%%%%%%%%%%%%%%%%%%%%%%%%%%%%%%%%%%%%%%%
\setcounter{equation}{0}
\section{Numerical resolution of strongly anisotropic elliptic problems}
\label{section_EL_strong_aniso}

%%%%%%%%%%%%%%%%%%%%%%%%%%%%%%%%%%%%%%%%%%%%%%%%%%%%%%%%%%%%%%%%%%%%%%%%%%%%%%%%%%%%%%%%%%%%%%%%
%%%%%%%%%%%%%%%%%%%%%%%%%%%%%%%%%%%%%%%%%%%%%%%%%%%%%%%%%%%%%%%%%%%%%%%%%%%%%%%%%%%%%%%%%%%%%%%%
\subsection{Introduction to strongly anisotropic elliptic problems}
\label{subsec_El_stron_aniso_intro}

This section is concerned with the numerical invesion of strongly anisotrpic discrete elliptic problems such as those undelying the FDAP-1 scheme (\ref{AP1_q_parallel}) or the FDAP-2 scheme (\ref{AP2_n_2}). 

More specifically, we are interested in:

\begin{definition}
The continuous anisotropic elliptic (AE) problem is as follows: find $n^\tau(x)$ defined on $\Omega$ as the solution of 
\begin{eqnarray}
& & \hspace{-1cm} \tau  n^\tau - \nabla_\parallel \cdot (\nabla_\parallel n^\tau - n^\tau E_\parallel ) = \tau F, \quad \mbox{ in } \, \Omega, \label{AEP_n} 
\end{eqnarray}
with boundary conditions:
\begin{eqnarray}
& & \hspace{-1cm} (b \cdot \nu) (\nabla_\parallel n^\tau - n^\tau E_\parallel)  = \tau g, \quad \mbox{ on } \, \Gamma_+ \cup \Gamma_-.  \label{AEP_n_bc} 
\end{eqnarray}
Here, the domain geometry is as defined in section \ref{subsection_EL_BC}. The vector field $E(x)$ and the function $F$ are known smooth functions defined on $\Omega$ while $g$ is a known function defined on $\Gamma_+ \cup \Gamma_-$. The parameter $\tau$ is positive. 
\label{def_AE_problem}
\end{definition}

\medskip
\noindent
This problem is the generic one which needs to be solved when using the SDAP-2 scheme, specifically when inverting (\ref{SD2L_n}), with boundary conditions (\ref{IEL_BC_n_pm}) up to obvious notational changes. 
Indeed, replacing $\delta^2 \tau$ by $\tau$ at the left-hand side of (\ref{SD2L_n}), and lumping all the terms at the right-hand side of (\ref{SD2L_n}) into F, we find (\ref{AEP_n}). We also have to assume an isothermal pressure relation $p(n) = n$ in order to get the linear AE problem. A nonlinear pressure relation would give rise to a nonlinear anisotropic elliptic problem:
\begin{eqnarray}
& & \hspace{-1cm} \tau  n^\tau - \nabla_\parallel \cdot (\nabla_\parallel p(n^\tau) - n^\tau E_\parallel ) = \tau F, \quad \mbox{ in } \, \Omega, \label{AEP_n_NL} \\
& & \hspace{-1cm} (b \cdot \nu) (\nabla_\parallel p(n^\tau) - n^\tau E_\parallel)  = \tau g, \quad \mbox{ on } \, \Gamma_+ \cup \Gamma_-.  \label{AEP_n_bc_NL} 
\end{eqnarray}
The resolution of this nonlinear problem by means of Newton's method would lead to solving a sequence of linear problems of the AE type. It may be desirable to design more elaborate methods in the small $\tau$ case but these developments will be the subject of future work. In these notes, we will restrict ourselves to the linear AE problem. 

The AE problem enters the class of strongly anisotropic problems because it can be recast in the form
\begin{eqnarray}
& & \hspace{-1cm} \tau  n^\tau - \nabla \cdot ( ( b \otimes b ) (\nabla n^\tau - n^\tau E ) )  = \tau F, \quad \mbox{ in } \, \Omega, \label{AEP_n_matrix} \\
& & \hspace{-1cm} (b \cdot \nu) b \cdot (\nabla n^\tau - n^\tau E)  = \tau g, \quad \mbox{ on } \, \Gamma_+ \cup \Gamma_-.  \label{AEP_n_bc_matrix} 
\end{eqnarray}
Thus, it appears as an elliptic problem with a diffusion matrix equal to $b \otimes b$. This matrix is a rank one matrix, and is therefore singular on a two-dimensional manifold (the orthogonal plane to $b$). This is highly degenerate elliptic problem.

The AE problem is also nothing but a one-dimensional elliptic problem posed along the field lines. However, we aim at designing methods which do not require the computation of these field lines, nor any integration along them. The difficulty with the AE problem is that it becomes singular when $\tau \to 0$. Indeed, letting $\tau \to 0$ in the AE problem leads to $\nabla_\parallel n^0 - n^0 E_\parallel = 0$. Supposing $E=0$ just for the sake of simplicity, this condition becomes $\nabla_\parallel n^0= 0$ and means that $n^0$ is constant along the magnetic field lines. However, this constant is undetermined by the leading order equations in the AE problem. To find the value of this constant, it is necessary to expand $n^\tau$ in powers of $\tau$: $n^\tau = n^0 + \tau n^1 = o(\tau)$ and to look for the existence condition for the first order perturbation $n^1$. From the numerical viewpoint, any standard discretization of the AE problem (like finite-difference or finite-element methods) will lead to inverting a matrix with condition number of the order $O(\tau^{-1})$, and is therefore unpractical if $\tau \ll 1$. The goal of the following discussion is to propose inversion methods with uniformly bounded condition number with respect to $\tau$ when $\tau \to 0$. We will again call these methods 'Asymptotic-Preserving' (AP) methods for highly anisotropic elliptic problems. 

These methods can be extended when the elliptic operator has a transverse part of order $O(1)$ (when $\tau \to 0$). Such problems are encountered for instance in ionosphere models (see \cite{}) and the presented material has been the subject of a \cite{}. The remainder of this section is organized as follows: we will first discuss a simple one-dimensional framework in the continuous case. Then, we extend it to the general three-dimensional framework. The, we discuss the discrete case, specifically aiming at solving the set of equations resulting to the application of the FDAP-2 scheme for the Euler-Lorentz model. Again, we start with a toy one-dimensional problem and then, in the last section, we discuss the fully discrete Three dimensional problem arising from the FDAP-2 scheme.

%%%%%%%%%%%%%%%%%%%%%%%%%%%%%%%%%%%%%%%%%%%%%%%%%%%%%%%%%%%%%%%%%%%%%%%%%%%%%%%%%%%%%%%%%%%%%%%%
%%%%%%%%%%%%%%%%%%%%%%%%%%%%%%%%%%%%%%%%%%%%%%%%%%%%%%%%%%%%%%%%%%%%%%%%%%%%%%%%%%%%%%%%%%%%%%%%
\subsection{A simple one-dimensional "anisotropic" elliptic problem: the continuous case}
\label{subsec_El_stron_aniso_1D_continuous}

This section is intended for introductory and illustration purposes. Basically, this one-dimensional example is illustrating what happens along each magnetic field line. The 1D-AE problem is defined as follows: 

\begin{definition}
The one-dimensional continuous anisotropic elliptic (1D-AE) problem is as follows: find $n^\tau(x)$ defined on $[0,1]$ as the solution of 
\begin{eqnarray}
& & \hspace{-1cm} \tau  n^\tau - \frac{d}{dx} \left( \frac{d}{dx} n^\tau - n^\tau E \right) = \tau F, \quad \mbox{ in } \, [0,1], \label{AEP_n_1D} 
\end{eqnarray}
with boundary conditions:
\begin{eqnarray}
& & \hspace{-1cm} \frac{d}{dx} n^\tau - n^\tau E  = \tau g, \quad \mbox{ at } \, x=0 \mbox{ and } x=1.  \label{AEP_n_bc_1D} 
\end{eqnarray}
The vector field $E(x)$ and the function $F$ are known smooth functions defined on $[0,1]$ while $g$ is a known function defined only at $x=0$ and $x=1$. The parameter $\tau$ is positive. 
\label{def_1DAE_problem}
\end{definition}

\medskip
\noindent
The first order derivative term can be removed by means of a simple transformation. Introducing the electric potential $\phi (x)$ such that 
\begin{eqnarray}
& & \hspace{-1cm} \phi (x) = - \int_0^x E(y) \, dy, 
\label{AEP_phi_1D} 
\end{eqnarray}
we remark that 
\begin{eqnarray}
& & \hspace{-1cm} \frac{d}{dx} n^\tau - n^\tau E = e^{- \phi} \frac{d u^\tau}{dx} 
, \quad u^\tau = e^\phi n^\tau.  
\label{AEP_phi_change} 
\end{eqnarray}
Then, the 1DAE problem is written as follows: 

\begin{proposition}
With $u^\tau$ defined by (\ref{AEP_phi_change}), problem 1DAE is equivalent to the following one-dimensional modified anisotropic elliptic (1D-MAE) problem:
\begin{eqnarray}
& & \hspace{-1cm} \tau  e^{-\phi} u^\tau - \frac{d}{dx} \left( e^{-\phi} \frac{d u^\tau}{dx} \right) = \tau F, \quad \mbox{ in } \, [0,1], \label{MPAEP_n_1D} \\
& & \hspace{-1cm}  e^{-\phi} \frac{d u^\tau}{dx}  = \tau g, \quad \mbox{ at } \, x=0 \mbox{ and } x=1.  \label{MAEP_n_bc_1D} 
\end{eqnarray}
\label{prop_1DMAE_problem_equiv}
\end{proposition}

\noindent
For the limit $\tau \to 0$ of the 1D-MAE problem, we have:

\begin{proposition}
Let $u^\tau$ be the solution of the 1D-MAE problem. Then, when $\tau \to 0$, $u^\tau$ converges to $u^0$ where $u^0$ is constant on $[0,1]$ and given by: 
\begin{eqnarray}
& & \hspace{-1cm} 
u^0 = \frac{\displaystyle \int_0^1 F(x) \, dx + [ g ]_0^1}{\displaystyle \int_0^1 e^{-\phi (y)} \, dy}
, \label{AEP_n_1D_u0} 
\end{eqnarray}
where $[ g ]_0^1 = g(1) - g(0)$. 
\label{prop_1DMAE_problem_solution}
\end{proposition}

\medskip
\noindent
{\bf Proof:} We expand $u^\tau = u^0 + \tau u^1 + o(\tau)$. Inserting this expansion in the 1D-MAE problem, we get, at leading order: 
\begin{eqnarray}
& & \hspace{-1cm} - \frac{d}{dx} \left( e^{-\phi} \frac{d u^0}{dx} \right) = 0, \quad \mbox{ in } \, [0,1], \label{PAEP_n_1D_tau=0} \\
& & \hspace{-1cm}  e^{-\phi} \frac{d u^0}{dx}  = 0, \quad \mbox{ at } \, x=0 \mbox{ and } x=1.  \label{AEP_n_bc_1D_tau=0} 
\end{eqnarray}
and at the next order:
\begin{eqnarray}
& & \hspace{-1cm} e^{-\phi} u^0 - \frac{d}{dx} \left( e^{-\phi} \frac{d u^1}{dx} \right) =  F, \quad \mbox{ in } \, [0,1], \label{PAEP_n_1D_tau=1} \\
& & \hspace{-1cm}  e^{-\phi} \frac{d u^1}{dx}  =  g, \quad \mbox{ at } \, x=0 \mbox{ and } x=1.  \label{AEP_n_bc_1D_tau=1} 
\end{eqnarray}
From (\ref{PAEP_n_1D_tau=0}), (\ref{AEP_n_bc_1D_tau=0}), we get that $u^0$ is a constant over $[0,1]$. Then, for problem (\ref{PAEP_n_1D_tau=1}), (\ref{AEP_n_bc_1D_tau=1}) to have a solution, a compatibility condition is required. Indeed, integrating (\ref{PAEP_n_1D_tau=1}) upon $x \in [0,1]$ and using the boundary conditions (\ref{AEP_n_bc_1D_tau=1}) leads to condition (\ref{AEP_n_1D_u0}). \endproof

\medskip
\noindent
Back to the $n^\tau$ variable, we find that $n^\tau \to n^0$ with $n^0 = u^0 e^{-\phi}$ with the constant $u^0$ given by (\ref{AEP_n_1D_u0}). 

\medskip
We now introduce a variational formulation. We denote by $V$ the space $H^1(0,1) = \{ u \in L^2(0,1) \, | \, u' \in L^2(0,1) \}$ where $L^2(0,1)$ is the space of square integrable functions on $[0,1]$ and $u'$ denotes the space derivative of $u$. We endow $V$ with the norm $|u|_V = (|u|^2 + |u'|^2)^{1/2}$ where $|u|= ( \int_0^1 |u(x)|^2 dx )^{1/2}$ is the norm on $L^2(0,1)$. The scalar products on $V$ and $L^2(0,1)$ are respectively denoted by $(u,v)_V$ and $(u,v)$. We also define the following bilinear forms on $V$ and $L^2(0,1)$ respectively:
\begin{eqnarray}
& & \hspace{-1cm} 
a_\phi(u,v) = \int_0^1 e^{-\phi} \, \frac{d u}{dx} \, \frac{d v}{dx} \, dx , \quad  (u,v)_\phi = \int_0^1 e^{-\phi} \, u \,  v \, dx. 
\label{MAEP_a_phi} 
\end{eqnarray}
Finally, we define the linear form $\Gamma$ on $V$ by: 
\begin{eqnarray}
& & \hspace{-1cm} 
\langle \Gamma , v \rangle = (F,v) + [g v ]_0^1
\label{MAEP_Gamma} 
\end{eqnarray}
Now, taking $v \in V$, multiplying (\ref{MPAEP_n_1D}) by $v$, integrating on $x \in [0,1]$ and using the boundary condition (\ref{MAEP_n_bc_1D}), we are led to the following

\begin{proposition}
A variational formulation of problem 1D-MAE is as follows: 
\begin{eqnarray}
& & \hspace{-1cm} 
\mbox{ Find $u^\tau \in V$ such that:}
\nonumber \\
& & \hspace{-1cm} 
a_\phi(u^\tau,v) + \tau (u^\tau,v)_\phi  = \tau \langle \Gamma , v \rangle, \quad \forall v \in V
. 
\label{MAEP_n_vf} 
\end{eqnarray}
\label{prop_MAEP1D_vf}
\end{proposition}

\medskip
\noindent
We note that the bilinear form $a_\phi(u,v) + \tau (u,v)_\phi$ is coercive on $V$. However, the best coercivity estimate is of the order of $\tau$: 
\begin{eqnarray}
& & \hspace{-1cm}  a_\phi(u,u) + \tau (u,u)_\phi \geq C \tau |v|_V^2.
\label{MAEP_coercivity} 
\end{eqnarray}
This leads to an estimate of the solution $u^\tau$ in terms of the right-hand side $\Gamma$ as follows
\begin{eqnarray}
& & \hspace{-1cm}  |u^\tau|_V \leq \frac{C}{\tau} |\Gamma|_{V'} .
\label{MAEP_estimate} 
\end{eqnarray}
This estimate deteriorates as $\tau \to 0$. Therefore, the condition number of any standard finite element method based on (\ref{MAEP_n_vf}) tends to $\infty$ as $\tau \to 0$. Such a method cannot be used when $\tau \ll 1$. We aim at finding a variational formulation which has bounded condition number when $\tau \to 0$. For this purpose, we adopt the method of \cite{Deg_Del_Loz_Nar_Neg_prep2} based on a micro-macro decomposition of the solution $u^\tau$. We indeed view the limit $u^0$ of $u^\tau$ when $\tau \to 0$ as the macroscopic component of $u^\tau$ and the difference $u^\tau - u^0$ as a microscopic correction. To define the functional spaces associated to the macro and micro components live, we introduce the following decomposition: 

\begin{definition}
We define the spaces $G$ and $A$ as follows: 
\begin{eqnarray}
& & \hspace{-1cm} 
G = \{ v \in V \, | \, v' = 0 \}, \quad A = \{ v \in V \, | \, v(0) = 0 \}. 
\label{MAEP_GA} 
\end{eqnarray}
\label{def_GA}
\end{definition}

\begin{proposition}
(i) We have: 
\begin{eqnarray}
& & \hspace{-1cm} 
V = G \oplus A. 
\label{MAEP_G+A} 
\end{eqnarray}
(ii) $G$ is characterized by 
\begin{eqnarray}
& & \hspace{-1cm} 
u \in G \quad \Longleftrightarrow \quad u \in V \quad \mbox{ such that } \quad a_\phi(u,v) = 0, \quad \forall v \in A. 
\label{MAEP_Gchar} 
\end{eqnarray}
\label{prop_G+A}
\end{proposition}

\medskip
\noindent
{\bf Proof:} (i) is obvious. \\
(ii): that $u \in G \, \Longrightarrow \, a_\phi(u,v) = 0, \, \forall v \in A$ is obvious (it is even true for all $v \in V$). Conversely, suppose that $u \in V$ such that $a_\phi(u,v) = 0, \, \forall v \in A$. Decompose $u = p + q$, with $p \in G$ and $q \in A$. Then, $0=a_\phi(u,v) = a_\phi(q,v), \, \forall v \in A$. Choosing $v = q$, we have $a_\phi(q,q)=0$. But, by Poincaré inequality, $\sqrt a_\phi$ is a norm on $v$ and is equivalent to $| \cdot |_V$. It follows that $q=0$ and that $u=p \in G$. \endproof

We now decompose the solution $u^\tau$ of (\ref{MAEP_n_vf}) into
\begin{eqnarray}
& & \hspace{-1cm} 
u^\tau = p^\tau + \tau q^\tau, \quad p^\tau \in G, \quad q^\tau \in A. 
\label{MAEP_decomp_utau} 
\end{eqnarray}
We have: 

\begin{proposition}
(i) $u^\tau$ is the solution of (\ref{MAEP_n_vf}) if and only if the pair $(p^\tau,q^\tau)$ given by (\ref{MAEP_decomp_utau}) is the solution of the following variational formulation: 
\begin{eqnarray}
& & \hspace{-1cm} 
\mbox{ Find $(p^\tau,q^\tau) \in V \times A$ such that:}
\nonumber \\
& & \hspace{-1cm} 
a_\phi(q^\tau,v) + (p^\tau + \tau q^\tau,v)_\phi  = \langle \Gamma , v \rangle, \quad \forall v \in V
, \label{MAEP_n_vfnew_1} \\
& & \hspace{-1cm} 
a_\phi(p^\tau,w)  = 0, \quad \forall w \in A
. \label{MAEP_n_vfnew_2} 
\end{eqnarray}
(ii) This variational formulation is well-posed and there exists a constant $C$, independent of $\tau$ such that 
\begin{eqnarray}
& & \hspace{-1cm} 
|(p^\tau,q^\tau)|_{V \times V} \leq C |\Gamma|_{V'} . \label{MAEP_n_vfnew_wellposed} 
\end{eqnarray}
(iii) When $\tau \to 0$, $(p^\tau,q^\tau) \to (p^0,q^0)$ and $u^\tau = p^\tau + \tau q^\tau \to p^0$ where $(p^0,q^0)$ is the solution of the following variational formulation: 
\begin{eqnarray}
& & \hspace{-1cm} 
\mbox{ Find $(p^0,q^0) \in V \times A$ such that:}
\nonumber \\
& & \hspace{-1cm} 
a_\phi(q^0,v) + (p^0 ,v)_\phi  = \langle \Gamma , v \rangle, \quad \forall v \in V
, \label{MAEP_n_vfnew_1_tau=0} \\
& & \hspace{-1cm} 
a_\phi(p^0,w)  = 0, \quad \forall w \in A
. \label{MAEP_n_vfnew_2_tau=0} 
\end{eqnarray}
\label{prop_MAEP1D_vfnew}
\end{proposition}

\noindent
{\bf Proof:} (i) We insert the decomposition (\ref{MAEP_decomp_utau}) into (\ref{MAEP_n_vf}) and get $(p^\tau,q^\tau) \in G \times A$ such that (\ref{MAEP_n_vfnew_1}) holds for any $v \in V$. Now, using the characterization (\ref{MAEP_Gchar}) of $G$, we get (\ref{MAEP_n_vfnew_2}). Conversely, let  $(p^\tau,q^\tau) \in V \times A$ be a solution of (\ref{MAEP_n_vfnew_1}), (\ref{MAEP_n_vfnew_2}). Constructing $u^\tau$ according to (\ref{MAEP_decomp_utau}) obviously leads to (\ref{MAEP_n_vf}). 

\noindent
(ii) We refer to \cite{Deg_Del_Loz_Nar_Neg_prep2}. 

\noindent
(iii) is obvious. \endproof

Now, we examine what is the PDE problem solved by the variational formulation (\ref{MAEP_n_vfnew_1}), (\ref{MAEP_n_vfnew_2}). Taking successively a smooth test function with compact support, followed by a smooth function with non-zero boundary values, we easily find that $(p^\tau,q^\tau)$ solve the following PDE problem: 
\begin{eqnarray}
& & \hspace{-1cm} e^{-\phi} (p^\tau + \tau q^\tau) - \frac{d}{dx} \left( e^{-\phi} \frac{d q^\tau}{dx} \right) = F, \quad \mbox{ in } \, [0,1], \label{MPAEP_n_1D_new} \\
& & \hspace{-1cm}  (e^{-\phi} \frac{d q^\tau}{dx})(0)  =  g(0), \quad (e^{-\phi} \frac{d q^\tau}{dx}) (1)  =  g(1),  \label{MAEP_n_bc_1D_new} \\
& & \hspace{-1cm}  q^\tau (0) = 0 ,  \label{MAEP_n_q_1D_new} \\
& & \hspace{-1cm}  p^\tau \mbox{ is a constant}.   \label{MAEP_n_p_1D_new}
\end{eqnarray}
The boundary value problem for $q^\tau$ is overdetermined, having one Dirichlet and one Neumann boundary condition at $x=0$. However, the presence of the unknown constant $p^\tau$ introduces another degree of freedom which compensates for this over-determination. 

More precisely, let us consider the map ${\mathcal F}$ which to $p^\tau \in {\mathbb R}$ associates ${\mathcal F}(p^\tau) = (e^{-\phi} \frac{d q^\tau}{dx}  -  g) (0) \in {\mathbb R}$, where $q^\tau$ is the unique solution of the mixed Dirichlet-Neumann boundary value problem: 
\begin{eqnarray}
& & \hspace{-1cm} e^{-\phi} (p^\tau + \tau q^\tau) - \frac{d}{dx} \left( e^{-\phi} \frac{d q^\tau}{dx} \right) = F, \quad \mbox{ in } \, [0,1], \label{MPAEP_n_1D_new_2} \\
& & \hspace{-1cm}  q^\tau (0) = 0, \quad (e^{-\phi} \frac{d q^\tau}{dx}) (1)  =  g(1),  \label{MAEP_n_bc_1D_new_2} 
\end{eqnarray}
This problem is uniquely solvable. Additionally, it is uniformly solvable when $\tau \to 0$ thanks to the Dirichlet boundary condition at $x=0$ which makes the elliptic operator $- \frac{d}{dx} \left( e^{-\phi} \frac{d }{dx} \right)$ invertible with these boundary conditions. Finding a solution to (\ref{MPAEP_n_1D_new}), (\ref{MAEP_n_p_1D_new}) is equivalent to finding a root to the equation ${\mathcal F}(p^\tau) = 0$. But ${\mathcal F}(p^\tau)$ is an affine function of $p^\tau$ with
$$ {\mathcal F}(p) = {\mathcal F}(0) + {\mathcal F}'(0) p , $$
and ${\mathcal F}'(0) = e^{-\phi} \frac{d Q^\tau}{dx} (0)$, with $Q^\tau$ the unique solution of the problem
\begin{eqnarray}
& & \hspace{-1cm} e^{-\phi} (1 + \tau Q^\tau) - \frac{d}{dx} \left( e^{-\phi} \frac{d Q^\tau}{dx} \right) = 0, \quad \mbox{ in } \, [0,1], \label{MPAEP_n_1D_deriv} \\
& & \hspace{-1cm}  Q^\tau (0) = 0, \quad (e^{-\phi} \frac{d Q^\tau}{dx}) (1)  =  0.  \label{MAEP_n_bc_1D_deriv} 
\end{eqnarray}
Then, the existence of $p^\tau$ is equivalent to ${\mathcal F}'(0)$ being non-zero. By contradiction, suppose that ${\mathcal F}'(0) = 0$. This means that $\frac{d Q^\tau}{dx} (0) = 0$. Then, denoting by $U^\tau = \frac{d Q^\tau}{dx}$ and differentiating (\ref{MPAEP_n_1D_deriv}) with respect to $x$ leads to 
\begin{eqnarray}
& & \hspace{-1cm} \tau U^\tau - \frac{d}{dx} \left( e^{\phi} \frac{d}{dx} ( e^{-\phi} U^\tau)  \right) = 0, \quad \mbox{ in } \, [0,1], \label{MPAEP_n_1D_deriv_2} \\
& & \hspace{-1cm}  U^\tau (0) = 0, \quad U^\tau (1)  =  0,  \label{MAEP_n_bc_1D_deriv_2} 
\end{eqnarray}
which implies that $U^\tau = 0$. It follows that $Q^\tau$ is a constant and by (\ref{MPAEP_n_1D_deriv}), this constant must be $Q^\tau = - 1/\tau$. But this is in contradiction with the first boundary condition (\ref{MAEP_n_bc_1D_deriv}). Therefore, $\frac{d Q^\tau}{dx} (0) \not = 0$ and consequently ${\mathcal F}'(0) \not = 0$. Finally, there exists a unique root $p^\tau$ of ${\mathcal F}(p^\tau) = 0$ given by $p^\tau = - \frac{{\mathcal F}(0)}{{\mathcal F}'(0)}$. Additionally, this root is uniformly bounded when $\tau \to 0$, because both ${\mathcal F}(0)$ and ${\mathcal F}'(0)$ remain bounded when $\tau \to 0$. Indeed, both elliptic problems associated to ${\mathcal F}(0)$ and ${\mathcal F}'(0)$ are uniquely solvable when $\tau = 0$, thanks to the Dirichlet boundary conditions and the Poincaré inequality and ${\mathcal F}'(0)$ cannot be zero also in that case. As a summary, we have proved by elementary means that problem (\ref{MPAEP_n_1D_new})-(\ref{MAEP_n_p_1D_new}) is uniquely solvable and that its inverse is uniformly bounded with respect to $\tau$ when $\tau \to 0$.

%%%%%%%%%%%%%%%%%%%%%%%%%%%%%%%%%%%%%%%%%%%%%%%%%%%%%%%%%%%%%%%%%%%%%%%%%%%%%%%%%%%%%%%%%%%%%%%%
%%%%%%%%%%%%%%%%%%%%%%%%%%%%%%%%%%%%%%%%%%%%%%%%%%%%%%%%%%%%%%%%%%%%%%%%%%%%%%%%%%%%%%%%%%%%%%%%
\subsection{The multi-dimensional strongly anisotropic elliptic problem: the continuous case}
\label{subsec_El_stron_aniso_multiD_continuous}

We recall some notations. We define $b = B/|B|$ where $B$ is a divergence free field ($nabla \cdot B = 0$) over the domain $\Omega$. As a consequence, $\nabla \cdot b = - b \cdot \nabla (\ln |B|)$. We define the boundaries $\Gamma_+$, $\Gamma_-$, $\Gamma_0$ like in section \ref{subsection_EL_BC}. The magnetic field lines are curves, solutions of the differential equation $ \frac{dX}{ds} (s) = b(X(s)) $. For simplicity, we suppose that all magnetic field lines intersect the boundaries  $\Gamma_+$ and $\Gamma_-$ at two points denoted by $x_0^+$ and $x_0^-$ respectively. In this case, we denote the magnetic field line starting from $x_0^-$: $X(x_0^-;s)$, or simply ${\mathcal B}_{x_0^-}$. If $b$ is regular, by the Cauchy-Lipschitz theorem, for any $x \in \Omega$, there exists a unique $x_0^-$ such that $x \in {\mathcal B}_{x_0^-}$. We count the arclength $s$ from its origin at $x_0^-$. Extensions of the present theory to more complex cases, where the magnetic field lines may intersect the boundary at one or zero points are possible but will be discarded here for simplicity. 

We introduce an 'arclength potential' $\phi$ by: 
\begin{eqnarray}
& & \hspace{-1cm} 
\phi(x) = - \int_{\overline{x_0^- x}} \,  E_\parallel (y) \, ds(y)
,  \label{MAEP_arcpot} 
\end{eqnarray}
where $x_0^-$ is such that $x \in {\mathcal B}_{x_0^-}$ and $\overline{x_0^- x}$ denotes the arc along the magnetic field line issued from $x_0^-$ and ending at $x$. Note that there is no condition on $E$ for the existence of $\phi$, because $\phi$ is not a 'true' potential, but only the primitive of the electric field along the direction of the magnetic field lines. The introduction of $\phi$ allows to remove the electric field by a simple transformation. Indeed, we remark that
\begin{eqnarray}
& & \hspace{-1cm} b \cdot ( \nabla n^\tau - n^\tau E)  = e^{- \phi} b \cdot \nabla u^\tau 
, \quad u^\tau = e^\phi n^\tau.  
\label{AEPMD_phi_change} 
\end{eqnarray}
Then, we have the

\begin{proposition}
With $u^\tau$ defined by (\ref{AEPMD_phi_change}), problem AE is equivalent to the following modified anisotropic elliptic (MAE) problem:
\begin{eqnarray}
& & \hspace{-1cm} \tau  e^{- \phi} u^\tau - \nabla \cdot ( e^{-\phi} (b \otimes b) \nabla u^\tau) = \tau F, \quad \mbox{ in } \, \Omega, \label{AEP_n_new} \\
& & \hspace{-1cm} (b \cdot \nu) e^{-\phi} b \cdot \nabla u^\tau  = \tau g, \quad \mbox{ on } \, \Gamma_+ \cup \Gamma_-.  
\label{AEP_n_bc_new} 
\end{eqnarray}
\label{prop_MAE_problem_equiv}
\end{proposition}

\noindent
We note the following lemma:

\begin{lemma} 
(i) Let $u$ be a solution of the problem:
\begin{eqnarray}
& & \hspace{-1cm} \nabla \cdot ( b u ) = 0, \quad \mbox{ in } \, \Omega, \label{AEP_nabla_b_u_=0} \\
& & \hspace{-1cm} (b \cdot \nu) u  = 0, \quad \mbox{ on } \, \Gamma_+ \cup \Gamma_-.  \label{AEP_nabla_b_u_=0_bc} 
\end{eqnarray}
Then, $u = 0$ identically on $\Omega$. 

\noindent
(ii) Let $f$ be defined on $\Omega$ and $g$ be defined on $\Gamma_+ \cup \Gamma_-$. Then, the problem
\begin{eqnarray}
& & \hspace{-1cm} \nabla \cdot ( b u ) = f, \quad \mbox{ in } \, \Omega, \label{AEP_nabla_b_u_=f} \\
& & \hspace{-1cm} (b \cdot \nu) u  = g, \quad \mbox{ on } \, \Gamma_+ \cup \Gamma_-.  \label{AEP_nabla_b_u_=g_bc} 
\end{eqnarray}
has a solution provided the following solvability condition is satisfied: 
\begin{eqnarray}
& & \hspace{-1cm} \int_{\overline{x_0^- x_0^+}} \frac{f}{|B|} (y) \, ds(y) - \left[ \frac{g}{B \cdot \nu} \right]_{x_0^-}^{x_0^+} = 0 ,
\label{AEP_nabla_b_u_compat} 
\end{eqnarray}
for all field lines, i.e. all $x_0^- \in \Gamma_-$. 
Under this condition, the system has a unique solution: 
\begin{eqnarray}
& & \hspace{-1cm} u(x) = |B|(x) \left\{ \int_{\overline{x_0^- x}} \frac{f}{|B|} (y) \, ds(y) +  \frac{g}{B \cdot \nu} (x_0^-) \right\} ,
\label{AEP_nabla_b_u_sol} 
\end{eqnarray}
where $x_0^-$ is such that $x \in {\mathcal B}_{x_0^-}$.
\label{lem_nabla_b_u}
\end{lemma}

\noindent
{\bf Proof:} (i) Elementary computations show that $b \cdot \nabla (\frac{u}{|B|}) = \frac{1}{|B|} \nabla \cdot (bu)$. Then, (\ref{AEP_nabla_b_u_=0}) is equivalent to saying that $\frac{u}{|B|}$ is constant along each magnetic field line. But (\ref{AEP_nabla_b_u_=0_bc}) implies that $u=0$. 

(ii) Eq. (\ref{AEP_nabla_b_u_=f}) implies that $ b \cdot \nabla (\frac{u}{|B|}) = \frac{f}{|B|}$. Integrating this equation along any magnetic field line leads to (\ref{AEP_nabla_b_u_sol}). Identifying (\ref{AEP_nabla_b_u_sol}) at $x_0^+$ with the boundary condition (\ref{AEP_nabla_b_u_=g_bc}) leads to the compatibility condition (\ref{AEP_nabla_b_u_compat}). If this relation is satisfied, then, formula (\ref{AEP_nabla_b_u_sol}) provides the unique solution to the problem. \endproof

\medskip
With this lemma, we can now prove the 

\begin{proposition}
Let $u^\tau$ be the solution of the MAE problem. Then, when $\tau \to 0$, $u^\tau$ converges to $u^0$ where $u^0$ is constant along the magnetic field lines (i.e. $b \cdot \nabla u^\tau =0$) and is given by: 
\begin{eqnarray}
& & \hspace{-1cm} 
u^0(x) = \frac{\displaystyle \int_{\overline{x_0^- x_0^+}} \frac{F}{|B|} (y) \, ds(y) - \left[ \frac{g}{B \cdot \nu} \right]_{x_0^-}^{x_0^+}}{\displaystyle \int_{\overline{x_0^- x_0^+}} \frac{e^{-\phi}}{|B|} (y) \, ds(y)}
, \label{AEP_n_u0_sol} 
\end{eqnarray}
where $x_0^-$ is such that $x \in {\mathcal B}_{x_0^-}$.
\label{prop_MAE_problem_solution}
\end{proposition}

\medskip
\noindent
{\bf Proof:} We expand $u^\tau = u^0 + \tau u^1 + o(\tau)$. Inserting this expansion in the MAE problem, we get, at leading order: 
\begin{eqnarray}
& & \hspace{-1cm} - \nabla \cdot ( e^{-\phi} (b \otimes b) \nabla u^0) = 0, \quad \mbox{ in } \, \Omega, \label{AEP_n_new_tau=0} \\
& & \hspace{-1cm} (b \cdot \nu) e^{-\phi} b \cdot \nabla u^0  = 0, \quad \mbox{ on } \, \Gamma_+ \cup \Gamma_-.  
\label{AEP_n_bc_new_tau=0} 
\end{eqnarray}
and at the next order:
\begin{eqnarray}
& & \hspace{-1cm} e^{- \phi} u^0 - \nabla \cdot ( e^{-\phi} (b \otimes b) \nabla u^1) = F, \quad \mbox{ in } \, \Omega, \label{AEP_n_new_tau=1} \\
& & \hspace{-1cm} (b \cdot \nu) e^{-\phi} b \cdot \nabla u^1  = g, \quad \mbox{ on } \, \Gamma_+ \cup \Gamma_-.  
\label{AEP_n_bc_new_tau=1} 
\end{eqnarray}
From (\ref{AEP_n_new_tau=0}), (\ref{AEP_n_bc_new_tau=0}) and Lemma \ref{lem_nabla_b_u} (i), we deduce that $b \cdot \nabla u^0 = 0 $. Then, from Lemma \ref{lem_nabla_b_u} (ii), problem (\ref{AEP_n_new_tau=1}), (\ref{AEP_n_bc_new_tau=1}) is solvable if and only if the compatibility condition (\ref{AEP_nabla_b_u_compat}) is satisfied (with $f$ replaced by $F - e^{-\phi} u^0$). This leads to condition (\ref{AEP_n_u0_sol}). \endproof

\medskip
\noindent
Back to the $n^\tau$ variable, we find that $n^\tau \to n^0 = u^0(x_0^-) e^{-\phi}$ with $u^0(x_0^-)$ given by (\ref{AEP_n_u0_sol}).

\medskip
We now introduce a variational formulation. We denote by $V$ the space $V = \{ u \in L^2(\Omega) \, \, | \, \, b \cdot \nabla u \in L^2(\Omega) \}$ where $L^2(\Omega)$ is the space of square integrable functions on $\Omega$. We endow $V$ with the norm $|u|_V = (|u|^2 + |b \cdot \nabla u |^2)^{1/2}$ where $|u|= ( \int_\Omega |u(x)|^2 dx )^{1/2}$ is the norm on $L^2(\Omega)$. The scalar products on $V$ and $L^2(\Omega)$ are respectively denoted by $(u,v)_V$ and $(u,v)$. We also define the following bilinear forms on $V$ and $L^2(0,1)$ respectively:
\begin{eqnarray}
& & \hspace{-1cm} 
a_\phi(u,v) = \int_\Omega e^{-\phi} \, (b \cdot \nabla u)  \, (b \cdot \nabla v) \, dx , \quad  (u,v)_\phi = \int_\Omega e^{-\phi} \, u \,  v \, dx. 
\label{MAE_a_phi} 
\end{eqnarray}
Finally, we define the linear form $\Gamma$ on $V$ by: 
\begin{eqnarray}
& & \hspace{-1cm} 
\langle \Gamma , v \rangle = (F,v) + \int_{\Gamma^- \cup \Gamma^+} g v \, d\Gamma(x)
\label{AEP_Gamma} 
\end{eqnarray}
Now, taking $v \in V$, multiplying (\ref{AEP_n_new}) by $v$, integrating on $x \in \Omega$ and using the boundary condition (\ref{AEP_n_bc_new}), we are led to the following

\begin{proposition}
A variational formulation of problem MAE is as follows: 
\begin{eqnarray}
& & \hspace{-1cm} 
\mbox{ Find $u^\tau \in V$ such that:}
\nonumber \\
& & \hspace{-1cm} 
a_\phi(u^\tau,v) + \tau (u^\tau,v)_\phi  = \tau \langle \Gamma , v \rangle, \quad \forall v \in V
. 
\label{MAE_n_vf} 
\end{eqnarray}
\label{prop_MAEP_vf}
\end{proposition}

\medskip
\noindent
Like in the one-dimensional case, we note that the bilinear form $a_\phi(u,v) + \tau (u,v)_\phi$ is coercive on $V$. However, the best coercivity estimate is of the order of $\tau$ and is given by (\ref{MAEP_coercivity}). This leads to an estimate of the solution $u^\tau$ in terms of the right-hand side $\Gamma$ of the form (\ref{MAEP_estimate}), which deteriorates as $\tau \to 0$. Therefore, a standard finite element method based on (\ref{MAEP_n_vf}) cannot be used when $\tau \ll 1$. We aim at finding a variational formulation which has bounded condition number when $\tau \to 0$. For this purpose, like in the one-dimensional case we adopt the method of \cite{Deg_Del_Loz_Nar_Neg_prep2} based on a micro-macro decomposition of the solution $u^\tau$.

\begin{definition}
We define the spaces $G$ and $A$ as follows: 
\begin{eqnarray}
& & \hspace{-1cm} 
G = \{ v \in V \, | \, b \cdot \nabla v = 0 \}, \quad A = \{ v \in V \, | \, v|_{\Gamma_-} = 0 \}. 
\label{MAE_GA} 
\end{eqnarray}
\label{def_GA_multid}
\end{definition}

\begin{proposition}
(i) We have: 
\begin{eqnarray}
& & \hspace{-1cm} 
V = G \oplus A. 
\label{MAE_G+A} 
\end{eqnarray}
(ii) $G$ is characterized by 
\begin{eqnarray}
& & \hspace{-1cm} 
u \in G \quad \Longleftrightarrow \quad u \in V \quad \mbox{ such that } \quad a_\phi(u,v) = 0, \quad \forall v \in A. 
\label{MAE_Gchar} 
\end{eqnarray}
\label{prop_G+A_multid}
\end{proposition}

\medskip
\noindent
{\bf Proof:} (i) First, suppose that $v \in G \cap A$. Then, $v$ is constant along the magnetic field lines and its restriction on $\Gamma_-$ is zero. By the assumption that all magnetic field lines contained in $\Omega$ intersect $\Gamma_-$ at one point, $v$ is identically zero in $\Omega$, which shows that $G \cap A = \emptyset$. Let us now take any function $v \in V$ and construct a function $p \in G$ by the formula $p(x) = v(x_0^-)$, where $x_0^-$ is the foot of the magnetic field line passing at $x$ (in other words, $w \in {\mathcal B}_{x_0^-}$). Then $q = v - p$ obviously belongs to $A$. This shows that $V = G + A$. 

\medskip
\noindent
(ii): the proof of proposition \ref{prop_G+A} can be reproduced in the multi-dimensional case without any change. \endproof

Like in the one-dimensional case, we decompose the solution $u^\tau$ of (\ref{MAE_n_vf}) into
\begin{eqnarray}
& & \hspace{-1cm} 
u^\tau = p^\tau + \tau q^\tau, \quad p^\tau \in G, \quad q^\tau \in A. 
\label{MAE_decomp_utau} 
\end{eqnarray}
Then, proposition \ref{prop_MAEP1D_vfnew} applies without any change to the multi-dimensional case. We state it in detail for the sake of completeness (the proof is similar and is omitted):

\begin{proposition}
(i) $u^\tau$ is the solution of (\ref{MAE_n_vf}) if and only if the pair $(p^\tau,q^\tau)$ given by (\ref{MAE_decomp_utau}) is the solution of the following variational formulation: 
\begin{eqnarray}
& & \hspace{-1cm} 
\mbox{ Find $(p^\tau,q^\tau) \in V \times A$ such that:}
\nonumber \\
& & \hspace{-1cm} 
a_\phi(q^\tau,v) + (p^\tau + \tau q^\tau,v)_\phi  = \langle \Gamma , v \rangle, \quad \forall v \in V
, \label{MAE_n_vfnew_1} \\
& & \hspace{-1cm} 
a_\phi(p^\tau,w)  = 0, \quad \forall w \in A
. \label{MAE_n_vfnew_2} 
\end{eqnarray}
(ii) This variational formulation is well-posed and there exists a constant $C$, independent of $\tau$ such that 
\begin{eqnarray}
& & \hspace{-1cm} 
|(p^\tau,q^\tau)|_{V \times V} \leq C |\Gamma|_{V'} . \label{MAE_n_vfnew_wellposed} 
\end{eqnarray}
(iii) When $\tau \to 0$, $(p^\tau,q^\tau) \to (p^0,q^0)$ and $u^\tau = p^\tau + \tau q^\tau \to p^0$ where $(p^0,q^0)$ is the solution of the following variational formulation: 
\begin{eqnarray}
& & \hspace{-1cm} 
\mbox{ Find $(p^0,q^0) \in V \times A$ such that:}
\nonumber \\
& & \hspace{-1cm} 
a_\phi(q^0,v) + (p^0 ,v)_\phi  = \langle \Gamma , v \rangle, \quad \forall v \in V
, \label{MAE_n_vfnew_1_tau=0} \\
& & \hspace{-1cm} 
a_\phi(p^0,w)  = 0, \quad \forall w \in A
. \label{MAE_n_vfnew_2_tau=0} 
\end{eqnarray}
\label{prop_MAEP_vfnew}
\end{proposition}

\medskip
\noindent
Now, we derive the PDE solved by the variational formulation (\ref{MAE_n_vfnew_1}), (\ref{MAE_n_vfnew_2}). Using the same method as in the one-dimensional case, we find that $(p^\tau,q^\tau)$ solve the following PDE problem: 
\begin{eqnarray}
& & \hspace{-1cm} e^{- \phi} (p^\tau + \tau q^\tau) - \nabla \cdot ( e^{-\phi} (b \otimes b) \nabla q^\tau) =  F, \quad \mbox{ in } \, \Omega, \label{MPAEP_q_new} \\
& & \hspace{-1cm} (b \cdot \nu) e^{-\phi} b \cdot \nabla q^\tau  = g, \quad \mbox{ on } \, \Gamma_+ \cup \Gamma_- , 
\label{MAEP_q_bc_new} \\
& & \hspace{-1cm}  q^\tau = 0 \quad \mbox{ on } \, \Gamma_-,   \label{MAEP_q_q_new} \\
& & \hspace{-1cm} - \nabla \cdot ( e^{-\phi} (b \otimes b) \nabla p^\tau) = 0, \quad \mbox{ in } \, \Omega, \label{MPAEP_p_new} \\
& & \hspace{-1cm} (b \cdot \nu) e^{-\phi} b \cdot \nabla p^\tau  = 0, \quad \mbox{ on } \, \Gamma_+ . 
\label{MAEP_p_bc_new} \\
\end{eqnarray}
The boundary value problem for $p^\tau$ is equivalent to saying that $p^\tau$ is constant along the magnetic field lines ($b \cdot \nabla p^\tau = 0$) and that $p^\tau$ is determined by its boundary value on $\Gamma_+$. The boundary value problem for $q^\tau$ is overdetermined, having one Dirichlet and one Neumann boundary condition on $\Gamma_-$. However, the presence of the unknown constant $p^\tau|_{\Gamma_+}$ introduces just the right number of degrees of freedom to compensate for this over-determination. 

We can also see intuitively that this formulation has a condition number which is independent of $\tau$. Indeed, for given $p^\tau$, constant along the magnetic field lines and determined by its boundary value on $\Gamma_+$, we can find $q^\tau$ by solving the mixed Dirichlet-Neumann boundary value problem:
\begin{eqnarray}
& & \hspace{-1cm} e^{- \phi} (p^\tau + \tau q^\tau) - \nabla \cdot ( e^{-\phi} (b \otimes b) \nabla q^\tau) =  F, \quad \mbox{ in } \, \Omega, \label{MPAEP_q_new_2} \\
& & \hspace{-1cm} (b \cdot \nu) e^{-\phi} b \cdot \nabla q^\tau  = g, \quad \mbox{ on } \, \Gamma_+, 
\label{MAEP_q_bc_new_2} \\
& & \hspace{-1cm}  q^\tau = 0 \quad \mbox{ on } \, \Gamma_-.   \label{MAEP_q_q_new_2} 
\end{eqnarray}
The operator $\, q \, \to \, \tau e^{- \phi} q \,  - \, \nabla \cdot ( e^{-\phi} (b \otimes b) \nabla q) \, $ with Dirichlet boundary conditions on $\Gamma_-$ and Neumann boundary conditions on $\Gamma_+$ is invertible whatever the value of $\tau$ is. Therefore, the inversion of (\ref{MPAEP_q_new_2})-(\ref{MAEP_q_q_new_2}) provides a solution $q^\tau$ as a function of $p^\tau$, whatever the value of $\tau$. The additional boundary condition $\, (b \cdot \nu) e^{-\phi} b \cdot \nabla q^\tau  = g$ on $\Gamma_- $ gives another condition which allows us to compute $p^\tau$. The solvability of this last condition is also indpendent of the value of $\tau$ as was explicitly seen in the one-dimensional case.

%%%%%%%%%%%%%%%%%%%%%%%%%%%%%%%%%%%%%%%%%%%%%%%%%%%%%%%%%%%%%%%%%%%%%%%%%%%%%%%%%%%%%%%%%%%%%%%%
%%%%%%%%%%%%%%%%%%%%%%%%%%%%%%%%%%%%%%%%%%%%%%%%%%%%%%%%%%%%%%%%%%%%%%%%%%%%%%%%%%%%%%%%%%%%%%%%
\subsection{The discrete anisotropic elliptic problem: back to the one-dimensional problem}
\label{subsec_El_stron_aniso_1D_discrete}

The discretization of the Euler-Lorentz model directly leads to an anisotropic elliptic problem in discrete form. Therefore, we are not free of choosing the numerical method for dicretizing it. Rather, the method is imposed by that of the discretization of the background Euler-Lorentz model. Despite this fact, we will see that we can apply the same ideas than those developed in the last two sections for the continuous problem, to these discrete formulations. Like for the continuous problem, we will start to develop the ideas in a simple one-dimensional framework, before applying them in full generality to the three (or more)-dimensional framework. 

The discretization of the Euler-Lorentz model by the FDAP-2 scheme (a similar study could be conducted for the FDAP-1 scheme) leads to a discretization of the form: 
\begin{eqnarray}
& & \hspace{-1cm} 
-  \frac{h^{-1}}{2} \left\{ \left( \frac{ h^{-1}}{2} \left( n_{k+2} - n_{K} \right) - n_{K+1} E_{K+1} \right) \right.  \nonumber \\
& & \hspace{1cm} 
\left. - \left( \frac{ h^{-1}}{2} \left( n_{K} - n_{K-2} \right) - n_{K-1} E_{K-1} \right) \right\} 
+ \tau n_K = \tau F_K, 
\label{DAEP1D_n} 
\end{eqnarray}
where again, we restrict ourselves to linear pressure-density relationships. We also have made $\delta = 1$ and collected all known terms at the right-hand side into the generic term $F_K$. $K$ is now a one-dimensional index ranging in ${\mathbb Z}$. The discrete electric field is also one-dimensional and denoted by $E_K$. Eq. (\ref{DAEP1D_n}) is a generic one-dimensional model for (\ref{AP2_n_2}).

Now, we introduce the boundary conditions which are an important aspect of this discussion. We suppose that the unknown $n_{K}$ is defined for $K$ ranging over a finite interval $K \in [-M, M] := \{ -M, -M+1, \ldots , M-1, M \}$. For the sake of the forthcoming developments, we write (\ref{DAEP1D_n}) in the form of a mixed problem and we highlight the dependence upon $\tau$: 
\begin{eqnarray}
& & \hspace{-1cm} 
-  \frac{h^{-1}}{2} \left( \gamma^\tau_{K+1} - \gamma^\tau_{K-1}  \right) + \tau n^\tau_K = \tau F_K, \quad K \in [-M, M], 
\label{DAEP1D_gam1} \\
& & \hspace{-1cm} 
\gamma^\tau_{K} =  \frac{ h^{-1}}{2} \left[ n^\tau_{k+1} - n^\tau_{K-1} \right] - n^\tau_{K} E_{K}, \quad K \in [-M+1, M-1] . 
\label{DAEP1D_gam2}
\end{eqnarray}
To complete this formulation, we need to impose boundary conditions in the last row of cells ($K = M$ or $K=-M$) and in an additional row of fictitious cells ($K = M+1$ or $K=-M-1$). We impose the following boundary conditions (which are the discrete counterpart of (\ref{IEL_BC_n_pm})): 
\begin{eqnarray}
& & \hspace{-1cm} 
\gamma^\tau_{M} = \gamma^\tau_{M+1} = \tau g_M, \quad \gamma^\tau_{-M} = \gamma^\tau_{-M-1} = \tau g_{-M}
, 
\label{DAEP1D_bc}
\end{eqnarray}
where $g_{\pm M}$ is supposed known. 

Since our solution method for strongly anisotropic elliptic problems relies on the introduction of an appropriate variational formulation, we first introduce the discrete counterpart of the original variational formulation (\ref{MAEP_n_vf}). Multiplying (\ref{DAEP1D_gam1}) by a test sequence $\{ v_K \}_{K \in [-M, M]}$, performing a discrete integration-by-parts and using (\ref{DAEP1D_gam2}) together with the boundary conditions (\ref{DAEP1D_bc}), we get the following

\begin{proposition}
The solution $n^\tau_{K}$ of problem (\ref{DAEP1D_gam1}), (\ref{DAEP1D_gam2}) with boundary conditions (\ref{DAEP1D_bc}) satisfies the following discrete variational formulation:
\begin{eqnarray}
& & \hspace{-1cm} 
\mbox{ Find $n^\tau \in V$ such that:}
\nonumber \\
& & \hspace{-1cm} 
a_E (n^\tau,v) + \tau (n^\tau,v)  = \tau \langle \Gamma , v \rangle, \quad \forall v \in V
,
\label{DAEP1D_vf} 
\end{eqnarray}
where $V = \ell^2([-M, M])$, the space of square integrable sequences on $[-M, M]$, and 
\begin{eqnarray}
& & \hspace{-1cm} 
a_E(u,v) = \sum_{k=-M+1}^{M-1}  \frac{h^{-1}}{2} \left( \frac{ h^{-1}}{2} \left[ u_{k+1} - u_{K-1} \right] - u_{K} E_{K}  \right) \left( v_{k+1} - v_{K-1} \right), \label{DAEP1D_aE} \\
& & \hspace{-1cm} 
(u,v) = \sum_{k=-M}^{M} u_k v_k  , \label{DAEP1D_()} \\
& & \hspace{-1cm} 
\langle \Gamma , v \rangle = (F,v) + \frac{h^{-1}}{2} \left[ g_M ( v_{M-1} + v_M ) - g_{-M} ( v_{-M+1} + v_{-M} ) \right]. \label{DAEP1D_Gamma} 
\end{eqnarray}
\label{prop_DAEP1D_vf}
\end{proposition}

We note that the form $a_E$ is not symmetric but its leading order part $a_0$ (corresponding to the discretization of the second order derivative) is. We also note that the for $a_E(u,u) + \tau (u,u)$ is coercive on $V$. However, we have no better estimate than 
\begin{eqnarray}
& & \hspace{-1cm} 
a_E(u,u) + \tau (u,u) \geq \tau |u|^2_V, 
\label{DAEP1D_coerc}
\end{eqnarray}
(with $|u|^2_V = (u,u)$) which implies that there is no better estimate for the solution $n^\tau$ of (\ref{DAEP1D_vf}) than $|n^\tau|_V \leq \tau^{-1} |\Gamma|_V$. The condition number of formulation (\ref{DAEP1D_vf}) tends to $\infty$ as $\tau \to 0$ and it cannot be used for solving for $n^\tau$ when $\tau$ is very small. We are going to remedy to this problem by introducing the discrete analog of formulation (\ref{MAEP_n_vfnew_1}), (\ref{MAEP_n_vfnew_2}).

\begin{definition}
We define the spaces $G_E$ and $A$ as follows: 
\begin{eqnarray}
& & \hspace{-1cm} 
G_E = \{ v \in V \, \,  | \, \, \frac{ h^{-1}}{2} \left[ v_{k+1} - v_{K-1} \right] - v_{K} E_{K} = 0 , \, \, \forall K \in [-M+1, M-1]  \}, \label{DAEP1D_GE} \\
& & \hspace{-1cm} 
A = \{ v \in V \, \, | \, \, v_{-M} = v_{-M+1} = 0 \}. 
\label{DAEP1D_A} 
\end{eqnarray}
\label{def_DAEP1D_GA}
\end{definition}

\begin{proposition}
(i) We have: 
\begin{eqnarray}
& & \hspace{-1cm} 
V = G_E \oplus A. 
\label{DAEP1D_G+A} 
\end{eqnarray}
(ii) $G_E$ is characterized by 
\begin{eqnarray}
& & \hspace{-1cm} 
u \in G_E \quad \Longleftrightarrow \quad u \in V \quad \mbox{ such that } \quad a_E(u,v) = 0, \quad \forall v \in A. 
\label{DAEP1D_Gchar} 
\end{eqnarray}
\label{prop_DAEP1D_G+A}
\end{proposition}

\medskip
\noindent
{\bf Proof:} (i) Let $v \in G_E \cap A$. Then, $v_K$ is the solution of a two-stages linear recursion where two successive steps are zero. This implies that $v$ is identically zero, proving that $G_E \cap A = \emptyset$. Now, let $v \in V$. Then, there exists a unique $p \in G_E$ which satisfies \, $p_{-M} = v_{-M}$, \,  $p_{-M+1} = v_{-M+1}$. Indeed, $p$ is the solution of two-stages linear recursion and has specified values at two successive points. Then, $p_K$ is uniquely defined for all $K$'s. Then, defining $q_K = v_K - p_K$, we obviously have $q \in A$, showing that $V = G_E + A$ and ending the proof of point (i). 

\medskip
\noindent
(ii): that $u \in G \, \Longrightarrow \, a_E(u,v) = 0, \, \forall v \in A$ is obvious (it is even true for all $v \in V$). Conversely, suppose that $u \in V$ such that $a_E(u,v) = 0, \, \forall v \in A$. Let $\psi_K$ for $K \in [-M+1, M-1]$ be arbitrary. There exists $v \in A$ such that $\psi_K = v_{K+1} - v_{K-1}$. Indeed, $v_K$ is defined by a two-stages recursion with initial conditions $v_{-M} = v_{-M+1} = 0$. Then, such a $v$ in $A$ exists and is unique. Replacing $v_{K+1} - v_{K-1}$ by $\psi_K$ in $a_E(u,v) = 0$, we are led to the following relation: 
\begin{eqnarray*}
& & \hspace{-1cm} 
\sum_{k=-M+1}^{M-1}  \frac{h^{-1}}{2} \left( \frac{ h^{-1}}{2} \left[ u_{k+1} - u_{K-1} \right] - u_{K} E_{K}  \right) \psi_K = 0, \quad \forall \{\psi_K\}_{K \in [-M+1, M-1]}. 
\end{eqnarray*}
This clearly implies that $u \in G_E$ and ends the proof of the proposition. \endproof

We now decompose the solution $n^\tau$ of (\ref{DAEP1D_vf}) into
\begin{eqnarray}
& & \hspace{-1cm} 
n^\tau = p^\tau + \tau q^\tau, \quad p^\tau \in G_E, \quad q^\tau \in A. 
\label{DAEP1D_decomp_ntau} 
\end{eqnarray}
We have:

\begin{proposition}
(i) $n^\tau$ is the solution of (\ref{DAEP1D_vf}) if and only if the pair $(p^\tau,q^\tau)$ given by (\ref{DAEP1D_decomp_ntau}) is the solution of the following variational formulation: 
\begin{eqnarray}
& & \hspace{-1cm} 
\mbox{ Find $(p^\tau,q^\tau) \in V \times A$ such that:}
\nonumber \\
& & \hspace{-1cm} 
a_E(q^\tau,v) + (p^\tau + \tau q^\tau,v)  = \langle \Gamma , v \rangle, \quad \forall v \in V
, \label{DAEP1D_vfnew_1} \\
& & \hspace{-1cm} 
a_E(p^\tau,w)  = 0, \quad \forall w \in A
. \label{DAEP1D_vfnew_2} 
\end{eqnarray}
(ii) This variational formulation is well-posed and there exists a constant $C$, independent of $\tau$ and $h$ such that 
\begin{eqnarray}
& & \hspace{-1cm} 
|(p^\tau,q^\tau)|_{V \times V} \leq C |\Gamma|_{V'} . \label{DAEP1D_vfnew_wellposed} 
\end{eqnarray}
(iii) When $\tau \to 0$, $(p^\tau,q^\tau) \to (p^0,q^0)$ and $u^\tau = p^\tau + \tau q^\tau \to p^0$ where $(p^0,q^0)$ is the solution of the following variational formulation: 
\begin{eqnarray}
& & \hspace{-1cm} 
\mbox{ Find $(p^0,q^0) \in V \times A$ such that:}
\nonumber \\
& & \hspace{-1cm} 
a_E(q^0,v) + (p^0 ,v)  = \langle \Gamma , v \rangle, \quad \forall v \in V
, \label{DAEP1D_vfnew_1_tau=0} \\
& & \hspace{-1cm} 
a_\phi(p^0,w)  = 0, \quad \forall w \in A
. \label{DAEP1D_vfnew_2_tau=0} 
\end{eqnarray}
\label{prop_DAEP1D_vfnew}
\end{proposition}

\noindent
{\bf Proof:} (i) We insert the decomposition (\ref{DAEP1D_decomp_ntau}) into (\ref{DAEP1D_vf}) and get $(p^\tau,q^\tau) \in G_E \times A$ such that (\ref{DAEP1D_vfnew_1}) holds for any $v \in V$. Now, using the characterization (\ref{DAEP1D_Gchar}) of $G_E$, we get (\ref{DAEP1D_vfnew_2}). Conversely, let  $(p^\tau,q^\tau) \in V \times A$ be a solution of (\ref{DAEP1D_vfnew_1}), (\ref{DAEP1D_vfnew_2}). Constructing $n^\tau$ according to (\ref{DAEP1D_decomp_ntau}) obviously leads to (\ref{DAEP1D_vf}). 

\noindent
(ii) We refer to \cite{Deg_Del_Loz_Nar_Neg_prep2}. 

\noindent
(iii) is obvious. \endproof

\medskip
\noindent
Another, more elementary view of this proposition is as follows. A given $p^\tau \in G_E$ depends on two arbitrary quantities: $p^\tau_{-M}$ and $p^\tau_{-M+1}$. For a given $p^\tau \in G_E$, we can solve for $q^\tau$ satisfying (\ref{DAEP1D_vfnew_1}). The form $a_0(u,v)$ is coercive on $V$. Indeed, if $a_0(u,u) = 0$, then $u \in G_0 \cap A$ and with (\ref{DAEP1D_G+A}), is such that $u=0$. Additinally, the coercivity constant can be proven uniform with respect to $h$. Therefore, using test sequences $v \in A$, (\ref{DAEP1D_vfnew_1}) can be solved uniquely for $q^\tau \in A$. Then, taking a sequence $v \in V$ such that $v_K = 0$, for all $K \in [-M+2,M]$ leads to two additional linear relations which allow to determine $p^\tau_{-M}$ and $p^\tau_{-M+1}$ uniquely. This procedure allows us to determine $(p^\tau, q^\tau)$ uniquely in a uniform way with respect to both $\tau$ and $h$. 

We now turn ourselves to the multi-dimensional case.

%%%%%%%%%%%%%%%%%%%%%%%%%%%%%%%%%%%%%%%%%%%%%%%%%%%%%%%%%%%%%%%%%%%%%%%%%%%%%%%%%%%%%%%%%%%%%%%%
%%%%%%%%%%%%%%%%%%%%%%%%%%%%%%%%%%%%%%%%%%%%%%%%%%%%%%%%%%%%%%%%%%%%%%%%%%%%%%%%%%%%%%%%%%%%%%%%
\subsection{The discrete strongly anisotropic elliptic problem: the multi-dimen\-sional problem}
\label{subsec_El_stron_aniso_MD_discrete}

We now assume that the multi-index $K$ belongs to a box ${\mathcal K} = \prod_{i=1}^3 [-M_i,M_i]$. The discretization of the Euler-Lorentz model by the FDAP-2 scheme (a similar study could be conducted for the FDAP-1 scheme) leads to a discretization of the form: 
\begin{eqnarray}
& & \hspace{-1cm} 
-  \sum_{i,j=1}^3 \frac{h_j^{-1}}{2} \left\{ b_j|_{K+e_j} \, b_i|_{K+e_j} \left( \frac{ h_i^{-1}}{2} \left[ n|_{K+e_j+e_i}- n|_{K+e_j-e_i} \right] - n|_{K+e_j} E_i|_{K+e_j} \right) \right.  \nonumber \\
& & \hspace{1cm} 
\left. - b_j|_{K-e_j} \,  b_i|_{K-e_j} \left( \frac{ h_i^{-1}}{2} \left[ n|_{K-e_j+e_i} - n|_{K-e_j-e_i} \right] - n|_{K-e_j} E_i|_{K-e_j} \right) \right\}
 \nonumber \\
& & \hspace{8cm} 
+ \tau n|_K = \tau F|_K, 
\label{DAEPMD_n} 
\end{eqnarray}
where we have introduced a linear pressure-density relationships. We also have made $\delta = 1$ and collected all known terms at the right-hand side into the generic term $F|_K$. 

Now, we write (\ref{DAEPMD_n}) in the form of a mixed problem and we highlight the dependence upon $\tau$: 
\begin{eqnarray}
& & \hspace{-1cm} 
-  \sum_{j=1}^3 \frac{h_j^{-1}}{2} \left( b_j|_{K+e_j} \, \gamma^\tau|_{K+e_j} - b_j|_{K-e_j} \,  \gamma^\tau|_{K-e_j}  \right) + \tau n^\tau|_K = \tau F|_K, \quad \forall K \in {\mathcal K}, 
\label{DAEPMD_gam1}  \\
& & \hspace{-1cm} 
\gamma^\tau|_K =   \sum_{i=1}^3  b_i|_{K} \left( \frac{ h_i^{-1}}{2} \left( n^\tau|_{K+e_i}- n^\tau|_{K-e_i} \right) - n^\tau|_{K} E_i|_{K} \right) , \quad  \forall K \in {\mathcal K}_{\mbox{\scriptsize int}} , 
\label{DAEPMD_gam2} 
\end{eqnarray}
where ${\mathcal K}_{\mbox{\scriptsize int}}$ denotes the internal cells ${\mathcal K}_{\mbox{\scriptsize int}} = \prod_{i=1}^3 [-M_i+1,M_i-1]$. To complete this formulation, we need to impose boundary conditions in the last row of cells ($K \in {\mathcal K}_I = \prod_{i=1}^3 [-M_i,M_i] \setminus \prod_{i=1}^3 [-M_i+1,M_i-1]$) and in an additional row of fictitious cells ($K \in {\mathcal K}_F = \prod_{i=1}^3 [-M_i-1,M_i+1] \setminus \prod_{i=1}^3 [-M_i,M_i]$). We impose the following boundary conditions (which are the discrete counterpart of (\ref{IEL_BC_n_pm})): 
\begin{eqnarray}
& & \hspace{-1cm} 
b_{j_K}|_K \, \, \gamma^\tau_{K} = \tau g_K, \quad \forall K \in {\mathcal K}_I \cup {\mathcal K}_F \, \,  \mbox{ and } \, j_K \mbox{ such that } \nu_K = \pm e_{j_K},  
\label{DAEPMD_bc}
\end{eqnarray}
where $g_{K}$ for $K \in {\mathcal K}_I \cup {\mathcal K}_F$ is supposed known. The condition on $j_K$ is that the direction $j_K$ is one of the directions to which the boundary of ${\mathcal K}$ at cell $K$ is normal. 

Multiplying (\ref{DAEPMD_gam1}) by a test sequence $\{ v_K \}_{K \in {\mathcal K}}$, performing a discrete integration-by-parts and using (\ref{DAEPMD_gam2}) together with the boundary conditions (\ref{DAEPMD_bc}), we get the following

\begin{proposition}
The solution $n^\tau_{K}$ of problem (\ref{DAEPMD_gam1}), (\ref{DAEPMD_gam2}) with boundary conditions (\ref{DAEPMD_bc}) satisfies the following discrete variational formulation:
\begin{eqnarray}
& & \hspace{-1cm} 
\mbox{ Find $n^\tau \in V$ such that:}
\nonumber \\
& & \hspace{-1cm} 
a_E (n^\tau,v) + \tau (n^\tau,v)  = \tau \langle \Gamma , v \rangle, \quad \forall v \in V
,
\label{DAEPMD_vf} 
\end{eqnarray}
where $V = \ell^2({\mathcal K})$, the space of square integrable sequences on ${\mathcal K}$, and 
\begin{eqnarray}
& & \hspace{-1cm} 
a_E(u,v) = \sum_{K \in {\mathcal K}_{\mbox{\scriptsize int}}}  
\left[ \sum_{i=1}^3  b_i|_{K} \left( \frac{ h_i^{-1}}{2} \left( u|_{K+e_i}- u|_{K-e_i} \right) - u|_{K} E_i|_{K} \right) \right] \nonumber \\
& & \hspace{5cm} 
\left[ \sum_{j=1}^3  b_j|_{K} \left( \frac{ h_j^{-1}}{2} \left( v|_{K+e_j}- v|_{K-e_j} \right) \right) \right] 
, \label{DAEPMD_aE} \\
& & \hspace{-1cm} 
(u,v) = \sum_{K \in {\mathcal K}} u_k v_k  , \label{DAEPMD_()} \\
& & \hspace{-1cm} 
\langle \Gamma , v \rangle = (F,v) + \sum_{K \in {\mathcal K}_I \cup {\mathcal K}_F} \varepsilon_{j_K} \frac{h_j^{-1}}{2}  g_K v_K , \label{DAEPMD_Gamma} 
\end{eqnarray}
where in the last term defining $\langle \Gamma , v \rangle$, $\varepsilon_{j_K}$ is defined by $ \nu_K = \varepsilon_{j_K} e_{j_K}$.
\label{prop_DAEPMD_vf}
\end{proposition}

We again note that $a_E(u,u) + \tau (u,u)$ is coercive on $V$. However, we have no better estimate than  (\ref{DAEP1D_coerc}), which implies that the condition number of formulation (\ref{DAEPMD_vf}) tends to $\infty$ as $\tau \to 0$. Therefore, this formulation cannot be used for solving for $n^\tau$ when $\tau$ is very small. We then introduce a new formulation in the spirit of what has been done in the one-dimensional case. To this aim, we assume that there is a space direction $i_0$ such that $b_{i_0}|_K \not = 0$ for all cells $K \in {\mathcal K}$. This condition can be removed at the expense of some additional work which will be detailed in future work. We suppose that $i_0 = 3$ for simplicity. We also define the set ${\mathcal K}_G$ by:
\begin{eqnarray}
& & \hspace{-1.5cm} 
{\mathcal K}_G = \prod_{i=1}^3 [-M_i,M_i] \, \setminus \, \prod_{i=1}^2 [-M_i+1,M_i-1] \times [-M_3+2,M_3]
. 
\label{DAEPMD_KB} 
\end{eqnarray}
We note that in the 3rd direction (corresponding to $i_0$ in the general case), the boundary cells are shifted and two layers of boundary cells are defined on the left-hand side and no boundary cells on the right-hand side. The complement ${\mathcal K} \setminus {\mathcal K}_G = \prod_{i=1}^2 [-M_i+1,M_i-1] \times [-M_3+2,M_3]$ is denoted by ${\mathcal K}_A$.

\begin{definition}
We define the spaces $G_E$ and $A$ as follows: 
\begin{eqnarray}
& & \hspace{-1cm} 
G_E = \left\{ v \in V \, \,  \left| \, \, \, \, \sum_{i=1}^3  b_i|_{K} \left(  (2h_i)^{-1} \left( u|_{K+e_i}- u|_{K-e_i} \right) - u|_{K} E_i|_{K} \right) = 0 \right. \right. , \nonumber \\
& & \hspace{9.5cm} \left. \phantom{\sum_{i=1}^3}
 \, \forall K \in {\mathcal K}_{\mbox{\scriptsize int}}  \right\}, \label{DAEPMD_GE} \\
& & \hspace{-1cm} 
A = \{ v \in V \, \, | \, \, v|_K = 0, \forall K \in {\mathcal K}_G \}. 
\label{DAEPMD_A} 
\end{eqnarray}
\label{def_DAEPMD_GA}
\end{definition}

\begin{proposition}
(i) We have: 
\begin{eqnarray}
& & \hspace{-1cm} 
V = G_E \oplus A. 
\label{DAEPMD_G+A} 
\end{eqnarray}
(ii) $G_E$ is characterized by 
\begin{eqnarray}
& & \hspace{-1cm} 
u \in G_E \quad \Longleftrightarrow \quad u \in V \quad \mbox{ such that } \quad a_E(u,v) = 0, \quad \forall v \in A. 
\label{DAEPMD_Gchar} 
\end{eqnarray}
\label{prop_DAEPMD_G+A}
\end{proposition}

\medskip
\noindent
{\bf Proof:} (i) We note that, for the elements $p$ of $G_E$, the components $p|_K$ for $K \in {\mathcal K}_G$ are free.  Similarly, for any element $q \in A$, the components $q|_K \in {\mathcal K}_A$ are free. Now, Let $v \in G_E \cap A$. Because $v \in A$, $v|_K$ is identically zero on ${\mathcal K}_G$. Then it is easy to see that the condition that $v \in G_E$, owing to the fact that $b_3|_K \not = 0$, allows to determine $v$ recursively in the layers corresponding to $K_3 = $ Constant, starting from the layer $K_3 = -M_3+2$ up to the layer $K_3 = M_3$. And this recursive calculation shows that $v|_K$ is identically zero on ${\mathcal K}$. This shows that $G_E \cap A = \emptyset$. Now, we consider $v \in V$. Using the same recursive procedure as above, we can define an element $p \in G_E$ such that $p|_K = v|_K$ for all $K \in  {\mathcal K}_G$. Once $p$ is found, $q = v - p$ is defined and obviously belongs to $A$, showing that $V=G_E+A$. 

\medskip
\noindent
(ii): that $u \in G \, \Longrightarrow \, a_E(u,v) = 0, \, \forall v \in A$ is obvious (it is even true for all $v \in V$). Conversely, suppose that $u \in V$ such that $a_E(u,v) = 0, \, \forall v \in A$. Let $\psi_K$ for $K \in {\mathcal K}_{\mbox{\scriptsize int}}$ be arbitrary. There exists $v \in A$ such that 
\begin{eqnarray}
& & \hspace{-1cm} 
\psi_K = \sum_{j=1}^3  b_j|_{K} \left( \frac{ h_j^{-1}}{2} \left( v|_{K+e_j}- v|_{K-e_j} \right) \right). 
\label{DAEPMD_construct_v_from_psi} 
\end{eqnarray}
Indeed, $v_K$ is defined by the same recursive procedure as above, starting from the zero values of $v|_K$ for $K \in {\mathcal K}_G$. Additionally, the so-constructed $v$ is the unique $v \in A$ satisfying  property (\ref{DAEPMD_construct_v_from_psi}). Inserting (\ref{DAEPMD_construct_v_from_psi}) into the expression (\ref{DAEPMD_aE}) of $a_E$ and using that  $a_E(u,v) = 0$, we are led to the following relation: 
\begin{eqnarray*}
\sum_{K \in {\mathcal K}_{\mbox{\scriptsize int}}}  
\left[ \sum_{i=1}^3  b_i|_{K} \left( \frac{ h_i^{-1}}{2} \left( u|_{K+e_i}- u|_{K-e_i} \right) - u|_{K} E_i|_{K} \right) \right] \psi_K = 0.
\end{eqnarray*}
This clearly implies that $u \in G_E$ and ends the proof of the proposition. \endproof

We now decompose the solution $n^\tau$ of (\ref{DAEPMD_vf}) into
\begin{eqnarray}
& & \hspace{-1cm} 
n^\tau = p^\tau + \tau q^\tau, \quad p^\tau \in G_E, \quad q^\tau \in A. 
\label{DAEPMD_decomp_ntau} 
\end{eqnarray}
We can copy proposition \ref{prop_DAEP1D_vfnew} 'mutatis mutandis': 

\begin{proposition}
(i) $n^\tau$ is the solution of (\ref{DAEPMD_vf}) if and only if the pair $(p^\tau,q^\tau)$ given by (\ref{DAEPMD_decomp_ntau}) is the solution of the following variational formulation: 
\begin{eqnarray}
& & \hspace{-1cm} 
\mbox{ Find $(p^\tau,q^\tau) \in V \times A$ such that:}
\nonumber \\
& & \hspace{-1cm} 
a_E(q^\tau,v) + (p^\tau + \tau q^\tau,v)  = \langle \Gamma , v \rangle, \quad \forall v \in V
, \label{DAEPMD_vfnew_1} \\
& & \hspace{-1cm} 
a_E(p^\tau,w)  = 0, \quad \forall w \in A
. \label{DAEPMD_vfnew_2} 
\end{eqnarray}
(ii) This variational formulation is well-posed and there exists a constant $C$, independent of $\tau$ and $h$ such that 
\begin{eqnarray}
& & \hspace{-1cm} 
|(p^\tau,q^\tau)|_{V \times V} \leq C |\Gamma|_{V'} . \label{DAEPMD_vfnew_wellposed} 
\end{eqnarray}
(iii) When $\tau \to 0$, $(p^\tau,q^\tau) \to (p^0,q^0)$ and $u^\tau = p^\tau + \tau q^\tau \to p^0$ where $(p^0,q^0)$ is the solution of the following variational formulation: 
\begin{eqnarray}
& & \hspace{-1cm} 
\mbox{ Find $(p^0,q^0) \in V \times A$ such that:}
\nonumber \\
& & \hspace{-1cm} 
a_E(q^0,v) + (p^0 ,v)  = \langle \Gamma , v \rangle, \quad \forall v \in V
, \label{DAEPMD_vfnew_1_tau=0} \\
& & \hspace{-1cm} 
a_\phi(p^0,w)  = 0, \quad \forall w \in A
. \label{DAEPMD_vfnew_2_tau=0} 
\end{eqnarray}
\label{prop_DAEPMD_vfnew}
\end{proposition}

This variational formulation has the following interpretation: First, take an arbitrary $p^\tau \in G_E$. Then, $p^\tau$ depends on independent components $p|_K$ for $K$ belonging to ${\mathcal K}_G$. Using the fact that $a_E$ is coercive on $A$ (the proof is similar as in the one-dimensional case), we find a unique $q^\tau \in A$ such that (\ref{DAEPMD_vfnew_1}) holds for any $v \in A$. But $q^\tau$ depends on the chosen element $p^\tau$ of $G$. Then, taking $v$ such that $v|_K \not = 0$ if and only if $K \in {\mathcal K}_G$ gives as many independent relations as needed to fully determine $p^\tau$. 

Variational formulation (\ref{DAEPMD_vfnew_1}), (\ref{DAEPMD_vfnew_2}) leads to a solution methods for the discrete anisotropic elliptic problem with a uniform condition number with respect to $\tau$ when $\tau \ll 1$. This method does not require the numerical determination of the magnetic field lines nor any integration along these lines.

%%%%%%%%%%%%%%%%%%%%%%%%%%%%%%%%%%%%%%%%%%%%%%%%%%%%%%%%%%%%%%%%%%%%%%%%%%%%%%%%%%%%%%%%%%%%%%%%
%%%%%%%%%%%%%%%%%%%%%%%%%%%%%%%%%%%%%%%%%%%%%%%%%%%%%%%%%%%%%%%%%%%%%%%%%%%%%%%%%%%%%%%%%%%%%%%%
\subsection{Strongly anisotropic elliptic problems: conclusion}
\label{subsec_El_stron_aniso_conclu}

The resolution of the Euler-Lorentz model when $\tau \ll 1$ (the limit $\tau \to 0$ corresponding to the so-called drift-fluid limit) leads to a strongly anisotropic discrete elliptic problem. We have seen that this problem degenerates when $\tau \to 0$ and leads to an ill-conditionned numerical resolution. We have proposed a new variational method. This method has a condition number which is independent of $\tau$ when $\tau \ll 1$ and is therefore efficient independently of the value of $\tau$. Additionally, it provides the  correct solution to the limit problem when $\tau  \to 0$. The knowledge of the magnetic field lines is not needed and no integration along these field lines need to be performed. Therefore, this method is particularly suitable to a context where the magnetic field is susceptible to vary with time.

%% file: conclu.tex
%%%%%%%%%%%%%%%%%%%%%%%%%%%%%%%%%%%%%%%%%%%%%%%%%%%%%%%%%%%%%%%%%%%%%%%%%%%%%%%%%%%%%%%%%%%%%%%%
%%%%%%%%%%%%%%%%%%%%%%%%%%%%%%%%%%%%%%%%%%%%%%%%%%%%%%%%%%%%%%%%%%%%%%%%%%%%%%%%%%%%%%%%%%%%%%%%

In these notes, we have described how to construct Asymptotic-Preserving schemes for plasma fluid models in a variety of situations. We have first considered the quasi-neutral limit and applied the methodogology to the Euler-Poisson and to the Euler-Maxwell problems. In a second part, we have focused on the Euler-Lorentz model in the drift-fluid limit which arises when the magnetic field is large and simultaneously the Mach number is small. In all cases, the same methodology has been applied. First, we find a reformulation of the original problem in such a way that it directly appears as a perturbation of the limit problem. Then, we focus on the time discretization by considering time semi-discrete schemes and we determine which terms must be evaluated implicitly in order to make the scheme AP. Once a proper time discretization has been found, we apply it to a fully-discrete version of the scheme. In this last step, issues like conservativity or numerical viscosity can be brought into the framework of AP schemes. 

We have focused on a presentation of the methodologies. We refer to the bibliography given in the introduction for applications to practical cases and performance tests. 

The AP methodology can be applied to a large variety of situations. Of particular interest are cases where where several limits must be taken independently. An important issue for instance in two-fluid models is to treat simultaneously the smallness of the Debye length (quasi-neutral limit) and the smallness of the electron mass (Low Mach-number limit in the electron fluid equations). Another issue is the design of schemes for the Euler-Lorentz model which are AP when both the Debye length and the cyclotron period may tend to zero independently. 

Other open problems concern the stability analysis of the schemes in the nonlinear settings, as well as the obtention of rigorous error estimates.